\documentclass[a4paper,fleqn,usenatbib]{mnras}

\usepackage{newtxtext,newtxmath}

\usepackage[T1]{fontenc}
\usepackage{ae,aecompl}
\usepackage{tabls}
\usepackage{lscape}
\usepackage{amsmath}

\usepackage[final]{changes}
\colorlet{Changes@Color}{red}

\usepackage{array}
\newcolumntype{H}{>{\setbox0=\hbox\bgroup}c<{\egroup}@{}}

\usepackage{graphicx}	
\usepackage{amsmath}	
\usepackage{amssymb}	
\usepackage{array}
\usepackage{mdwmath}
\usepackage{mdwtab}
\usepackage[caption=false,font=footnotesize]{subfig}
\usepackage{longtable}

\title[Variability in Low-Gravity Brown Dwarfs]{A Search for Variability in Exoplanet Analogues and Low-Gravity Brown Dwarfs}

\author[Vos et al.]{
Johanna M. Vos$^{1,2,3}$\thanks{E-mail: jvos@amnh.org },
Beth A. Biller$^{2,3}$,
Mariangela Bonavita$^{2,3}$,
Simon Eriksson$^{4}$,
\newauthor Michael C. Liu$^{5}$,
William M. J. Best$^{5}$,
Stanimir Metchev$^{6}$,
Jacqueline Radigan$^{7}$,
\newauthor Katelyn N. Allers$^{8}$,
Markus Janson$^{4}$,
Esther Buenzli$^{9}$,
Trent J. Dupuy$^{10}$,
\newauthor Micka\"el Bonnefoy$^{11}$,
Elena Manjavacas$^{12}$,
Wolfgang Brandner$^{13}$,
Ian Crossfield$^{14}$,
\newauthor Niall Deacon$^{13,15}$,
Thomas Henning$^{13}$,
Derek Homeier$^{16}$,
\newauthor Taisiya Kopytova$^{17}$
and Joshua Schlieder$^{18}$\\
$^{1}$ American Museum of Natural History Department of Astrophysics, Central Park West at 79th Street, New York, NY 10034, USA\\
$^{2}$SUPA, Institute for Astronomy, University of Edinburgh, Royal Observatory, Edinburgh EH9 3HJ, UK\\
$^{3}$Centre for Exoplanet Science, University of Edinburgh, Edinburgh EH9 3HJ, UK\\
$^{4}$Department of Astronomy, Stockholm University, AlbaNova University Center, SE-106 91 Stockholm, Sweden\\
$^{5}$Institute for Astronomy, University of Hawaii at Manoa, Honolulu, HI 96822, USA\\
$^{6}$Department of Physics \& Astronomy and Centre for Planetary Science and Exploration, The University of Western Ontario, London, Ontario N6A 3K7, Canada \\
$^{7}$Utah Valley University, 800 West University Parkway, Orem, UT 84058, USA\\
$^{8}$Department of Physics and Astronomy, Bucknell University, Lewisburg, PA 17837, USA\\
$^{9}$Institute for Astronomy, ETH Zurich, Wolfgang-Pauli-Strasse 27, CH-8093 Zurich, Switzerland\\
$^{10}$Gemini Observatory, Northern Operations Center, 670 N. A'ohoku Place, Hilo, HI 96720 USA\\
$^{11}$Univ. Grenoble Alpes, IPAG; CNRS, IPAG, 38000 Grenoble, France\\
$^{12}$Department of Astronomy/Steward Observatory, The University of Arizona, 933 N. Cherry Avenue, Tucson, AZ 85721, USA\\
$^{13}$Max-Planck-Institut für Astronomie, Königstuhl 17, D-69117 Heidelberg, Germany\\
$^{14}$Department of Physics, Massachusetts Institute of Technology, 77 Massachusetts Avenue, Cambridge, MA 02139, USA\\
$^{15}$Centre for Astrophysics Research, University of Hertfordshire, College Lane, Hatfield, UK\\
$^{16}$Zentrum für Astronomie der Universität Heidelberg, Landessternwarte, Königstuhl 12, D-69117 Heidelberg, Germany\\
$^{17}$School of Earth \& Space Exploration, Arizona State University, Tempe AZ 85287, USA\\ 
$^{18}$Exoplanets and Stellar Astrophysics Laboratory, Code 667, NASA Goddard Space Flight Center, Greenbelt, MD, USA\\
}

\date{Accepted XXX. Received YYY; in original form ZZZ}

\pubyear{2018}

\begin{document}
\label{firstpage}
\pagerange{\pageref{firstpage}--\pageref{lastpage}}
\maketitle

\begin{abstract}
We report the results of a $J$-band survey for photometric variability in a sample of young, low-gravity objects using the New Technology Telescope (NTT) and the United Kingdom InfraRed Telescope (UKIRT). Surface gravity is a key parameter in the atmospheric properties of brown dwarfs and this is the first large survey that aims to test the gravity dependence of variability properties. We do a full analysis of the spectral signatures of youth and assess the group membership probability of each target using membership tools from the literature. This results in a 30 object sample of young low-gravity brown dwarfs.
Since we are lacking in objects with spectral types later than L9, we focus our statistical analysis on the L0-L8.5 objects. We find that the variability occurrence rate of L0-L8.5 low-gravity brown dwarfs in this survey is $30^{+16}_{-8}\%$. We reanalyse the results of \citet{Radigan2014a} and find that the field dwarfs with spectral types L0-L8.5 have a variability occurrence rate of $11^{+13}_{-4}\%$. 
\added{We determine a probability of $98\%$ that the samples are drawn from different distributions.}
This is the first quantitative indication that the low-gravity objects are more likely to be variable than the field dwarf population. Furthermore, we present follow-up $J_S$ and $K_S$ observations of the young, planetary-mass variable object PSO 318.5--22 over three consecutive nights. We find no evidence of phase shifts between the $J_S$ and $K_S$ bands and find higher $J_S$ amplitudes. We use the $J_S$ lightcurves to measure a rotational period of  $8.45\pm0.05~$hr for PSO 318.5--22.
\end{abstract}




\section{Introduction}

Time-resolved photometric variability monitoring is a key probe of atmospheric inhomogeneities in brown dwarf atmospheres, as it is sensitive to the spatial distribution of condensates as a brown dwarf rotates. Photometric variability has been well-studied in the more massive field L and T spectral type dwarfs, but the variability properties of the population of younger, low-gravity objects are less understood.

\citet{Radigan2014} reported the results of a large, ground-based search for $J$-band variability in L and T dwarfs, finding that 9 out of 57 ($16\%$) objects showed significant variability above photometric noise. Furthermore, the authors report enhanced variability frequency and amplitudes at the L/T transition, supporting the hypothesis that cloud holes contribute to the abrupt decline in condensate opacity and $J$-band brightening observed at the L/T transition \citep[however subsequent variability studies have shown that varying cloud layers, as opposed to holes, are responsible for observed variability;][]{Apai2013, Buenzli2015a}.
\citet{Wilson2014} presented a similar ground-based variability survey, the \textit{Brown Dwarf Atmospheric Monitoring} (BAM) survey, which monitored 69 brown dwarfs spanning L0 to T8.
Significant variability was reported in 14 of 69 objects ($20\%$), with no evidence for an enhancement in frequency or amplitude across the L/T transition. However, \citet{Radigan2014a} carried out a reanalysis of the the 13 highly variable objects reported by \citet{Wilson2014} and found significant variability in only 4 from 13. Combining the revised BAM survey with the \citet{Radigan2014} survey, \citet{Radigan2014a} found that $24^{+11}_{-9}\%$ of objects in the L9-T3.5 range exhibit $J$-band variability, in contrast to $2.9^{+4.1}_{-2.1}\%$ of L0-L8.5 brown dwarfs and $3.2^{+4.4}_{-2.3}\%$ of T4-T9.5 brown dwarfs.

\citet{Buenzli2014} presented a 22 target \textit{HST} grism spectroscopy survey at wavelengths of $1.1-1.7~\mu$m, attaining point-to-point precision of $0.1-0.2\%$ during $\sim40~$min observations. Low-level ($\sim1\%$) variability trends were detected in 6 brown dwarfs ($27\%$), with no evidence for enhanced frequency across the L/T transition, suggesting that low-level heterogeneities are a frequent characteristic of brown dwarf atmospheres across the entire L-T spectral range.
\citet{Metchev2015a} reported results from a \textit{Spitzer} program to search for photometric variability in a larger sample of 44 L3-T8 dwarfs at $3.6 ~\mu$m and $4.5 ~\mu$m, reaching $0.2 - 0.4\%$ precision. 
\citet{Metchev2015a} reach a similar conclusion, finding that photometric variability is common among L and T dwarfs. 
The survey included eight low or intermediate gravity brown dwarfs to probe the effects of low surface gravity on the variability properties of brown dwarfs. A tentative correlation was found between low-gravity and high amplitude variability, however a larger sample is necessary to confirm this potential relation \citep{Metchev2015a}.

For the majority of directly-imaged exoplanets, the contrast between host star and planet make it difficult to obtain sufficiently high S/N photometry to allow detailed studies of their variability, thus only a handful are amenable to variability studies. In fact, \citet{Apai2016} explored the rotational variability of the HR8799 planets, reaching a photometric precision of $\sim 10\% $, thus insufficient to detect variability on levels of a few percent. However, young brown dwarfs provide an excellent analogue to directly-imaged exoplanets. Recently, a handful of young brown dwarfs with colours and magnitudes similar to directly-imaged planets have been discovered \citep[see compilation of young objects made by ][]{Faherty2016, Liu2016}. The atmospheres of these young brown dwarfs can provide insight into the atmospheres of directly-imaged planets. 
Like their higher-mass brown dwarf counterparts \citep{ZapateroOsorio2006}, young companion exoplanets and free-floating objects appear to be fast rotators with measured rotational periods of $\sim7-11$ hours \citep{Snellen2014, Biller2015,Allers2016,Zhou2016}.  This makes them excellent targets for photometric variability monitoring.

Variability has now been detected in a small sample of low-gravity objects.
As part of this survey, variability was detected in the planetary-mass object PSO J318.5338--22.8603 \citep[PSO 318.5--22;][]{Biller2015}. With a variability amplitude  of $7-10\%$, PSO 318.5--22 displays a very high variability amplitude compared to most objects in the field population.
This was swiftly followed by a variability detection in the $3~M_\mathrm{Jup}$ companion 2MASSW J1207334--393254 (2M1207b), which displayed $\sim1.36\%$ variability in the F125W filter during a $9~$hr observation with \textit{HST} \citep{Zhou2016}. The $19~M_\mathrm{Jup}$ object WISEP J004701.06+680352.1 (W0047) was found to exhibit $\sim8\%$ variability during a $9~$hr \textit{HST} observation \citep{Lew2016}. \citet{Vos2018},  reported results from a \textit{Spitzer} program to monitor variability on the intermediate gravity late-L dwarfs W0047 and 2MASS J2244316+204343 (2M2244) and the planetary-mass T5.5 object SDSS 111010+011613 (SDSS1110). W0047 and 2M2244 were both found to be variable in the mid-IR, with fairly high amplitudes compared to the sample of higher-mass field dwarfs that have been studied. 
There is also tentative evidence that the low-gravity T dwarfs exhibit higher variability amplitudes compared to field objects. \citet{Gagne2017} find that the highly variable object SIMP0136 is a likely member of the $\sim200$ Myr Carina-Near moving group. \citet{Gagne2018a} confirm the variable object 2M1324 as a member of the AB Doradus moving group and estimate a mass of $11-12~M_{\mathrm{Jup}}$.
\citet{Naud2017} obtained three $5-6~$hr epochs of variability monitoring observations of the young T-type companion GU Psc b in AB Doradus. The authors detect marginal variability in one epoch but do not detect significant variability in the other two epochs. The high amplitudes observed in this small sample of low-gravity variable objects adds to the growing evidence that there is a link between low-gravity and enhanced variability.

Here we present the results of the first  photometric monitoring survey of young, low-gravity L and T dwarfs, with the goal of investigating the gravity dependence of variability properties. Observations were carried out at the {$3.5~$m New Technology Telescope} (NTT) and the $3.8~$m UK Infrared Telescope (UKIRT).

\section{Sample Selection}

From Autumn 2014 to Spring 2017 we observed a sample of $36$ brown dwarfs that are candidate members of young moving groups in the literature and/or show signatures of youth in their spectra. Our survey targets are primarily sourced from the BANYAN catalogues \citep{Gagne2014a, Gagne2015c} and \citet{Best2015}. We additionally include the wide companions HN Peg B and GU Psc b \citep{Luhman2007,Naud2014a}. The full survey sample is shown in Table \ref{tab:sampleprops}. 
We consider the following young moving groups in this paper:
TW Hydra \citep[TWA, $10\pm3~$Myr;][]{Bell2015},
$\beta$ Pictoris \citep[$\beta$ Pic, $22\pm6~$Myr;][]{Shkolnik2017},
Columba \citep[Col, $42^{+6}_{-4}~$Myr;][]{Bell2015},
Tucana-Horologium \citep[THA, $45\pm4~$Myr;][]{Bell2015},
Carina \citep[Car, $45^{+11}_{-7}~$Myr;][]{Bell2015}, 
Argus \citep[Arg, $30-50~$Myr;][]{Torres2008},
AB Doradus \citep[AB Dor, $110-150~$Myr;][]{Barenfeld2013, Luhman2007}
and Carina-Near \citep[CarN, $200\pm50~$Myr;][]{Zuckerman2006}. 
Our targets show signs of low-gravity in their spectra and/or are candidate members of nearby young moving groups. We reassess the spectral and kinematic evidence of low-gravity/youth for each object in Section \ref{sec:assess_sample}.

To obtain high signal-to-noise (S/N) measurements that could be robustly compared to previous surveys \citep{Radigan2014,Wilson2014}, targets were limited to objects with magnitudes brighter than $J_{2MASS}=17.0~$mag (apart from one target, GU Psc b). We observed our targets at airmasses $<1.5$ to maximise the S/N.

The sample consists of spectral types L0 and later, as these are less likely to exhibit magnetic spot activity due to the increasingly neutral atmospheres present in objects with $T_\mathrm{eff}$ below $\sim2100~$K. \citet{Gelino2002} and \citet{Miles-Paez2017} find no correlation between magnetic activity (in the form of H$\alpha$ emission) and photometric variability in a sample of L and T dwarfs. We attempt to cover the entire L-T spectral range uniformly, however few young T dwarfs sufficiently bright for ground-based IR photometric monitoring are known, preventing us from fully covering the T spectral type.
Thus our sample is predominantly comprised of L-type objects.

There is only one known binary in our sample, {2MASS J03572695--4417305} \citep{Bouy2003}. The binary separation ($\approx0.1\arcsec$) is less than the seeing so the photometry in this study records the combined flux from both components. The variability of one component in an unresolved binary will be diluted by flux from the non-variable component, making it more difficult to detect the variability. Alternatively, if both components of the binary are variable \citep[as is the case for the Luhman 16AB binary system;][]{Biller2013, Buenzli2015a}, their differing variability amplitudes and rotational periods will be combined in the observed lightcurve, likely resulting in a complex and/or rapidly evolving lightcurve.

\section{Observations and Data Reduction}
\subsection{NTT SofI}\label{reduc1}
The observations took place between October 2014 and March 2017 with the SofI (Son of Isaac) instrument, mounted on the 3.6 m New Technology Telescope (NTT) at La Silla Observatory.  Observations were carried
out in large field imaging mode, which has a pixel scale of $0.288''$ and a $4.92'\times4.92'$
field of view. Targets were observed using the $J_{S}$ band ($1.16-1.32~\mu$m). The $J_{S}$
filter was chosen as it avoids contamination from the water band at $1.4~\mu$m.
Two targets were observed each night, alternating between nods in an ABBA pattern, with 3 exposures
at each position. At each nod we ensured the target was accurately placed on the same 
original pixel in order to preserve photometric precision. $2-5~$hr observations were obtained for each target. The flux of the target 
was kept below 10,000 ADU to prevent any non-linearity effects.

The data reduction steps are outlined in the SofI manual, and an IRAF pipeline was provided by ESO. We processed our images using both the standard IRAF routine as well as an IDL 
version. Here we detail our data reduction process.

\begin{table}
\centering
\caption{Observing log}
\label{my-label}
\renewcommand{\arraystretch}{1.0}
\begin{tabular}{llllll}
\hline \hline
Target      		& Telescope  			& Band 	& Date       		& $\Delta t$ & FWHM \\
\hline
2M0001+15   	& UKIRT 				& $J$    	& 2016-10-12 		& 3.75  	& 1.01        \\
2M0045+16   	& NTT	  			& $J_S$  	 & 2014-11-11 		& 4.05  	& 1.30         \\
2M0045+16   	& NTT	    			& $J_S$  	& 2015-08-17 		& 3.36  	& 0.56        \\
2M0045+16   	& UKIRT 				& $J$    	& 2016-11-13 		& 4.36  	& 0.94        \\
2M0103+19   	& NTT	   			& $J_S$   	& 2014-11-03 		& 5.28  	& 0.40         \\
GU Psc b    	& NTT	   			& $J_S$   	& 2014-10-11 		& 3.46  &	 0.83        \\

2M0117--34   	& NTT	   			& $J_S$  	& 2014-11-08 		& 4.44  	& 0.44        \\ 
2M0117--34   	& NTT	    			& $J_S$  	& 2016-10-18 		& 1.92  	& 2.26        \\
2M0234--64   	& NTT	   			 & $J_S$   & 2014-11-10 		& 5.59 	 & 0.63        \\
2M0303--73   	& NTT	  			 & $J_S$   & 2014-11-09 		& 5.50   	& 0.49        \\
2M0310--27   	& NTT	  			 &$J_S$  	& 2014-11-08 		& 3.00     	& 0.46        \\
2M0323--46   	& NTT	    			& $J_S$   	& 2014-11-07 		& 5.32  	& 1.07        \\
2M0326--21   	& NTT	   		 	& $J_S$  	& 2014-11-04 		& 4.68  	& 0.51        \\
2M0342--68  	 & NTT	   			 & $J_S$   & 2014-11-03 		& 2.88  	& 0.44        \\
PSO 057+15 	 & UKIRT 				& $J$    	& 2016-12-23 		& 3.59  	& 1.26        \\
2M0355+11   	& NTT	    			& $J_S$   	& 2014-10-07 		& 4.73  	& 0.83      \\
2M0357--44   	& NTT	    			& $J_S$   & 2014-10-10 		& 4.13  	& 1.10         \\
2M0418--45   	& NTT	    			& $J_S$   	& 2017-03-14 		& 2.11  	& 0.68        \\
2M0421--63   	& NTT	   			&$J_S$ 	& 2014-10-08 		& 5.50   	&  1.01           \\
PSO071.8--12 	& NTT	    			& $J_S$  	& 2017-10-18 		& 3.31  	& 1.71        \\
PSO071.8--12 	& UKIRT 		 		& $J$   		&2017-12-08		&4.29	& 1.16	     \\
2M0501--00   	& NTT	    			& $J_S$   	& 2014-11-11 		& 4.03  	& 1.06        \\
2M0501--00   	& NTT	    			& $J_S$   	& 2015-08-16 		& 2.01  	& 2.39        \\
2M0501--00   	& NTT	   			& $J_S$   	& 2016-10-19 		& 4.99  	& 1.49        \\
2M0501--00   	& NTT	    			&$J_S$   	& 2017-03-12 		& 1.85  	& 0.43        \\
2M0512 --27     	& NTT	  	 		& $J_S$   	& 2017-03-13 		& 3.00     	& 0.40         \\
2M0518--27   	& NTT	    			& $J_S$   	& 2014-11-05 		& 3.98  	& 1.52        \\
2M0536--19   	& NTT	    			& $J_S$   	& 2014-10-11 		& 2.88  	& 0.90         \\
SDSS1110+01 & NTT	    			& $J_S$   	& 2017-03-12 		& 5.40 	 & 0.37        \\
2M1207--39   	& NTT	   			& $J_S$  	& 2017-03-13 		& 4.49  	& 0.34        \\
2M1256--27   	& NTT	    			& $J_S$   	& 2017-03-14 		& 2.54 	 & 0.40         \\
2M1425--36   	& NTT	   			&$J_S$   	& 2015-08-17 		& 2.52  	& 0.42    \\
2M1425--36   	& NTT	    			&$J_S$   	& 2017-03-14 		& 4.10   	& 0.39        \\
2M1615+49  	 & UKIRT 				& $J$    	& 2016-07-10 		& 4.33  	& 1.14        \\
W1741      	 & NTT	    			& $J_S$   	& 2014-10-11 		& 2.37  	&0.67             \\
PSO 272.4-04   & NTT	    			& $J_S$   	& 2017-03-12 		& 2.28  	& 0.34        \\
2M2002--05   	& UKIRT 				&$ J$    	& 2016-07-09 		& 4.37  	& 1.22        \\
2M2011--05   	& NTT	    			& $J_S$  	& 2015-08-15 		& 3.43  	& 0.48        \\
SIMP J2154  	& NTT	    			&$J_S$  	& 2014-11-07 		& 3.44  	& 1.12        \\
HN Peg B    	& NTT	    			&$J_S$   	& 2014-10-08 		& 3.88  	& 1.17        \\
HN Peg B    	& NTT	    			& $J_S$   	& 2015-08-17 		& 1.42	 & 0.58        \\
HN Peg B    	& UKIRT 				& $J$    	& 2016-07-11 		& 4.97  	& 1.00           \\
HN Peg B    	& UKIRT 				& $J$    	& 2016-07-13 		& 5.01  	& 1.00           \\
PSO J318-22 	& NTT	  			& $J_S$   	& 2014-10-09 		& 5.13  	& 0.48        \\
PSO J318--22 & NTT  				& $J_S$   	& 2014-11-09 		& 2.83  	& 0.42        \\
PSO J318--22 & NTT 				& $K_S$   	& 2014-11-10 		& 3.10  	& 0.52        \\
PSO J318--22 & NTT     				&$J_S$  	& 2015-08-16 		& 4.99  	& 0.38        \\
PSO J318--22 & NTT		   			& $J_S$   	& 2016-08-09 		& 9.12  	& 1.26        \\
PSO J318--22 & NTT	    				& $K_S$ 	& 2016-08-10 		& 9.65  	& 1.44        \\
PSO J318--22 & NTT	    				& $J_S$  	& 2016-08-11 		& 10.46 	& 1.37        \\
PSO J318--22 & NTT		    			& $K_S$   & 2016-08-11 		& 9.84  	& 1.22        \\
2M2244+20   & UKIRT				& $J$    	& 2016-07-21 		& 4.10   	& 1.13 \\
2M2322--61   	& NTT	   			& $J_S$   	& 2014-10-10 		& 4.37  	& 1.28        \\

 \hline 
     
\end{tabular}
\end{table}

\subsubsection{Inter-quadrant Row Crosstalk}\label{sec:reducstep1}
The SofI detector suffers from inter-quadrant row crosstalk, where a bright target imaged in one
quadrant can cause a faint glow in equivalent rows of the other quadrants. The intensity of the 
crosstalk feature scales with the total intensity along a given row by an empirically 
determined value of $1.4\times10^{-5}$ and can be removed easily.

\subsubsection{Flat-fielding}
The shade pattern on the array is a function of the incident flux, so the method of creating flat-fields by subtracting lamp-off
from lamp-on dome flats leaves a residual shade pattern across the centre of the array. For this
reason ``special" dome flats are taken using standard frames along with frames in which the array is
partially obscured to estimate the illumination dependent shade pattern of the array. The shade pattern
can be removed as described by the ESO documentation.

\subsubsection{Illumination Correction}
Illumination correction removes the difference between the illumination pattern of the dome 
flat screen and the sky. This correction is determined from a grid of 16 observations of a standard 
star across the field of view. The illumination correction is created by fitting a 2D surface
to the fluxes of the star after flat-fielding.

\subsubsection{Sky Subtraction}
The sky subtraction of images obtained with SofI serves to remove the dark current as well as 
the illumination-dependent shade pattern. Sky frames are created by median combining normalised 
frames of different nods which are closest in time. These are then re-scaled to the science frame before
being subtracted from the science frame.

\subsubsection{Aperture Photometry}\label{sec:reducsteplast}
The positions of the target star as well as a set of reference stars in the field of view were
found in each frame using IDL {\sc find.pro} followed by {\sc gcntrd.pro} to measure the 
centroids. Aperture photometry was performed on the target as well as the set of reference stars.
Fixed apertures of sizes similar to the median FWHM of all stars on the chip were used. The final aperture was chosen to minimise the photometric noise.

The standard deviation provides a good estimate of noise for a non-variable lightcurve.
However the standard deviation of a variable lightcurve measures both noise and intrinsic variations, and hence overestimates the noise. We estimate the typical photometric error for each lightcurve, $\sigma_\mathrm{pt}$, using a method described by \citet{Radigan2014}.  This is the standard deviation of the lightcurve subtracted from a shifted version of itself, $f_{i+1} - f_{i}$, divided by $\sqrt2$. This quantity is sensitive to high frequency noise in the data and is insensitive to the low frequency trends we expect from variable brown dwarfs. Thus it provides a more accurate estimate of the photometric noise for variable lightcurves.

\subsection{UKIRT WFCAM}
Observations of 8 targets were taken with the infrared Wide-Field Camera \citep[WFCAM;][]{Casali2007}. WFCAM is a wide-field imager on the $3.8~$m UK Infrared Telescope on Mauna Kea, with a pixel scale of $0.4''$. The observations were carried out in the $J$-band. Each target was observed using an ABBA nod pattern, as before. Frames were reduced using the WFCAM reduction pipeline \citep{Irwin2008, Hodgkin2009} by the Cambridge Astronomy Survey Unit. The pipeline reduction steps include linearity correction, dark correction, flat-fielding, gain-correction, decurtaining, defringing, sky subtraction and crosstalk removal \citep{Irwin2004}.
We performed aperture photometry on the target and reference stars in the field of similar brightness, using a range of aperture sizes similar to the median FWHM of all stars in the field.

\subsection{Lightcurve Analysis}\label{sec:lcanalysis}
The raw light curves obtained from aperture photometry display fluctuations in brightness due to 
changing atmospheric transparency, airmass and residual instrumental effects. To a very good approximation
these changes are common to all stars in the field of view and can be removed via division of a 
calibration curve calculated from a set of iteratively chosen, well-behaved reference stars \citep{Radigan2012}.  Firstly, reference stars with peak flux values below 10 or greater than 10,000 ADU were discarded. Different nods were normalised via division by their median flux before being combined to give a relative flux light curve. For each star a calibration curve was created by median combining all other reference stars (excluding that of the target and star in question).
The standard deviation and linear slope for each light curve was calculated and stars with a standard deviation or slope $\sim1.5-3$ times 
greater than that of the target were discarded. This process was iterated a number of times, until a set of 
well-behaved reference stars was chosen. Final detrended lightcurves were obtained by dividing the raw curve for each
star by its calibration curve. Lightcurves shown in this paper have been binned by a factor of $1-3$.

 \begin{figure}
 \subfloat{
\includegraphics[width=0.5\textwidth]{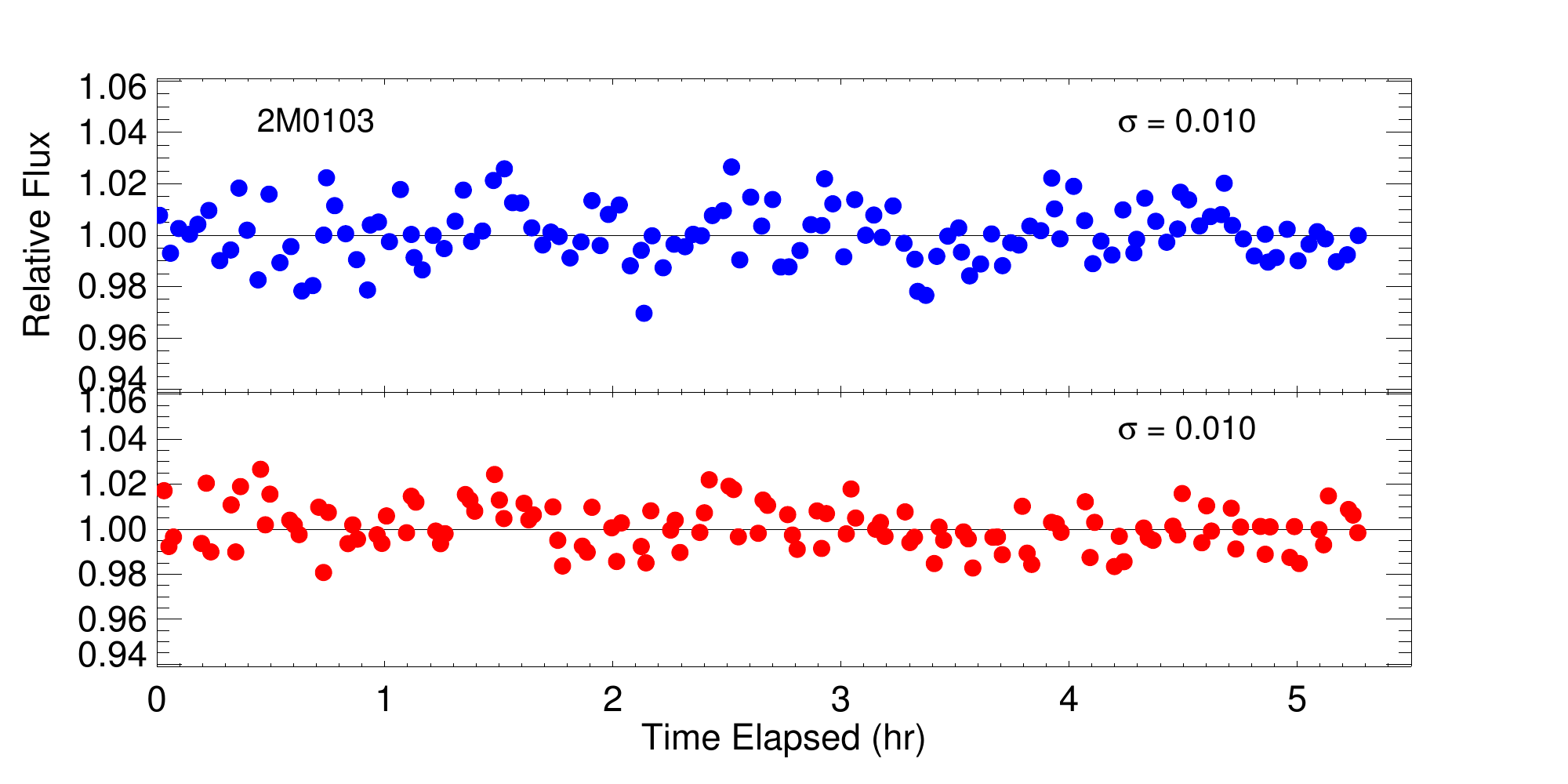}} \\
 \subfloat{
\includegraphics[width=0.5\textwidth]{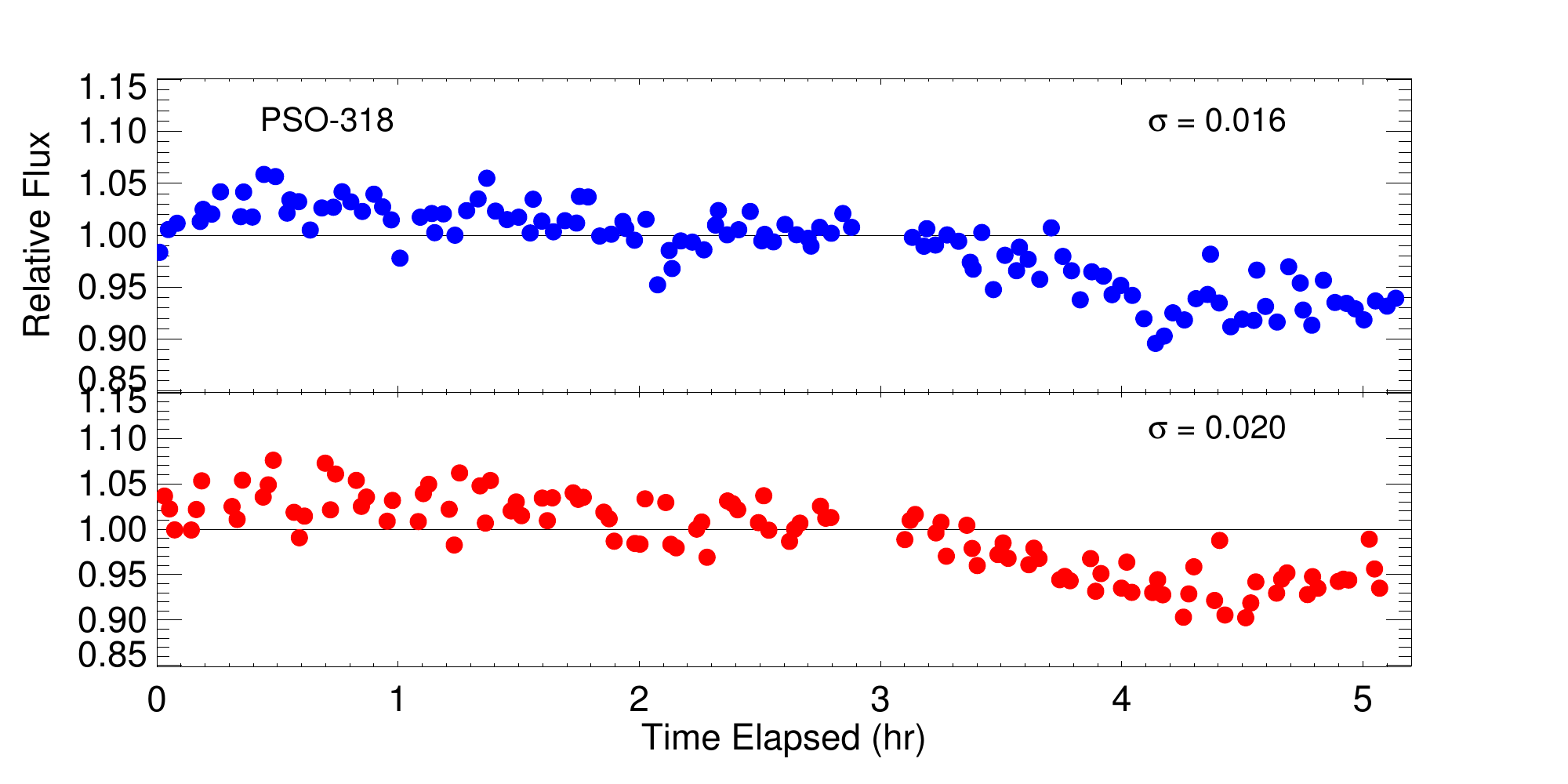}}\\
\caption{Lightcurves of 2M0103 and PSO-318 reduced and analysed using two methods. The blue points show the lightcurve obtained from the method described in Section \ref{reduc1} and the red points show the lightcurve obtained from the independent reduction described in Section \ref{reduc2}. Both lightcurves are binned by a factor of 2. The same lightcurve shape and a similar photometric error $\sigma$ is recovered for each observation.}
\label{fig:SOFI_comparison}
 \end{figure}
 
\subsection{Independent Reduction of NTT/SofI Data} \label{reduc2}
We additionally present the results of an independent 
data reduction process outlined in the MSc thesis of Simon Eriksson and supervisor Markus Janson \citep{Eriksson2016}.
20 of the 21 targets observed with the NTT in 2014 were independently reduced and analysed. A further 10 observations from 2015-2017, mainly follow-ups, were investigated in late 2017 in the same way. Overall, 24 out of 30 NTT targets underwent reduction. The reduction steps previously outlined in Section \ref{sec:reducsteplast} were performed, with the addition of a dark subtraction. The SofI pipeline provided by ESO was used for dark, flat-field and crosstalk corrections and sky subtraction. The process resulted in combined images of two different nods closest in time, and subsequent photometry was obtained using {\sc phot} in IRAF. Errors were estimated from {\sc{phot}} output together with a polynomial fitting to the light curves. 

In Figure \ref{fig:SOFI_comparison} we compare the lightcurves of two objects in our survey that were analysed using both reductions -- the non-variable object 2M0103 (although \citet{Metchev2015a} report low-amplitude mid-IR variability in this object) and the variable object PSO 318.5--22. The blue points show the lightcurve obtained from the method described in Section \ref{reduc1} and the red points show the lightcurve obtained from the independent reduction described above. Both methods produce the same lightcurve shape and a similar photometric error $\sigma$. 
Since the results were consistent between reductions, for the rest of the paper we present lightcurves obtained using the method described in Sections \ref{reduc1} and \ref{sec:lcanalysis}.

\subsection{Identification of Variables}\label{sec:idvar}

 \begin{figure*}
 \subfloat{\label{beta1}
\includegraphics[width=0.52\textwidth]{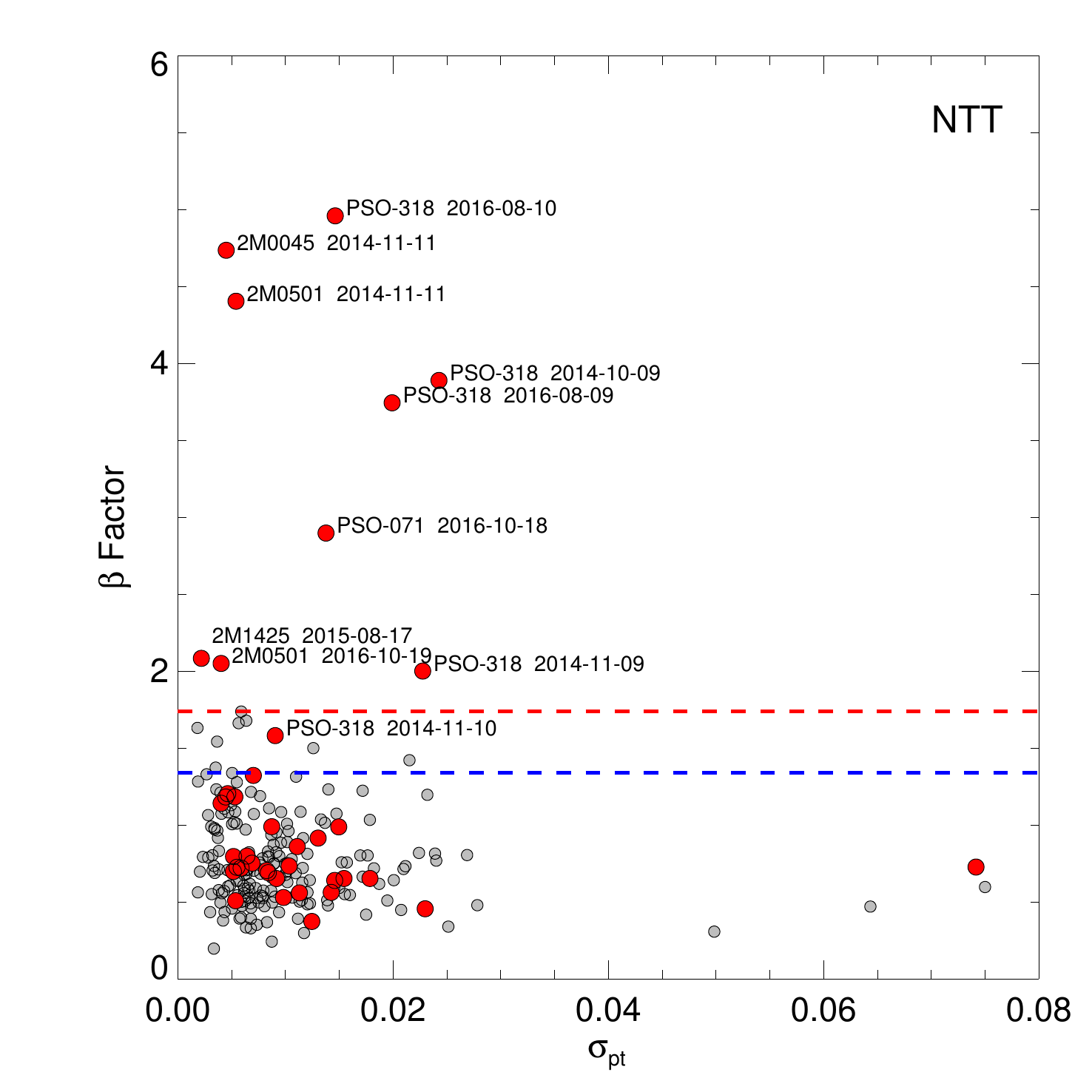}}
\subfloat{\label{beta2}
\includegraphics[width=0.52\textwidth]{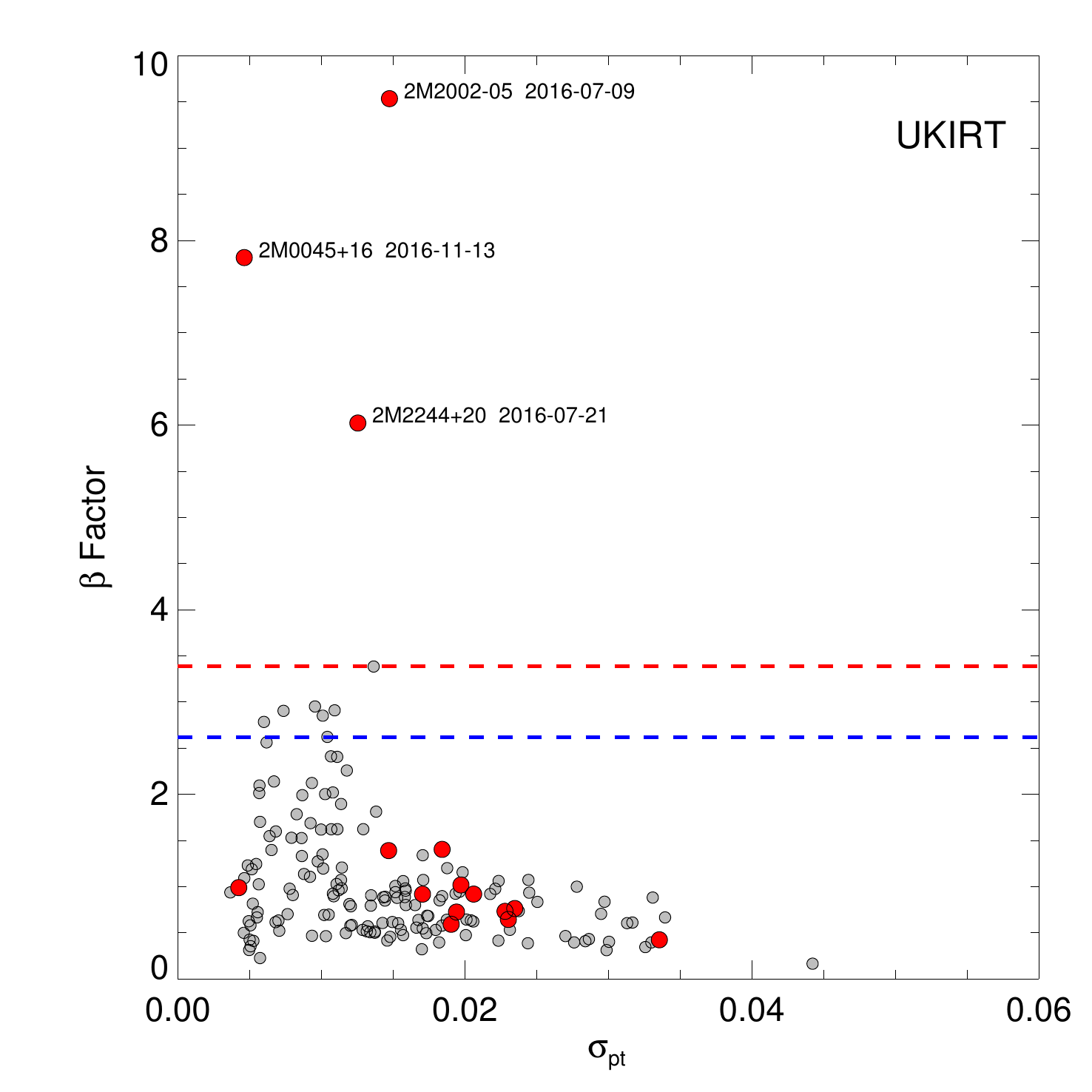}}\\
\caption{$\beta$ factor plotted against the photometric error, $\sigma_{\mathrm{pt}}$ for reference stars (grey circles) and targets (red circles) for the NTT (left) and UKIRT (right) samples. The $\beta$ factor is defined as the periodogram peak power of each reference star divided by $99\%$ significance as calculated from our simulations. The updated, empirical $99\%$ and $95\%$ significance thresholds are shown by the red and blue dashed lines respectively.}
\label{fig:SOFI_beta}
 \subfloat{\label{power1}
\includegraphics[width=0.52\textwidth]{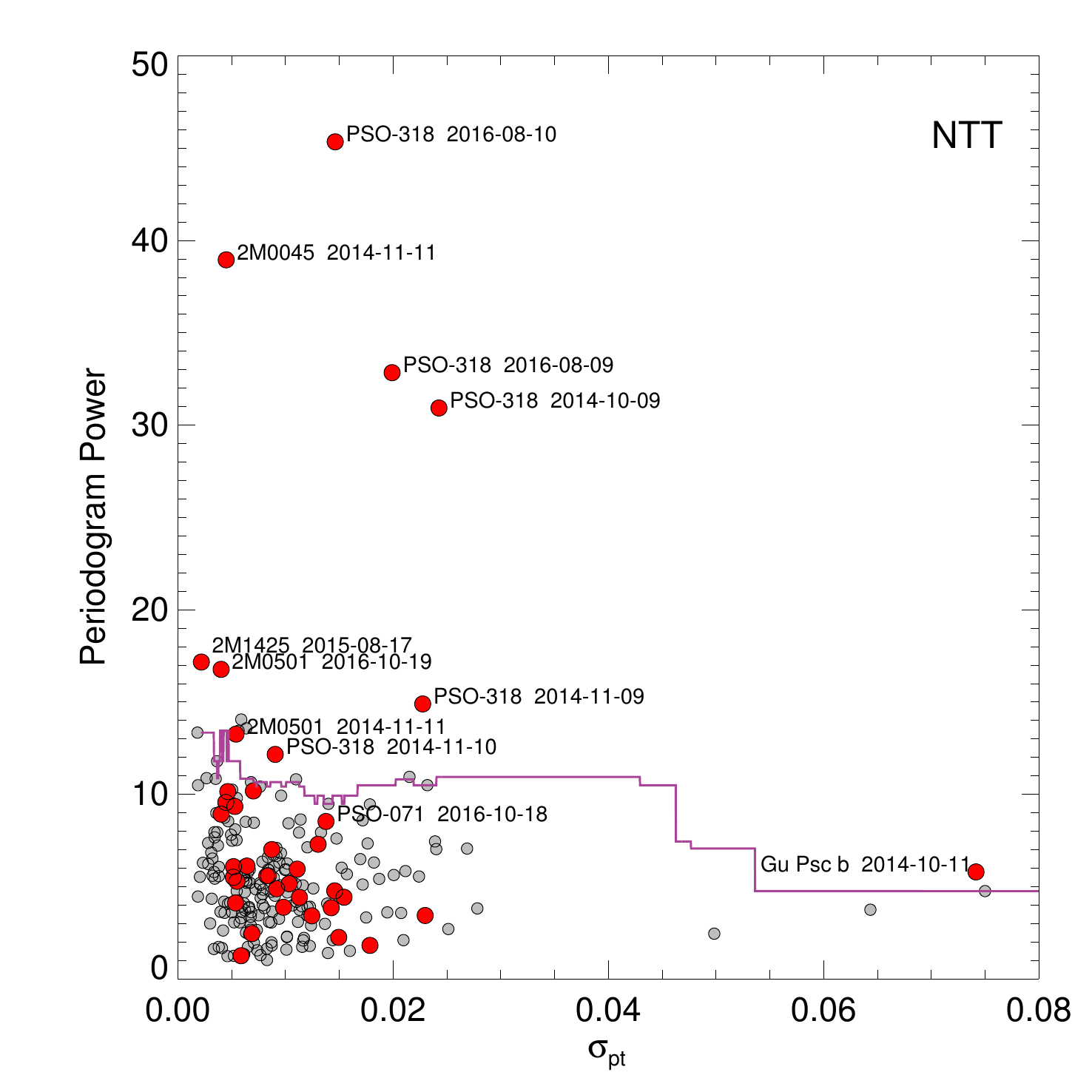}}
\subfloat{\label{power2}
\includegraphics[width=0.52\textwidth]{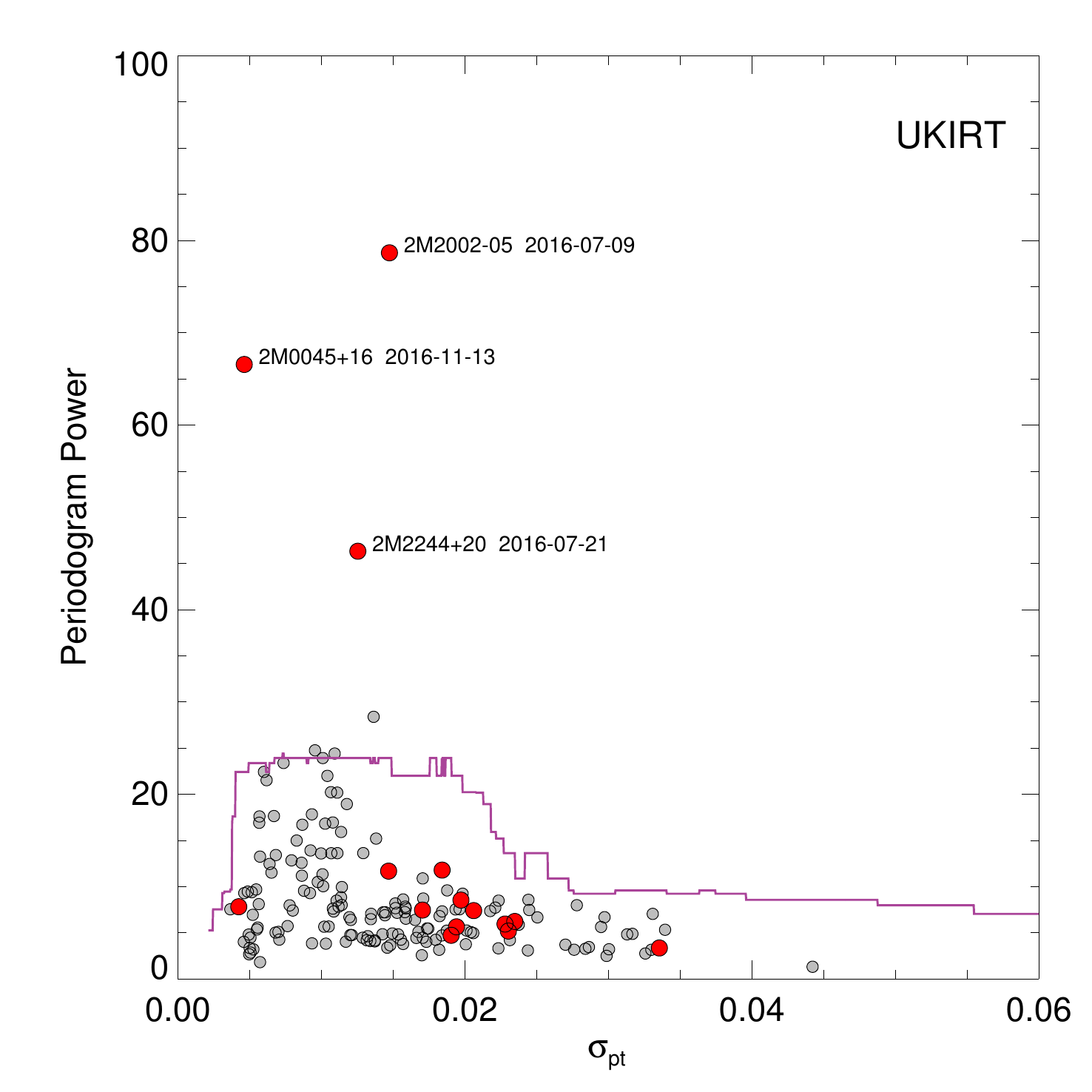}}\\
\caption{Maximum periodogram power plotted against the photometric error, $\sigma_{\mathrm{pt}}$, for reference stars (grey circles) and targets (red circles) for the NTT (left) and UKIRT (right) samples. The $\sigma_\mathrm{pt}$-dependent $95\%$ significance is shown by the purple line.  }
\label{fig:SOFI_power}
\end{figure*}

Variable targets were identified using a method similar to the periodogram analysis outlined in \citet{Vos2018}. The periodograms of each target and its respective reference stars are plotted to identify periodic variability. For each observation, the $1\%$ false-alarm probability (FAP) is calculated from 1000 simulated light curves. These light curves are produced by randomly permuting the indices of the reference star lightcurves \citep{Radigan2014}. The $1\%$ FAP value is the periodogram power above which only $1\%$ of the simulated lightcurves fall. 
This method assumes Gaussian-distributed noise in the reference stars, however to assess the significance of residual correlated noise in the reference star lightcurves we measure the $\beta$ factor of every light curve, which is the peak periodogram power of each reference star divided by the $1\%$ FAP power \citep{Radigan2014}. Figure \ref{fig:SOFI_beta} shows the $\beta$ factor of reference stars and targets for NTT (top) and UKIRT (bottom) observations. We display these separately as each instrument has unique systematics. For reference stars exhibiting Gaussian-distributed noise, we would expect that $1\%$ of reference star peak powers would fall above $\beta=1$, however for both samples more than $1\%$ of reference star peak powers fall above $\beta = 1$, and this is likely due to residual correlated noise in the lightcurves. To account for this excess noise, we find the empirical $1\%$ FAP by finding the $\beta$ factor above which $1\%$ of reference star peak powers fall.  Blue and red dashed lines indicate the new, empirical $95\%$ and $99\%$ significance thresholds. This increases the $99\%$  significance thresholds by a factor of 1.7 and 3.4 for the NTT and UKIRT samples respectively. 

We additionally explore an alternative method for identifying significantly variable objects, following a method described in \citet{Heinze2015} in a survey for optical variability in T-type brown dwarfs. In this study, the authors find a weak dependence of their variability metric on the RMS of each target. We can also see this in Figure \ref{fig:SOFI_beta}, where reference stars with a higher photometric error tend to have lower $\beta$ factors. We thus investigate the dependence of the periodogram power on $\sigma_{\mathrm{pt}}$ (defined in Section \ref{sec:reducsteplast}). We can expect some dependence because if two lightcurves vary with the same amplitude but different noise levels, the lightcurve with lower photometric error produces a periodogram with a higher power.  Thus we take this into account in our significance threshold criteria. We show the  peak periodogram power of targets and reference stars in Figure \ref{fig:SOFI_power}.
 We calculate a $\sigma$-dependent $95\%$ threshold using a sliding box as described in \citet{Heinze2015}. The box width was chosen such that $>50$ reference star points were available to calculate the $95\%$ threshold up to $0.02~\sigma_{\mathrm{pt}}$ and $0.03~\sigma_{\mathrm{pt}}$ for the NTT and UKIRT data respectively. We find that a box width of $0.02~\sigma_{\mathrm{pt}}$ is suitable for both. We show the noise-dependent $95\%$ significance threshold by the purple line in Figure  \ref{fig:SOFI_power}. 
 Both methods identify the same variable objects with the exception of PSO 071.8--12 and  GU Psc b. PSO 071.8--12 is identified as variable in the $\beta$ factor method but is identified as non-variable in the noise-dependent periodogram power method. We count this object as variable since the lightcurve shows high-amplitude modulation. It is likely that the periodogram power is low because PSO 071.8--12 has a  rotational period that is significantly longer than the observation duration. GU Psc b is identified as variable in the noise-dependent method shown in Figure \ref{fig:SOFI_power}. With a magnitude of $J=18.12$, GU Psc b is at least an order of magnitude fainter than the other targets in our survey and as such, has a much higher photometric error than the other survey targets. Additionally, we have very few reference stars at $\sigma_{\mathrm{pt}}>0.03$, so calculating a $95\%$ threshold at values greater than this is not valid. For these reasons we do not count GU Psc b as a detection. Thus, we have detected variability in thirteen epochs of observations, finding seven variable objects in the survey. We show the lightcurves of each variable object and three reference stars in Figure \ref{fig:variables1}.
 
\section{Periodogram Analysis and Rotational Periods}

\citet{Manjavacas2018} find that the Lomb-Scargle periodogram method is sensitive to gaps in their \textit{HST} lightcurve of a brown dwarf companion. The authors find that the publicly available Bayesian Generalised Lomb-Scargle method \citep{Mortier2015} is insensitive to these gaps, and produces a strong peak at the true rotational period of the brown dwarf. As some of our variable lightcurves have gaps in the data due to bad weather and/or instrumental difficulties at the telescope, we use the BGLS periodogram method to confirm that the detected trends are real, and not due to gaps in the data. The BGLS method produces periodograms with strong peaks at periods that are consistent with those of the Lomb-Scargle method for each variability detection. Thus we can conclude that the periodicity observed in the periodograms is not due to gaps in the lightcurve.

While periodogram analysis can be used to provide an estimate of the rotational periods of brown dwarfs \citep[e.g.][]{Croll2016a, Manjavacas2018}, we caution that the observation duration of our lightcurves are too short to robustly measure a period for most cases. Many of the variable lightcurves shown in Figure \ref{fig:variables1} do not exhibit a local maximum or minimum and thus we can only place lower limits on their rotational periods. For lightcurves that do appear to exhibit local maxima or minima, the rotational periods cannot be confidently measured because double-peaked and evolving lightcurves have been observed in many variable brown dwarfs to date \citep[e.g.][]{Apai2017,Vos2018}. Longer duration follow-up observations are necessary to robustly measure the rotational periods of the variable objects detected in this survey.
 
  \section{Sensitivity to Variability Signals}\label{sec:sensplot}
We construct a sensitivity plot for each observation to determine our sensitivity to variability signals of given amplitudes and periods.
We inject simulated sinusoidal curves into random permutations of each target lightcurve. For targets found to be variable in the survey we divide the lightcurve by a polynomial fit to the lightcurve before injecting the simulated sinusoidal signals.
The 1000 simulated sine curves have peak-to-peak amplitudes of $0.5 - 10\%$ and periods of $1.5 - 20~$hr, with randomly assigned phase shifts. Each simulated lightcurve is put through our periodogram analysis, which allows us to produce a sensitivity plot, showing the percentage of recovered signals as a function of amplitude and period. Sensitivity plots for all light curves are shown in the bottom panel of Figure \ref{fig:3panels} for variable objects and Appendix 3 (available online) for non-variables in the survey.

\section{Significant Detections of Variability}
\label{sec:detections}
\begin{figure*}
\subfloat{
\includegraphics[width=0.5\textwidth]{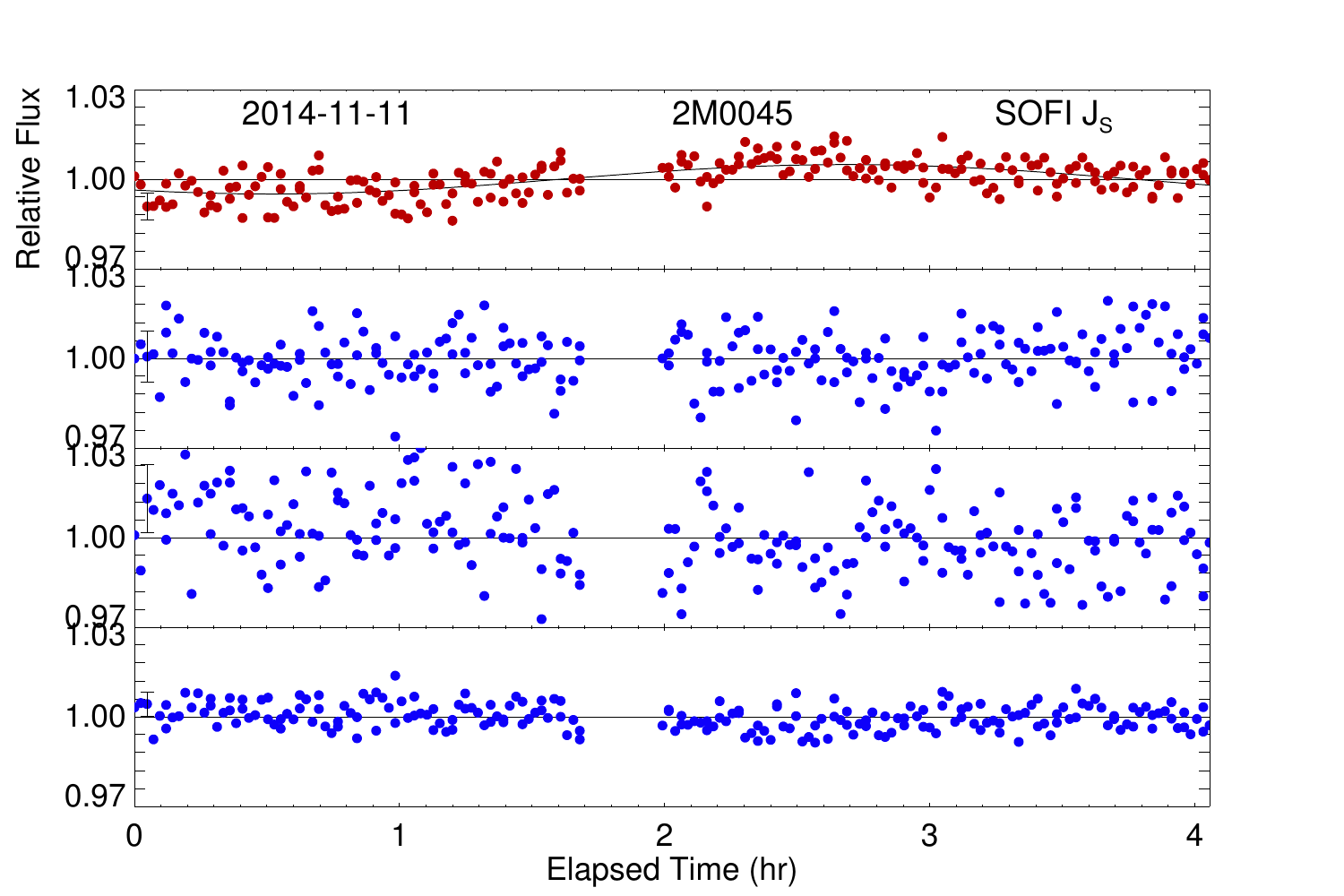}}
\subfloat{
\includegraphics[width=0.5\textwidth]{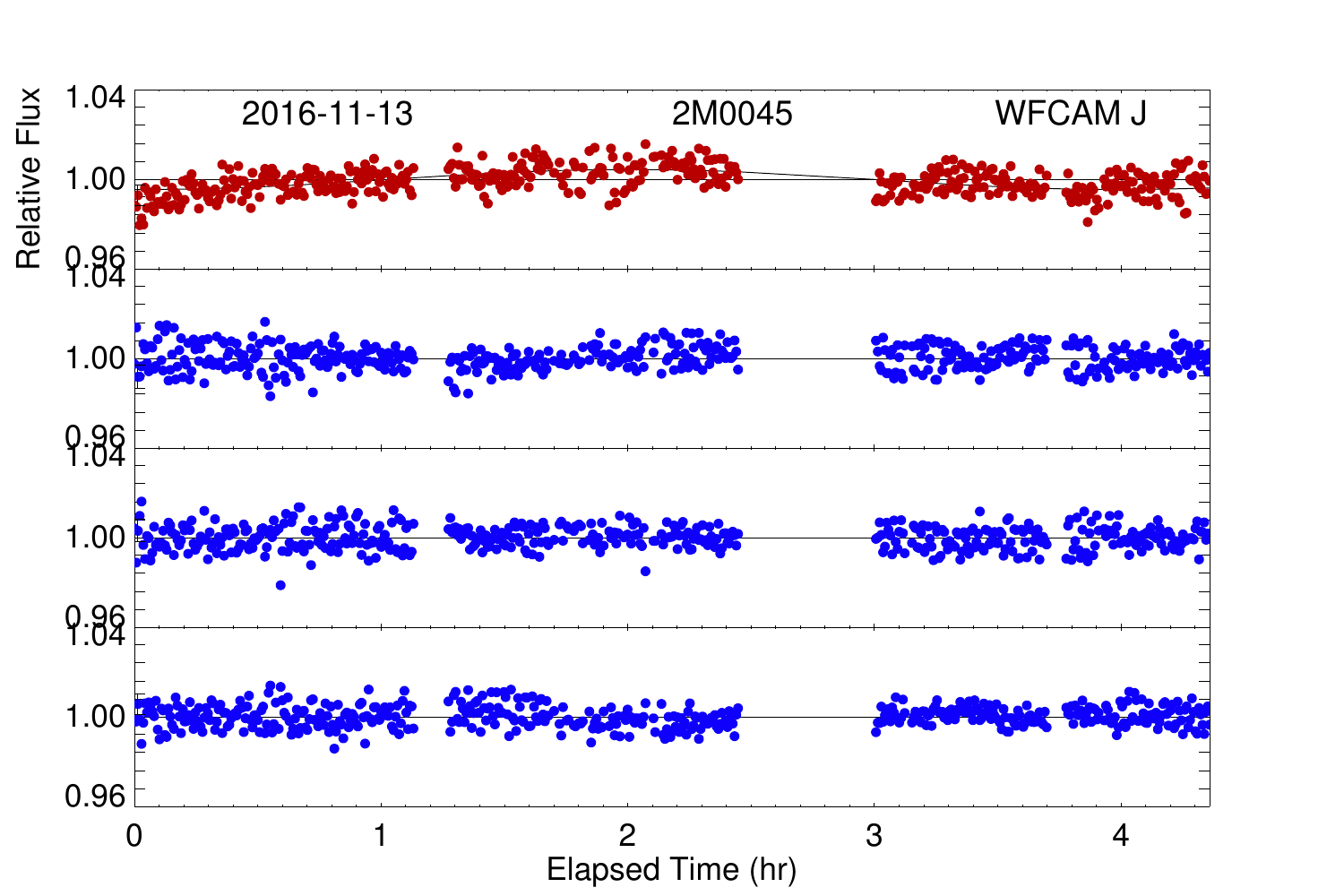}}\\
\subfloat{
\includegraphics[width=0.5\textwidth]{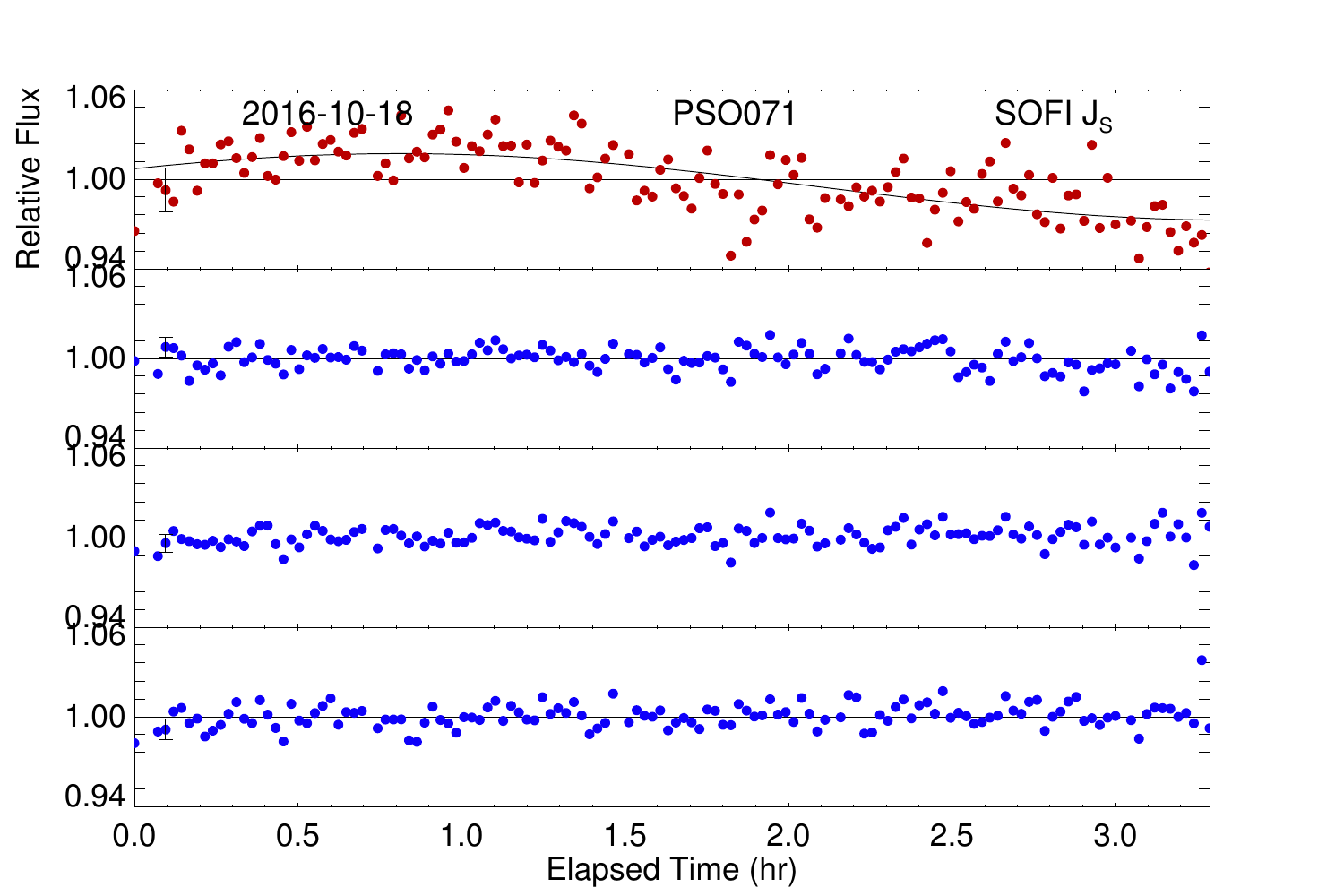}}
\subfloat{
\includegraphics[width=0.5\textwidth]{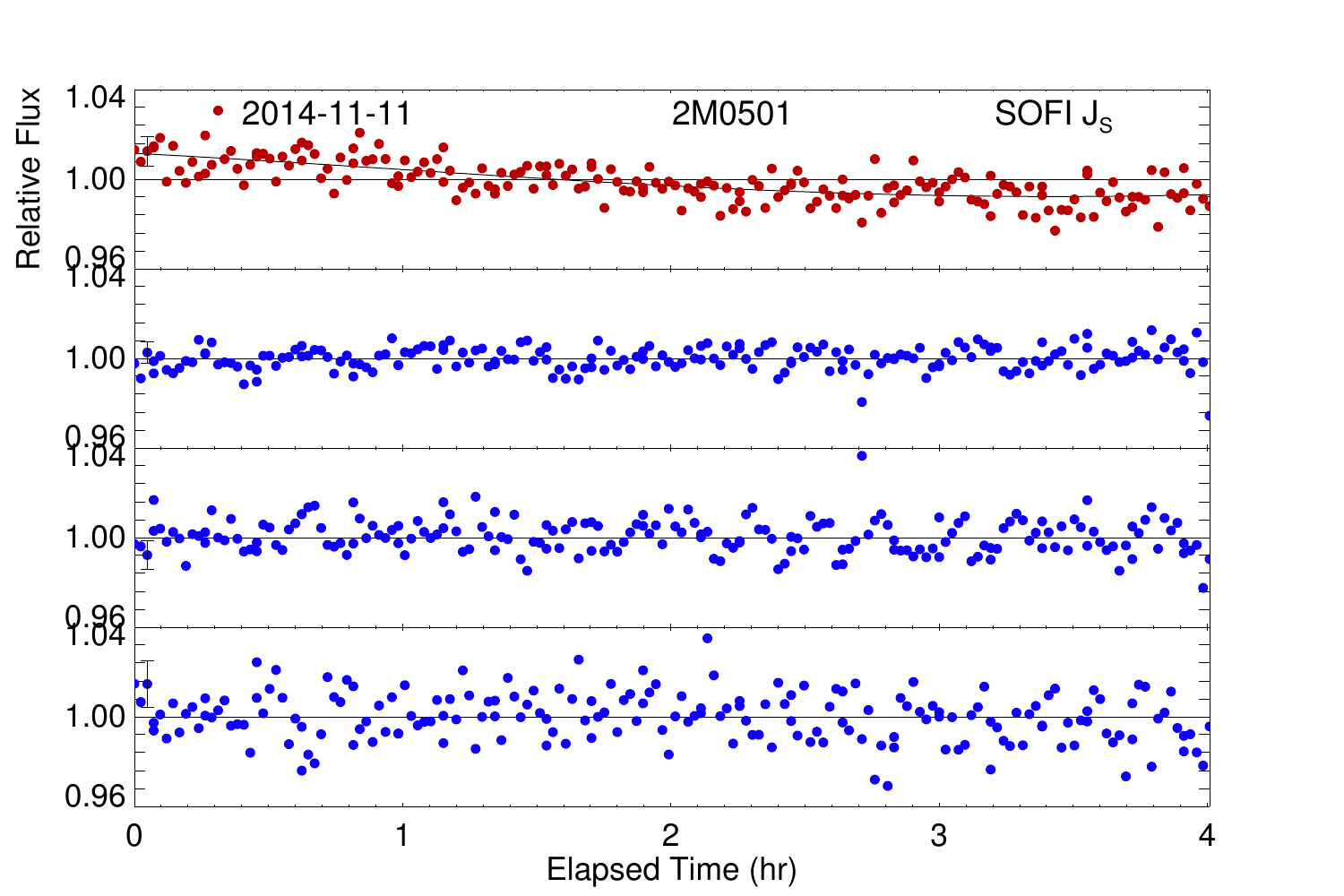}}\\
\subfloat{
\includegraphics[width=0.5\textwidth]{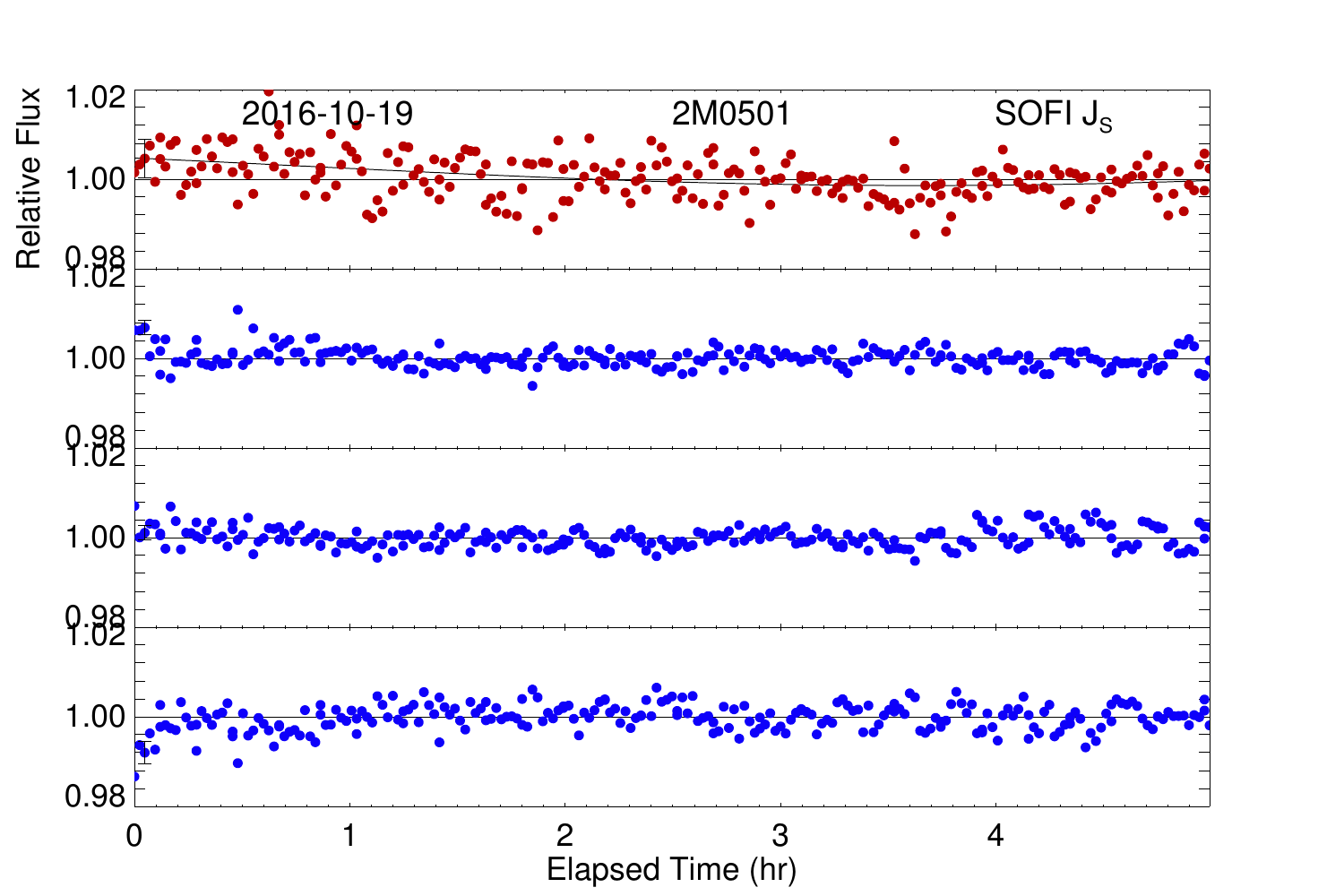}}
\subfloat{
\includegraphics[width=0.5\textwidth]{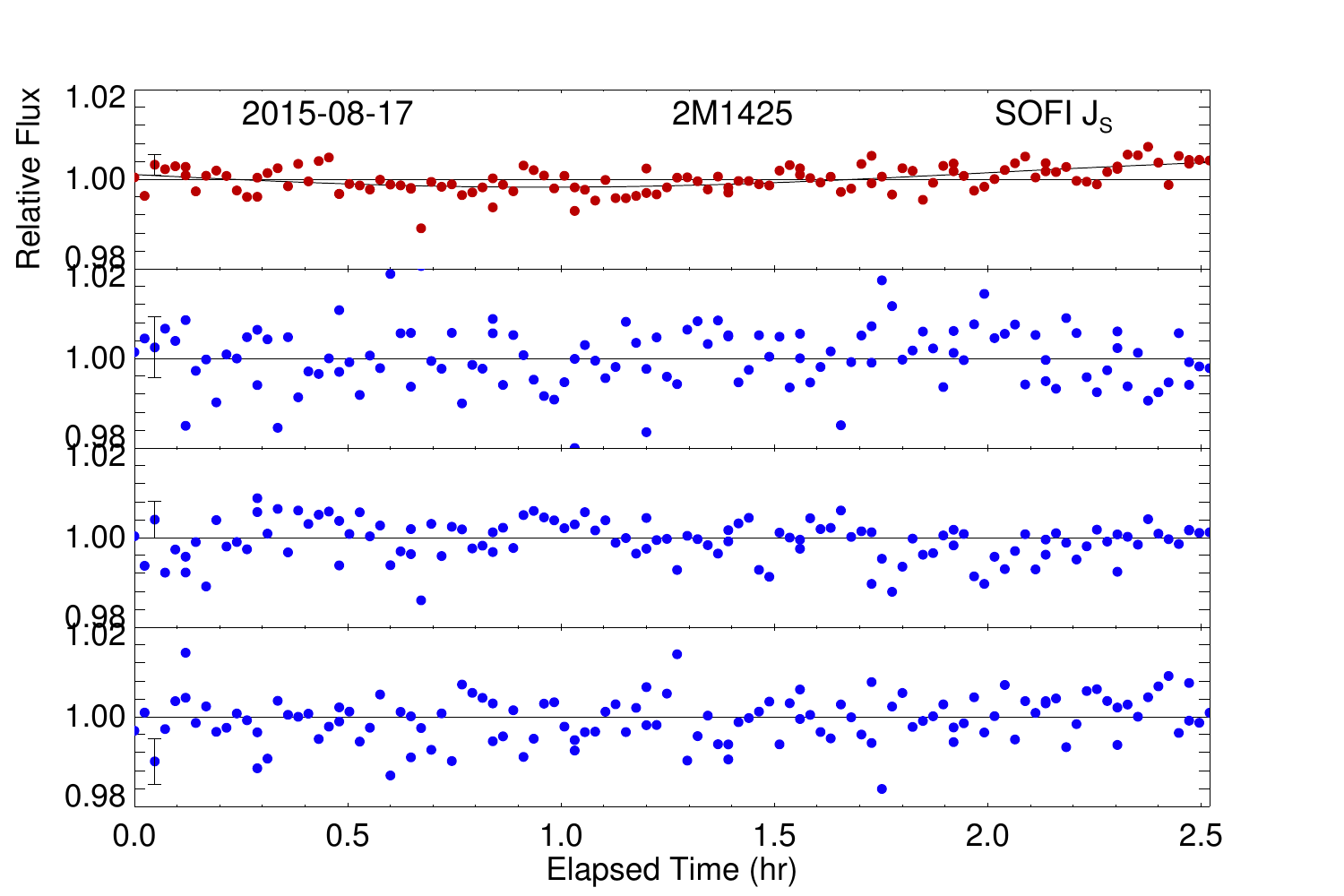}}\\
\caption{Lightcurves of variable targets (red) compared to a sample of reference stars in the field (blue). The black line shows the least-squares best fit sinusoidal model to the lightcurve.}
\label{fig:variables1}
\end{figure*}

\begin{figure*}\ContinuedFloat
\subfloat{
\includegraphics[width=0.5\textwidth]{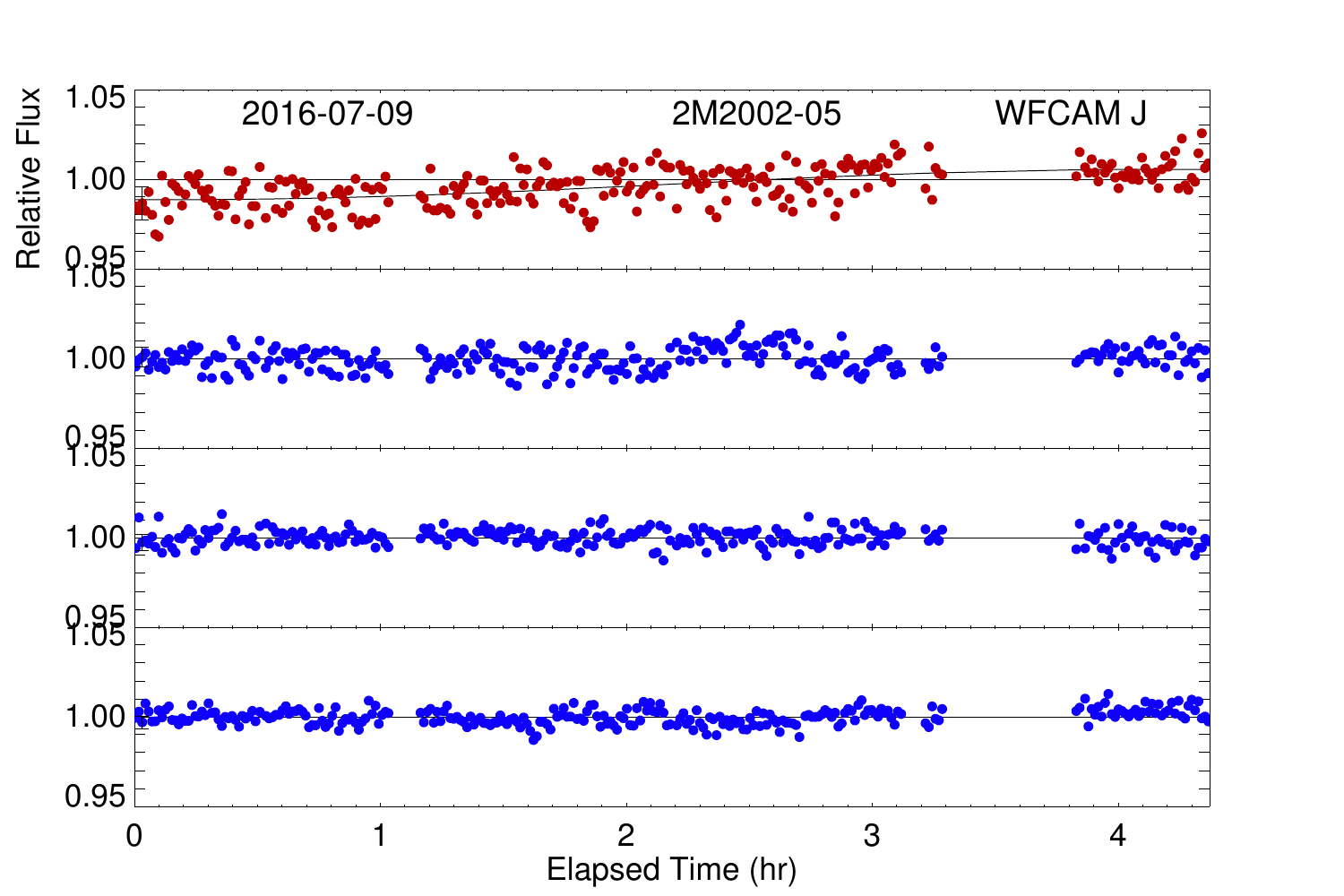}}
\subfloat{
\includegraphics[width=0.5\textwidth]{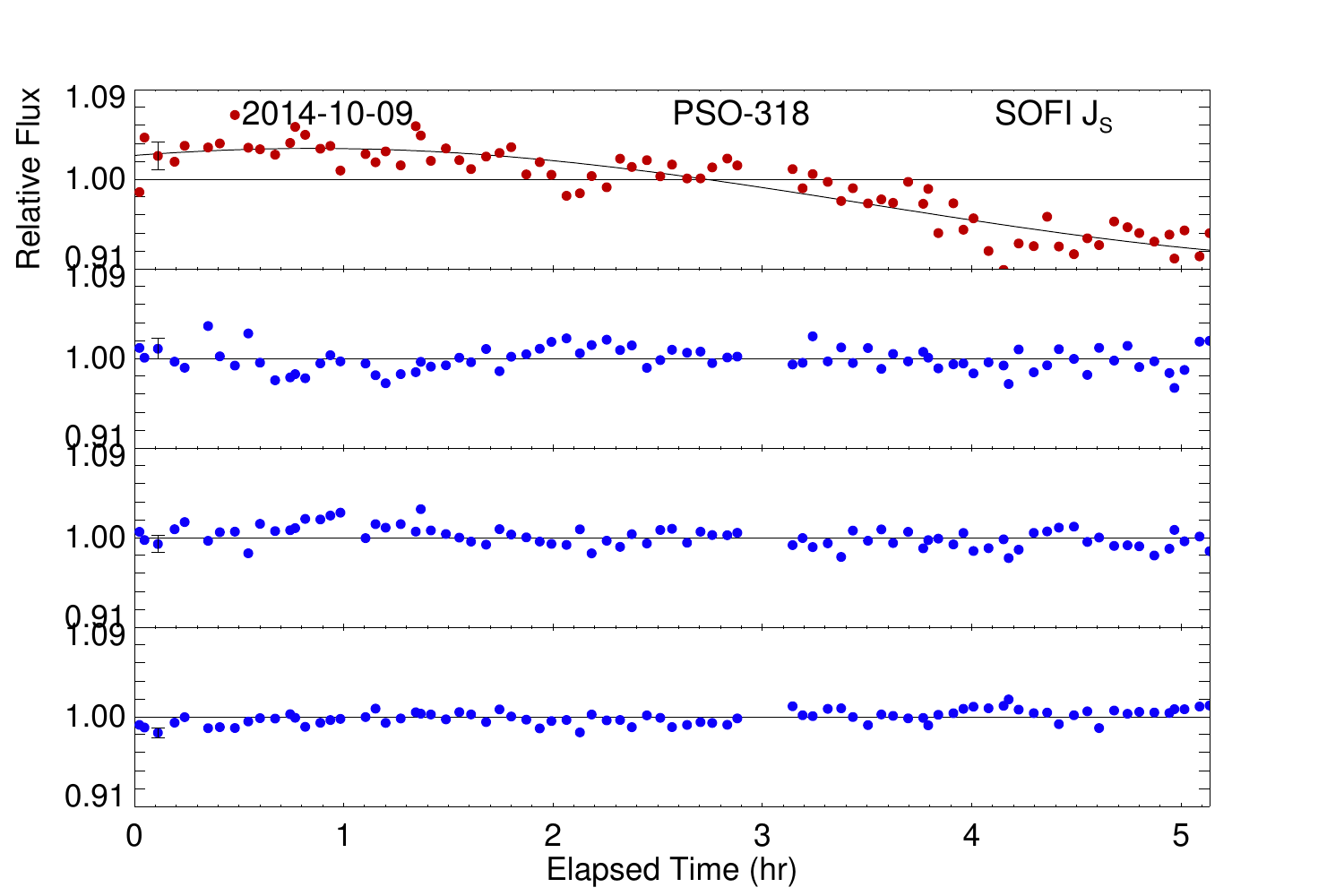}}\\
\subfloat{
\includegraphics[width=0.5\textwidth]{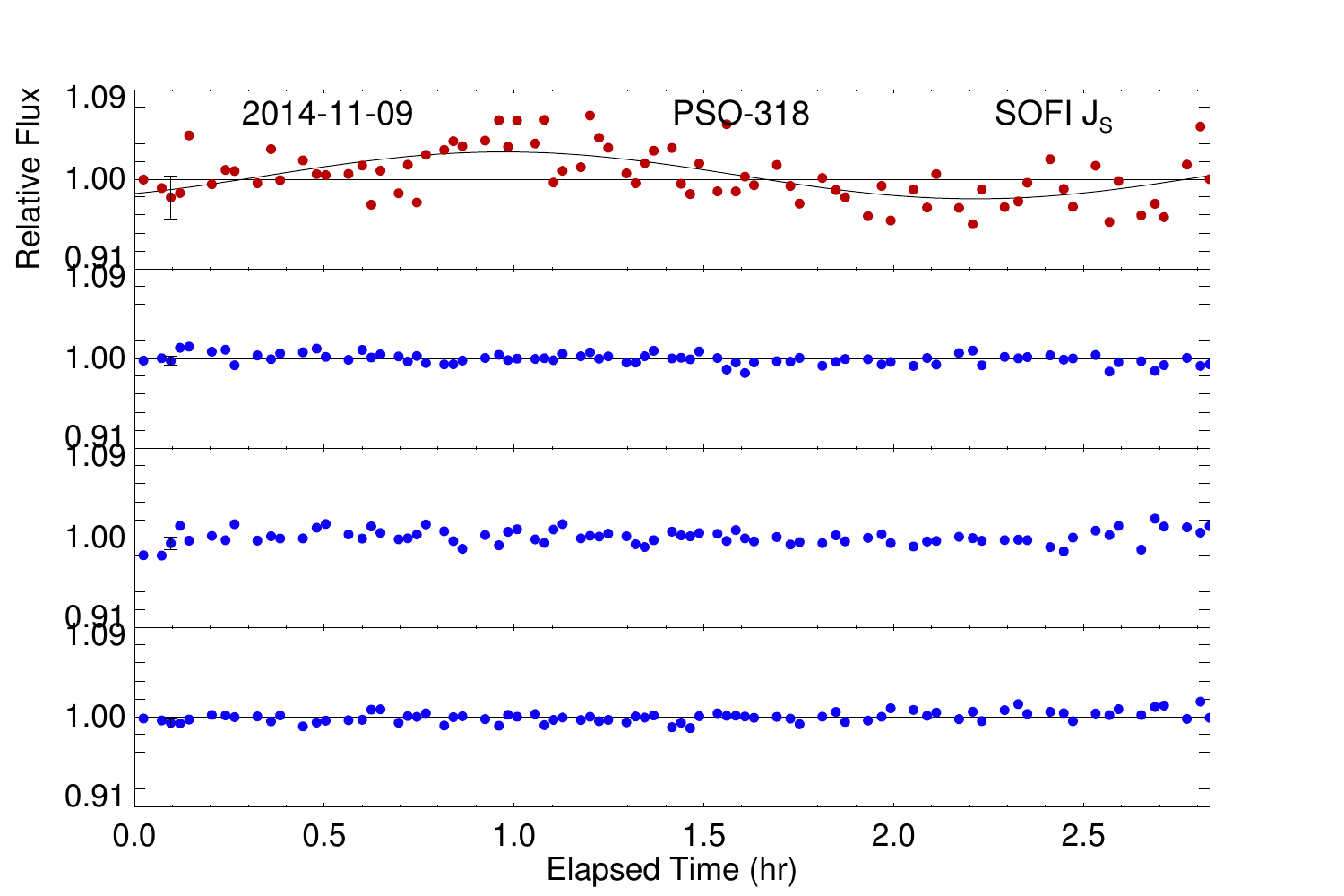}}
\subfloat{
\includegraphics[width=0.5\textwidth]{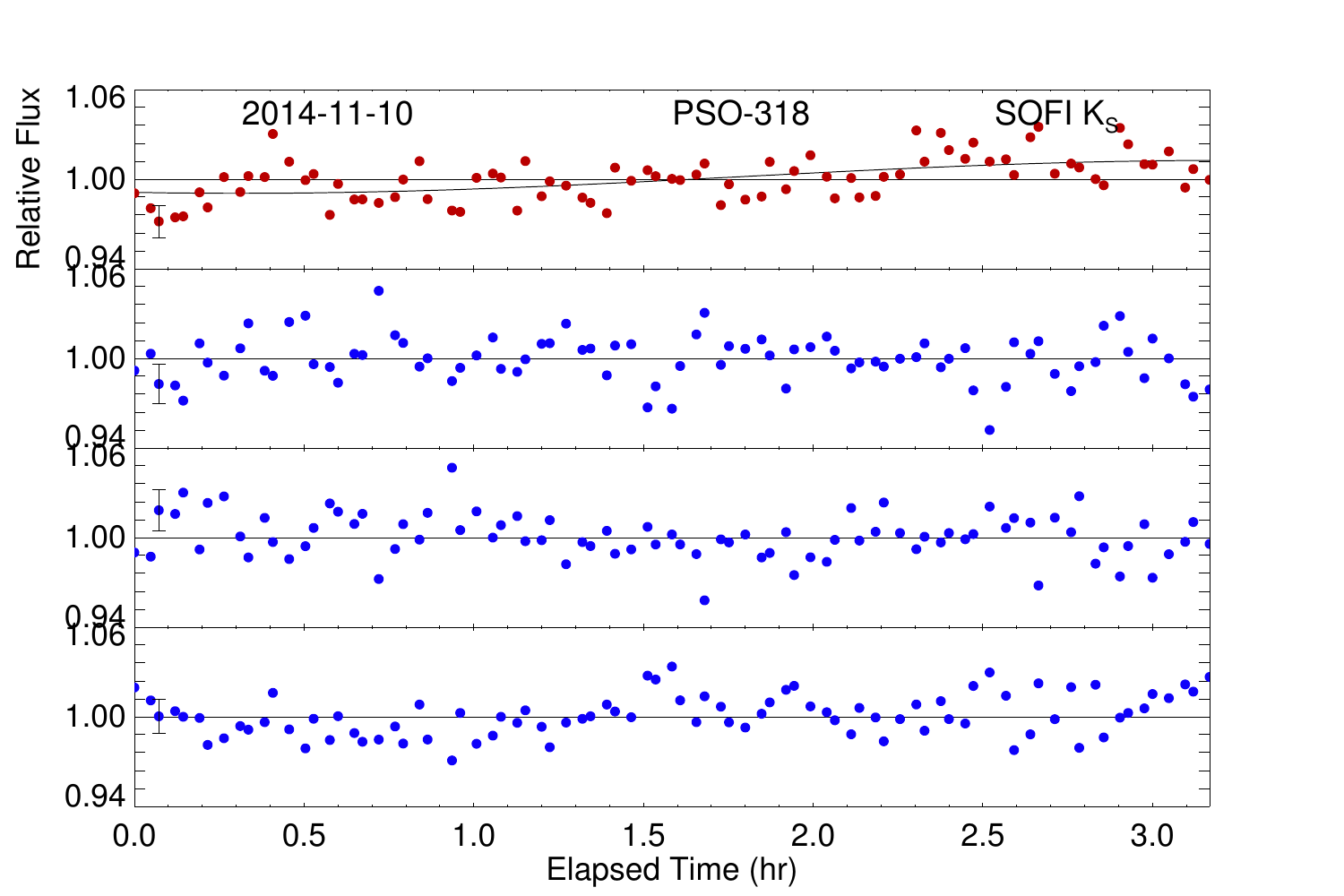}}\\
\subfloat{
\includegraphics[width=0.5\textwidth]{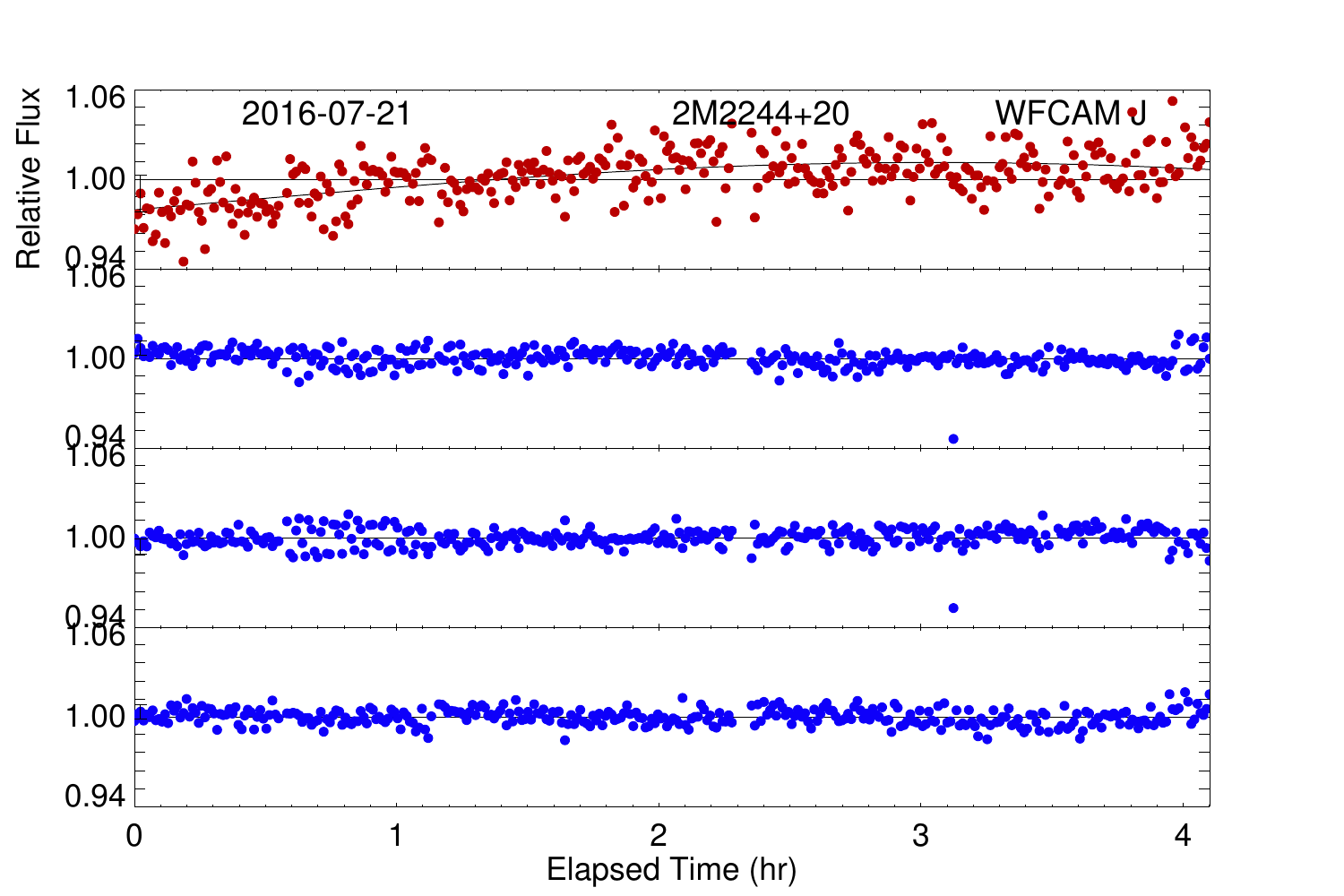}}\\
\caption{Lightcurves of variable targets (red) compared to a sample of reference stars in the field (blue). The black line shows the least-squares best fit sinusoidal model to the lightcurve.}
\label{fig:variables2}
\end{figure*}

\begin{figure*}
\subfloat{ 
\includegraphics[width=0.5\textwidth]{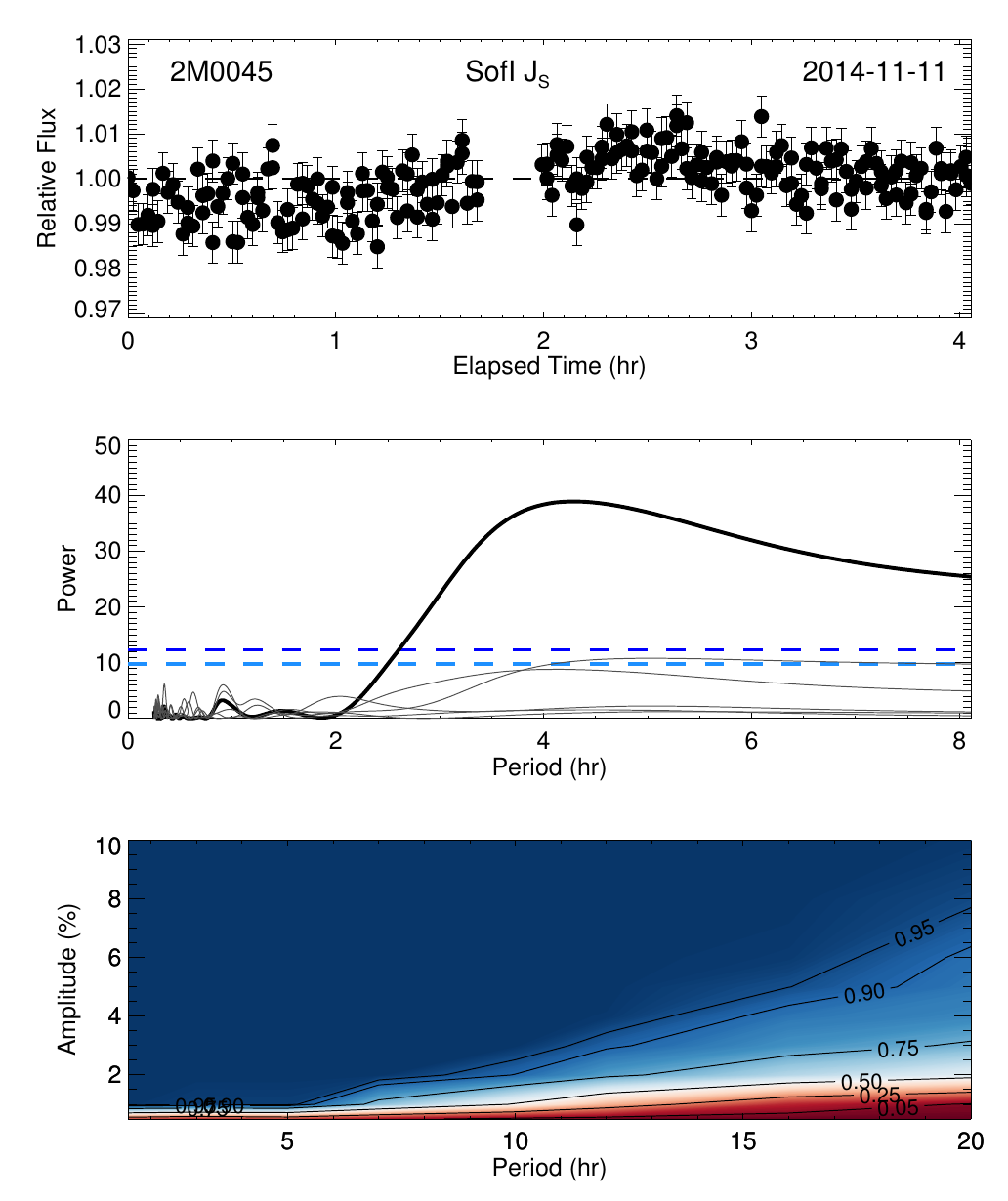}}
\subfloat{  \hspace*{-0.65cm}
\includegraphics[width=0.5\textwidth]{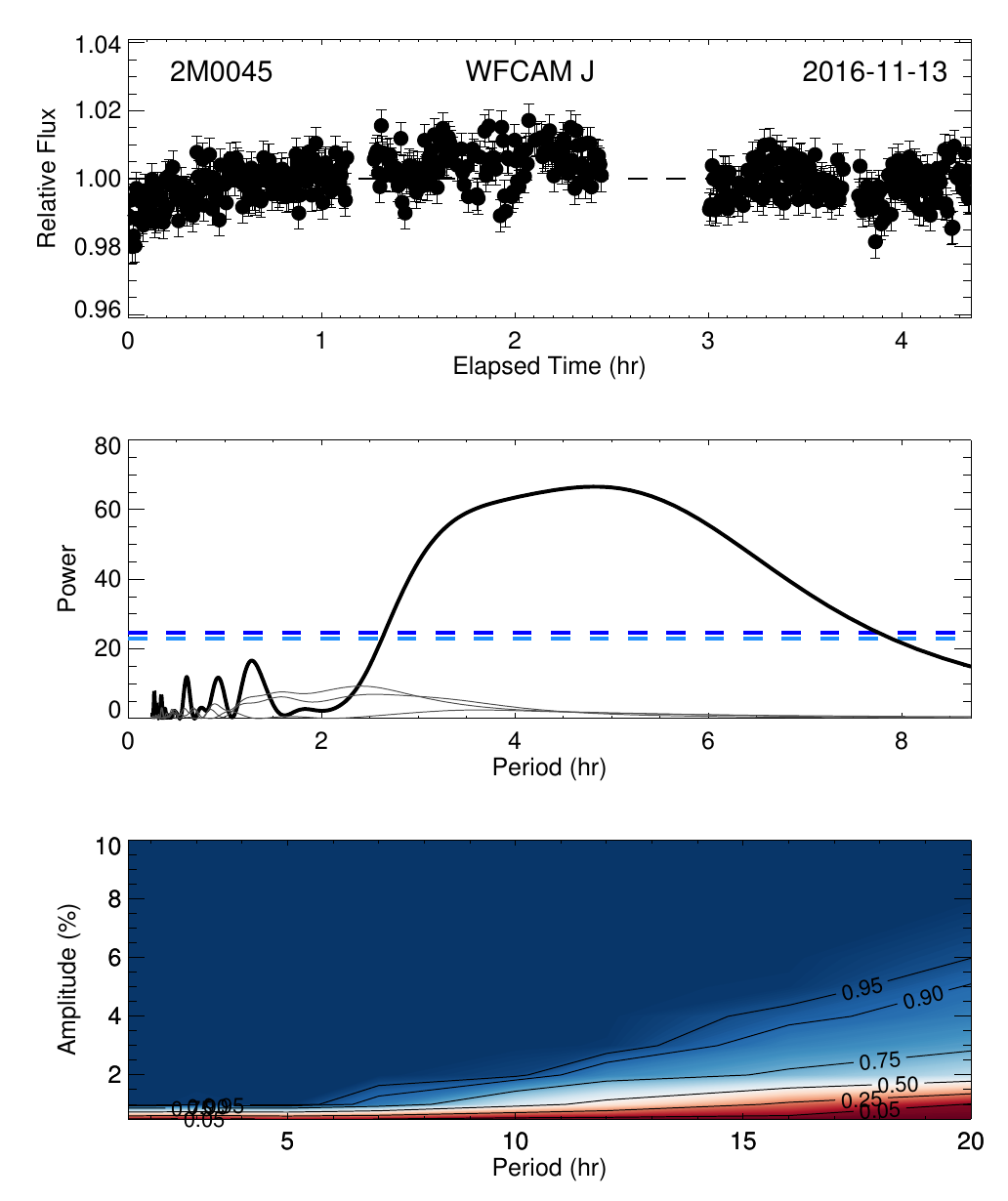}}\\
\subfloat{ 
\includegraphics[width=0.5\textwidth]{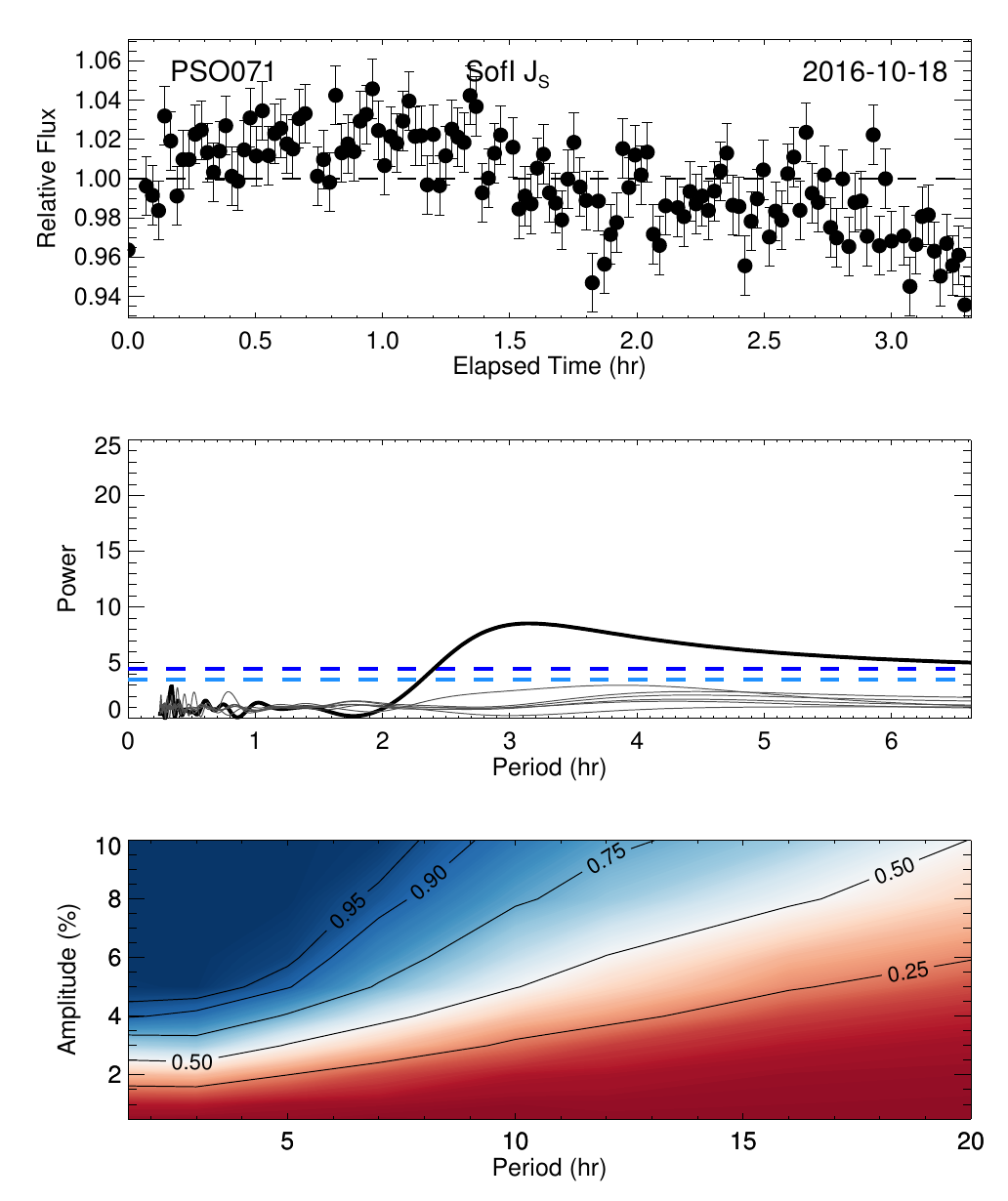}}
\subfloat{  \hspace*{-0.65cm}
\includegraphics[width=0.5\textwidth]{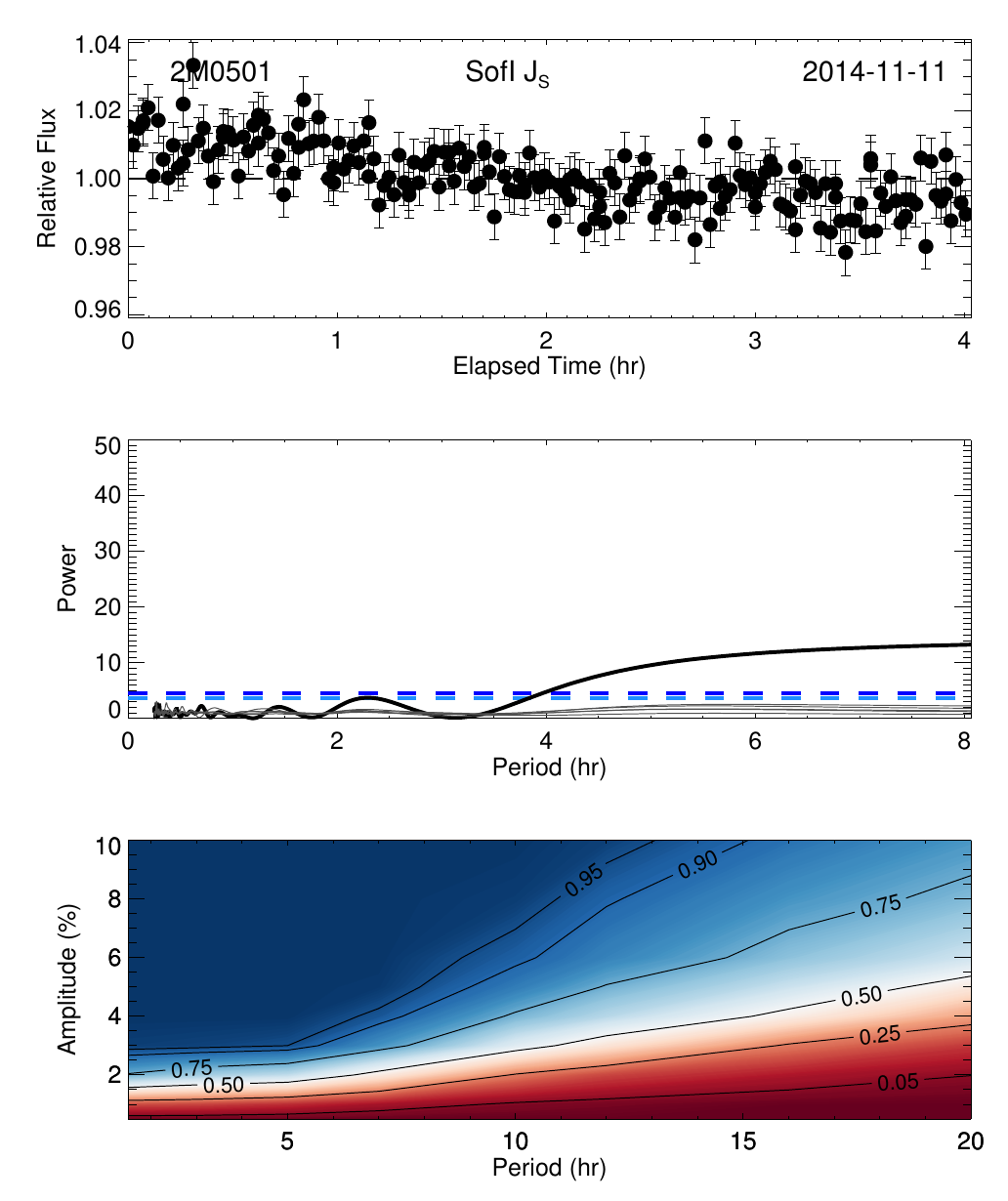}}
\caption{Lightcurves, periodograms and sensitivity plots for variable objects. \textit{Top panel:} Relative photometry of target. \textit{Middle panel:} Periodogram of target lightcurve (black) and periodograms of reference stars (grey). Blue dashed lines show the $95\%$ and $99\%$ significance thresholds. \textit{Bottom panel:} Sensitivity plot showing the percentage of recovered signals for injected sinusoidal signals of various variability amplitude and periods.}
\label{fig:3panels}
\end{figure*}

\begin{figure*}\ContinuedFloat
\subfloat{ 
\includegraphics[width=0.5\textwidth]{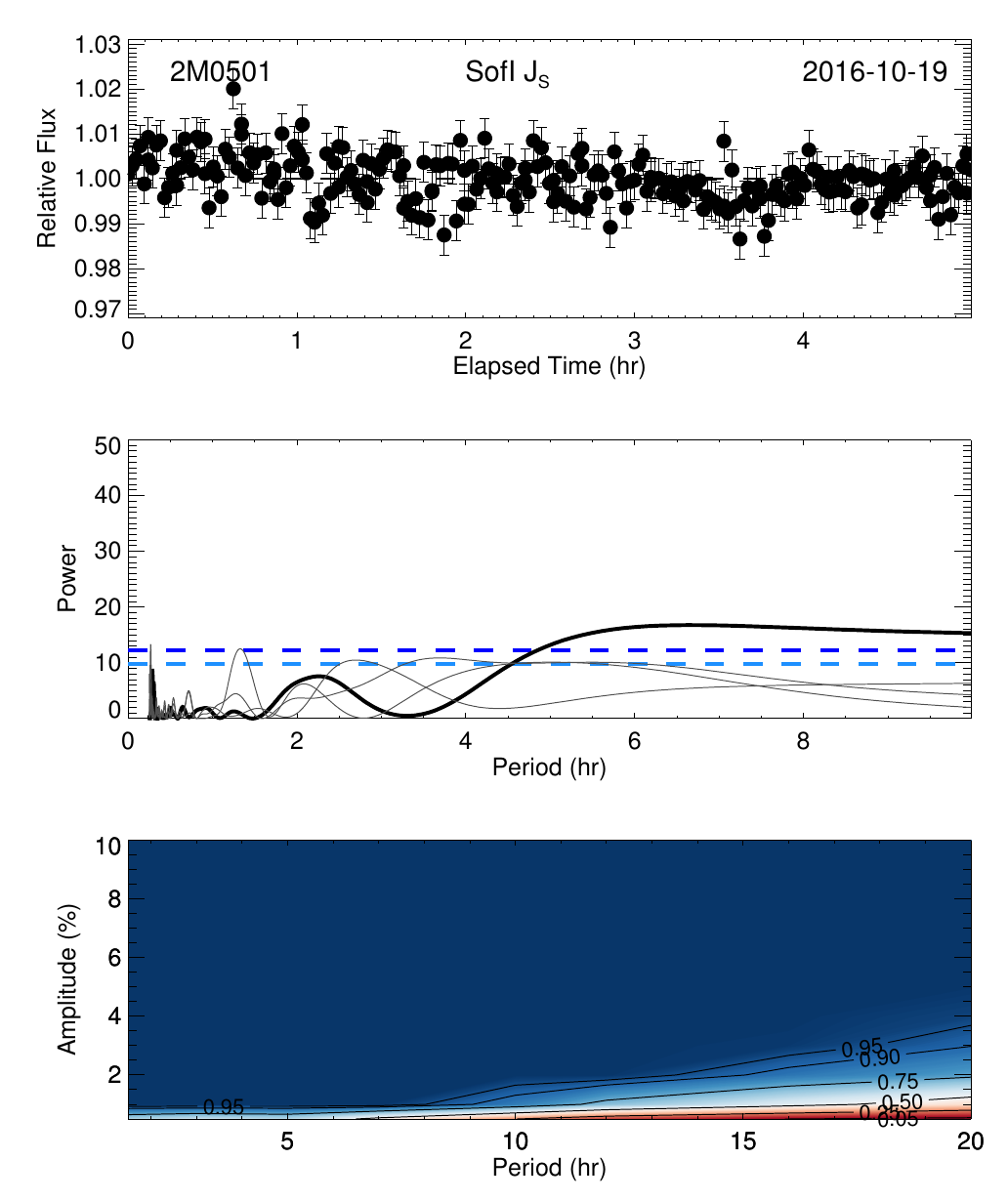}}
\subfloat{  \hspace*{-0.65cm}
\includegraphics[width=0.5\textwidth]{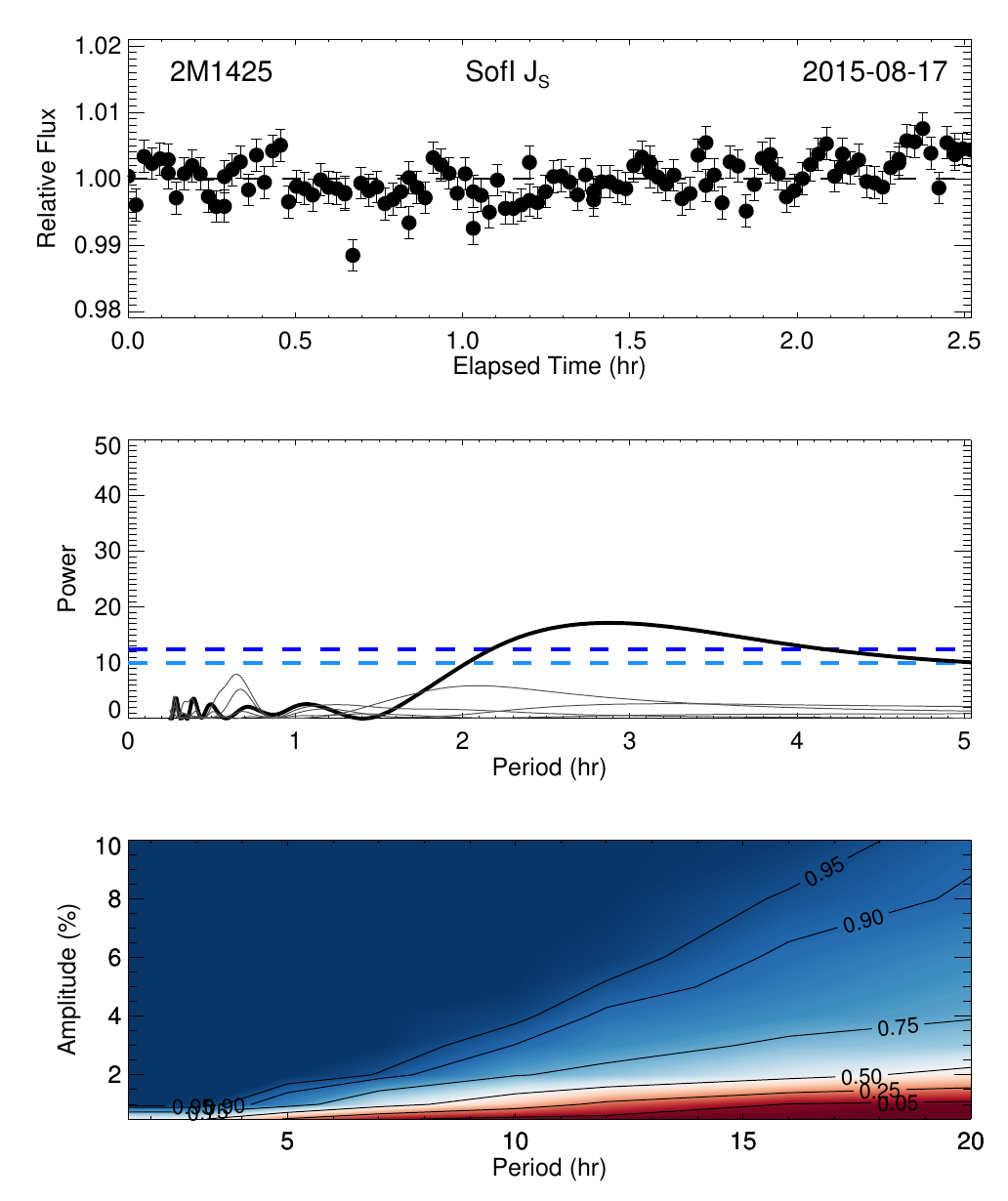}}\\
\subfloat{ 
\includegraphics[width=0.5\textwidth]{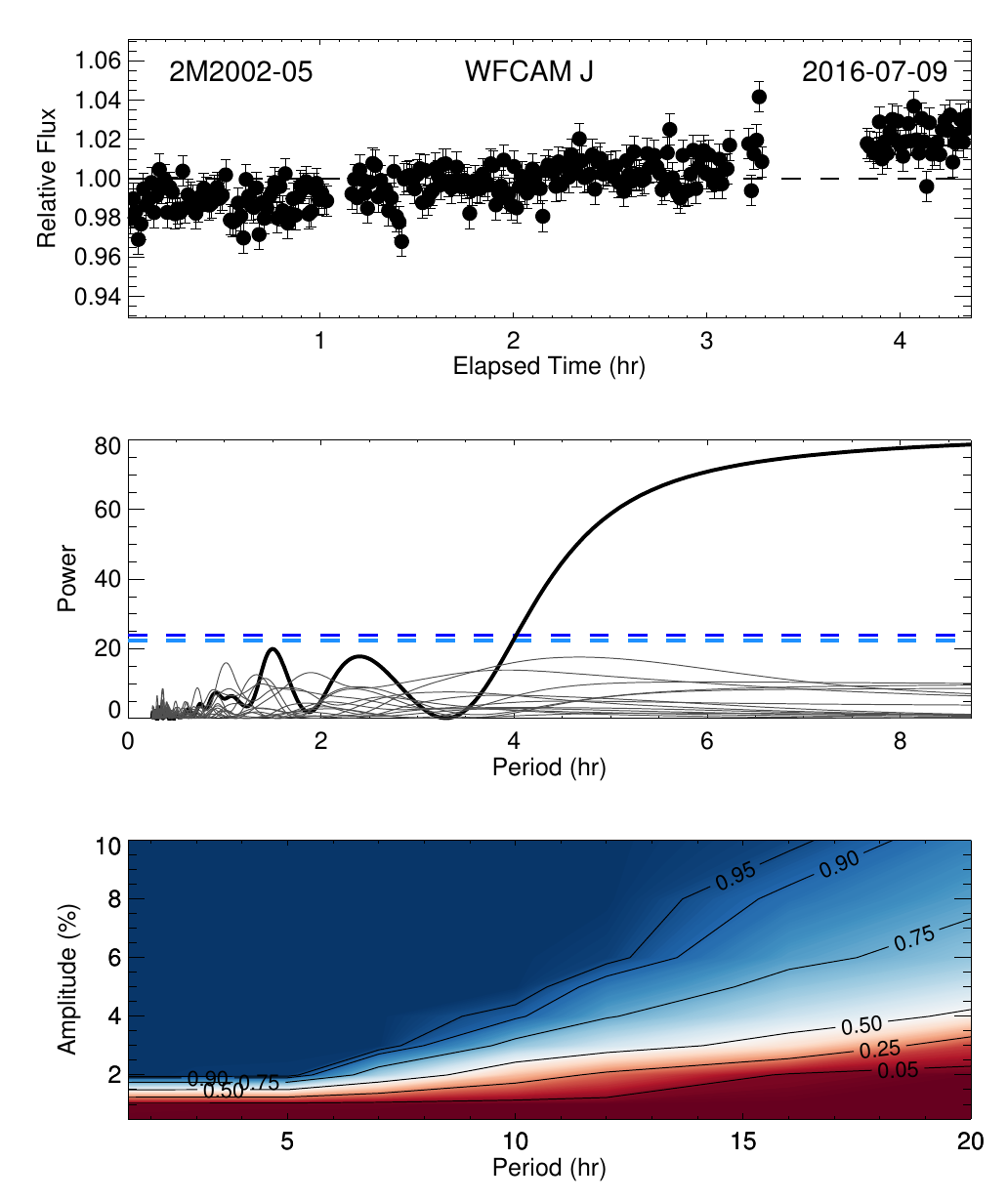}}
\subfloat{  \hspace*{-0.65cm}
\includegraphics[width=0.5\textwidth]{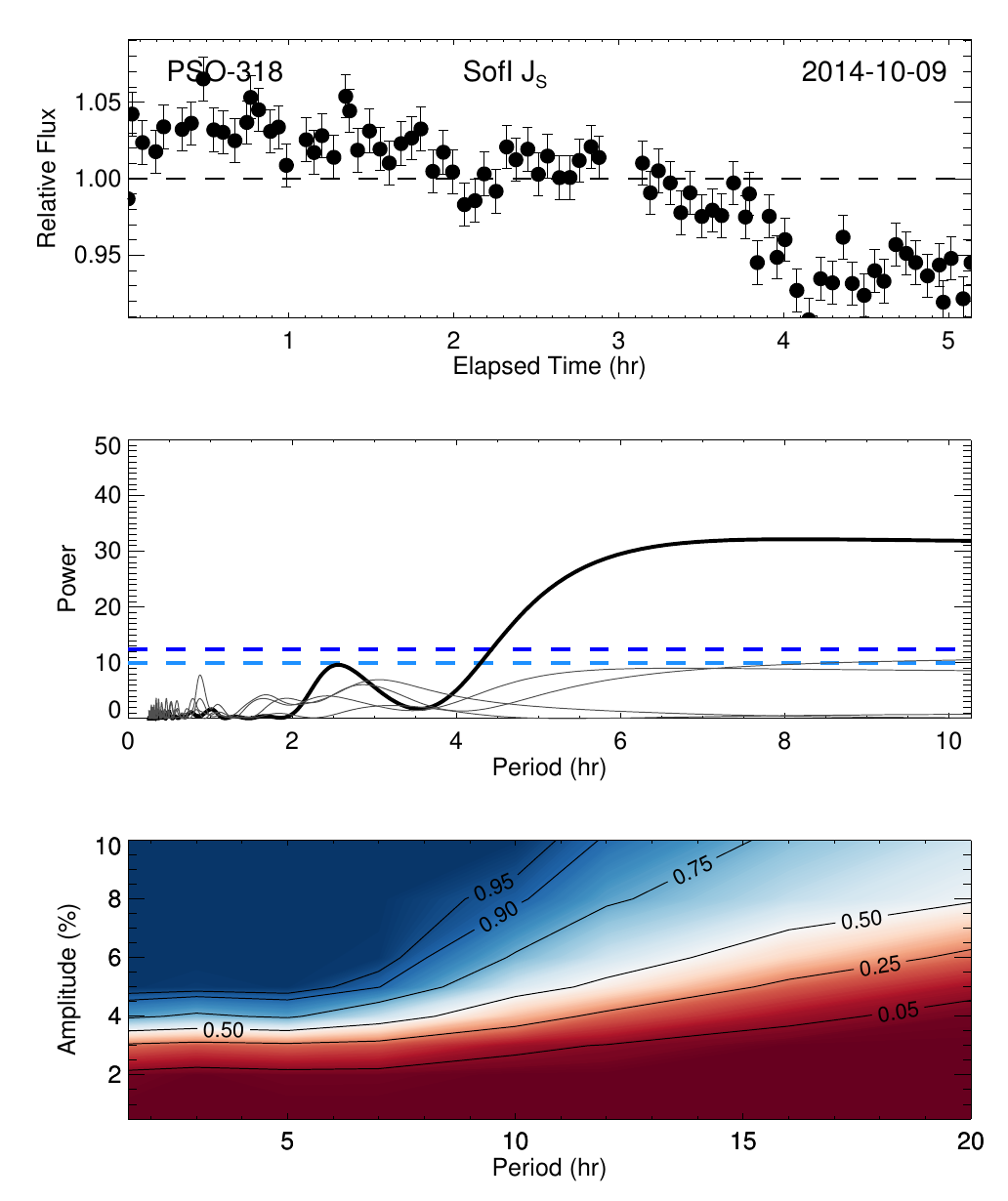}}
\caption{Lightcurves, periodograms and sensitivity plots for variable objects. \textit{Top panel:} Relative photometry of target. \textit{Middle panel:} Periodogram of target lightcurve (black) and periodograms of reference stars (grey). Blue dashed lines show the $95\%$ and $99\%$ significance thresholds. \textit{Bottom panel:} Sensitivity plot showing the percentage of recovered signals for injected sinusoidal signals of various variability amplitude and periods.}
\label{fig:3panels}
\end{figure*}

\begin{figure*}\ContinuedFloat\subfloat{
\includegraphics[width=0.5\textwidth]{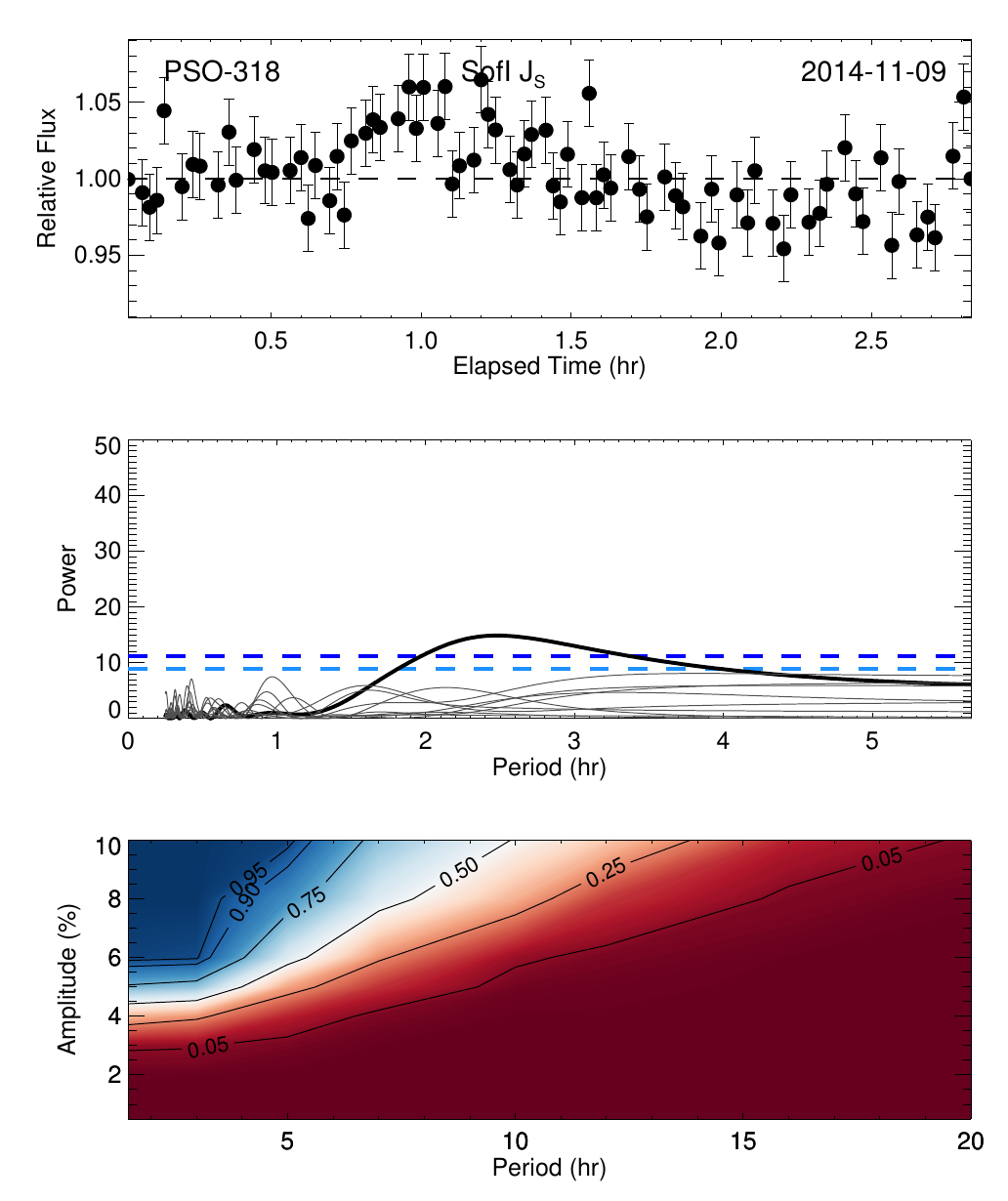}}
\subfloat{ 
\includegraphics[width=0.5\textwidth]{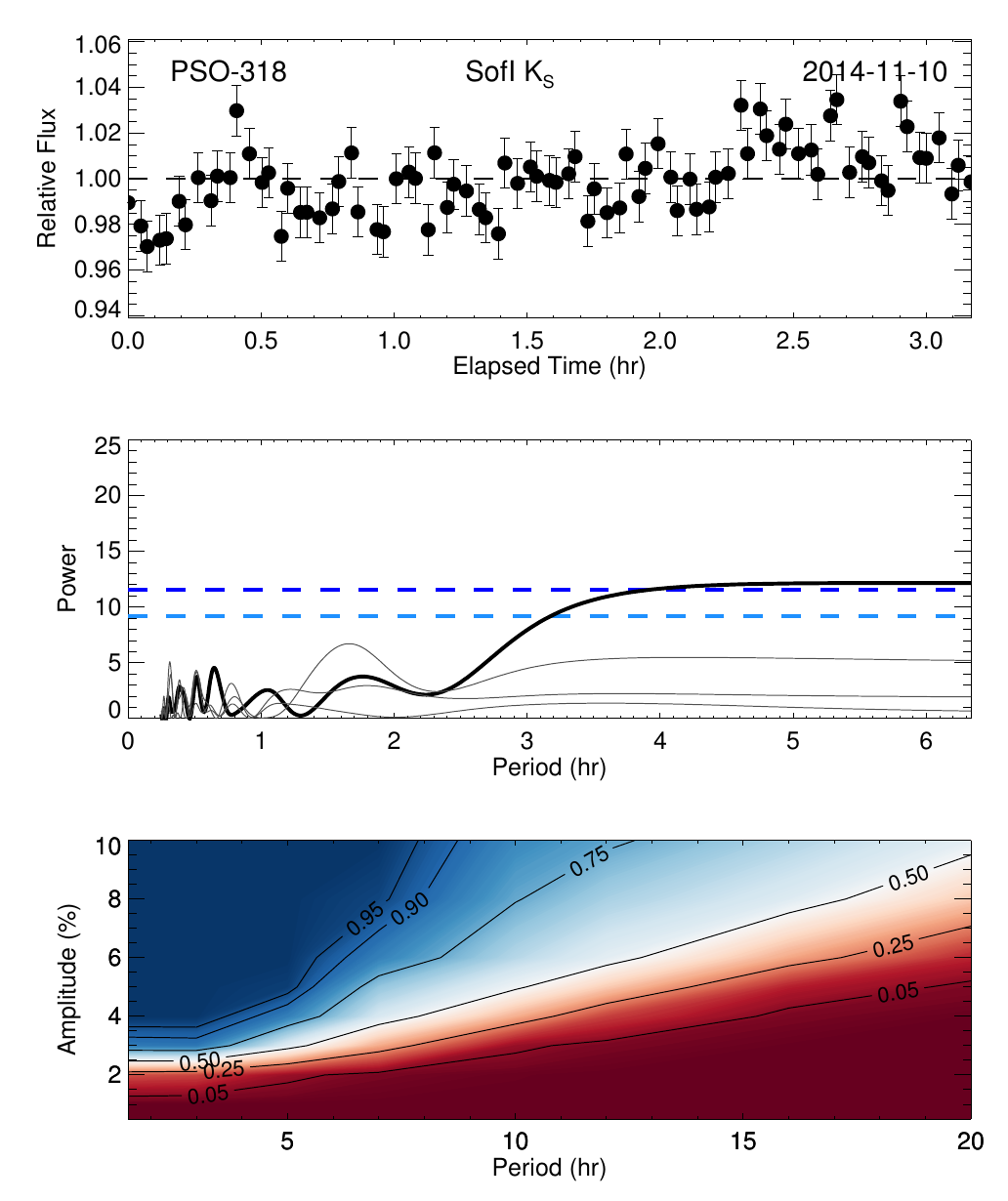}}\\
\subfloat{  \hspace*{-0.65cm}
\includegraphics[width=0.5\textwidth]{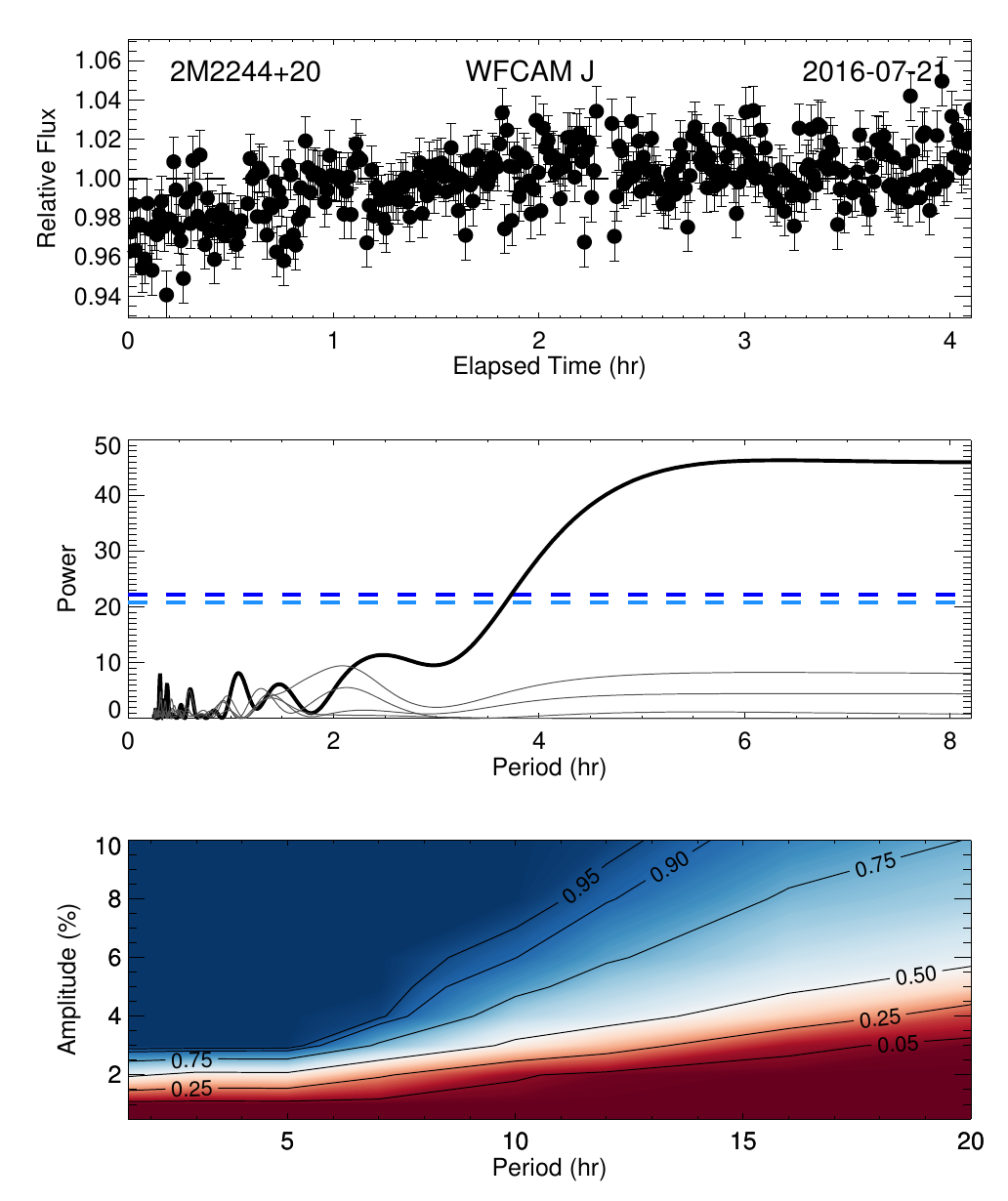}}\\
\caption{Lightcurves, periodograms and sensitivity plots for variable objects. \textit{Top panel:} Relative photometry of target. \textit{Middle panel:} Periodogram of target lightcurve (black) and periodograms of reference stars (grey). Blue dashed lines show the $95\%$ and $99\%$ significance thresholds. \textit{Bottom panel:} Sensitivity plot showing the percentage of recovered signals for injected sinusoidal signals of various variability amplitude and periods.}
\label{fig:3panels}
\end{figure*}

We detect significant variability in seven objects in the survey. Variable objects and their estimated variability amplitudes are presented in Table \ref{tab:variables}.
We present the light curves of these variable objects along with their reference stars in Figure \ref{fig:variables1}. We show their periodograms and sensitivity plots in Figure \ref{fig:3panels}. We discuss the first epoch variability detection of each object below.

 \begin{table}
\centering
\caption{Measured first epoch variability amplitudes for variability detections.}
\label{tab:variables}
\begin{tabular}{lll}
\hline \hline
Target       & Spt   & Amp ($\%$)   \\
\hline
2M0045+16    & L2    & $1.0\pm0.1$    \\
PSO 071.8--12 & T2    & $4.5\pm0.6$     \\
2M0501--00       & L3    & $2\pm1$  \\
2M1425--36       & L4    & $0.7\pm0.3$     \\
2M2002--05    & L5.5 & $1.7\pm0.2$     \\
PSO 318.5--22 & L7    & $10\pm1.3$    \\
2M2244+20    & L6--L8    & $5.5\pm0.6$     \\
\hline
\end{tabular}
\end{table}

\textit{2MASS J00452143+1634446} ---
\citet{Gagne2014a} classify the L2 object 2M0045+16 as a very low gravity brown dwarf, with H$\alpha$ emission and unusually red colours.
It has been identified as a bona fide member of  the Argus association ($30-50$ Myr), giving it an estimated mass of $14.7\pm0.3~M_{\mathrm{Jup}}$ \citep{Gagne2015c}, however recent work by \citet{Bell2015} has called into question the validity of the Argus association.
We observed  2M0045+16 on Nov 11 2014 using NTT SofI.
We detect highly significant variability in this object at this epoch. The lightcurve of 2M0045+16 and the reference stars used for detrending are shown in Figure \ref{fig:variables1} and the periodogram and sensitivity plots are shown in Figure \ref{fig:3panels}. A lack of stars in the field resulted in only 3 references stars suitable for detrending the lightcurve. We fit a sinusoid to the lightcurve using  a Levenberg-Marquardt least-squares algorithm to estimate the amplitude of the modulation in the first epoch. This gives an amplitude of $1.0\pm0.1\%$ and a period $4.3\pm0.3~$hr. While it appears that we have covered a full rotational period, additional longer duration observations are necessary to rule out the possibility of a double-peaked lightcurve with a longer rotational period \citep[e.g.][]{Vos2018}. We discuss followup observations of 2M0045+16 in Section \ref{sec:followup}.

\textit{PSO J071.8769 $-$12.2713} ---
PSO 071.8--12 was identified as a high-probability candidate member of $\beta$ Pictoris by \citet{Best2015}. Assuming membership of the $\beta$ Pictoris moving group, it has an estimated mass of $6.1 \pm0.7~M_{\mathrm{Jup}}$ \citep{Best2015}, making it the lowest mass object to date to exhibit photometric variability. 
The lightcurve shown in Figure \ref{fig:variables1} displays high-amplitude ($4.5\pm0.6\%$) variability. The periodogram shown in Figure \ref{fig:3panels} shows a highly significant ($>99\%$) peak. Since we did not cover a full rotational period we can only estimate a period $>3~$hr. We discuss subsequent follow-up observations of this object in Section \ref{sec:followup}.

\textit{2MASS J05012406$-$0010452} ---
\citet{Gagne2015c} categorise 2M0501--00 as L$3\gamma$, and an ambiguous candidate member of Columba or Carina (both moving groups are coeval at $20-40$ Myr). If 2M0501--00 is indeed a member of Columba or Carina it has an estimated mass of $10.2^{+0.8}_{-1.0}~M_{\mathrm{Jup}}$.
We detect significant variability in 2M0501--00 on Nov 11 2014 with NTT SofI (shown in Figure \ref{fig:variables1}). The periodogram shown in Figure \ref{fig:3panels} shows a highly significant peak at periods $>4~$hr. A sinusoidal fit to the lightcurve gives a peak-to-peak amplitude of $2.0\pm0.1\%$ and a period of $>4~$hr.  Since we did not cover a full period of rotation in either epoch, our amplitude measurement is a lower limit and our period estimate is very uncertain.
We obtained additional follow-up monitoring of 2M0501--00 and discuss these observations in Section \ref{sec:followup}.

\textit{2MASS J14252798$-$36502295.23} ---
2M1425--36 is classified as a bona fide member of AB Doradus \citep{Gagne2015c}. \citet{Radigan2014} previously reported 2M1425--36 as a marginal variable, displaying low-level variability over a $\sim2.5~$hr observation. 
We detect low-amplitude variability in the L4 object 2M1425--36 on Aug 17 2015. Figure \ref{fig:3panels} shows a highly significant periodogram peak that favours a period of $\sim3~$hr. Using our least-squares algorithm we estimate a variability amplitude of $0.7\pm0.3\%$ for this epoch. We place a lower limit of  $2.5~$h on the rotational period since we did not cover a full period. We observed 2M1425--36 a second time in 2017 and discuss this observation in Section \ref{sec:followup}.

\textit{2MASS J20025073$-$0521524} ---
2M2002--05 is classified as a L5-L7 $\gamma$ object by \citet{Gagne2015c}, but has not been identified as a candidate of a young moving group \citep{Faherty2016}. Our UKIRT/WFCAM observation of this object taken on July 09 2016 shows significant variability. Fitting a sinusoid to the lightcurve we estimate an amplitude of $1.7\pm0.2\%$ and a period of $8\pm2~$hr, however these are very uncertain as we did not cover a full rotational period in this epoch.

 \textit{PSO J318.5338$-$22.8603} ---
 \citet{Allers2016} confirm the L7 PSO 318.5--22 as a member of the $23\pm3~$Myr \citep{Mamajek2014} $\beta$ Pictoris moving group. This implies a mass estimate of $8.3\pm0.5~M_{\mathrm{Jup}}$, placing PSO 318.5--22 clearly in the planetary-mass regime. The lightcurves of PSO 318.5--22 from 2014 are presented in \citet{Biller2015} and are also included in this paper in Figure \ref{fig:variables1}.
As discussed in \citet{Biller2015}, we detect significant, high-amplitude variability in PSO 318.5--22 on October 9 2014. The periodogram in Figure \ref{fig:3panels} shows a highly significant peak at periods $>4.5~$hr. A sinusoidal fit to the lightcurve gives a peak-to-peak amplitude of $10\pm1.3\%$ and a period of $10\pm2~$hr. We obtained two additional epochs of follow-up monitoring for PSO 318.5--22 as part of the variability survey and discuss these observations in Section \ref{sec:followup}.

 \textit{2MASS J2244316+204343} ---
 2M2244+20 is a confirmed member of AB Doradus, with an estimated mass of $\sim 19~M_{\mathrm{Jup}}$ \citep{Vos2018}. \citet{Morales-Calderon2006} and \citet{Vos2018} detect variability in the \textit{Spitzer} $4.5~\mu$m and $3.6~\mu$m bands respectively. The $J$-band lightcurve obtained in this survey was initially presented in \citet{Vos2018}, where we measured a period of $11\pm2~$hr for this object using \textit{Spitzer} data.
The UKIRT/WFCAM lightcurve obtained on July 21 2016 shows significant variability. The periodogram shown in Figure \ref{fig:3panels} shows a significant peak for periods $>4~$hr. We set the period to $11\pm2$ hr \citep[as measured in][]{Vos2018} in our least-squares sinusoidal fit and find an amplitude of $5.5\pm0.6\%$ for this epoch.


\section{Non-detections}

We present light curves and reference star light curves of non-variables in Appendix 1 (available online). Periodograms and sensitivity plots are shown in Appendix 3 (available online). We discuss some of the noteworthy non-detections below. 

 \textit{2MASS J01033203+1935361} --- 2M0103+19 has been assigned $\beta$ and {\sc int-g} gravity classifications \citep{Kirkpatrick2000, Faherty2012, Allers2013}, however has not been assigned membership of a young moving group. \citet{Metchev2015a} obtained $21~$hr of \textit{Spitzer} monitoring, detecting variability in both the $3.6~\mu$m and $4.5~\mu$m bands. 2M0103+19 was observed to exhibit a regular periodic modulation with a period of $2.7\pm0.1$~hr. This short rotational period combined with variability amplitudes of $0.56\pm0.03\%$ and $0.98\pm0.09\%$ in the $3.6~\mu$m and $4.5~\mu$m bands respectively, would suggest that a $J$-band detection is likely for this object. Our observation taken on November 3 2014 shows no evidence of variability over a $5.3~$hr observation. According to the sensitivity plot shown in Appendix 3 (available online), we would have detected variability with an amplitude $>2\%$ for a $2.7~$hr period with a $90\%$ probability. 

 \textit{GU Psc b} --- GU Psc b is a wide separation T3.5 planetary-mass companion to the young M3 star, a likely member of the AB Doradus moving group \citep{Naud2014a}. Recently, \citet{Naud2017} reported results from a $J$-band search for variability in this object. Photometric variability with an amplitude of $4\pm1\%$ was marginally detected during one $\sim6~$hr observation, with no significant variations observed at two additional epochs. The authors estimate a period $>6~$hr as the lightcurve does not appear to repeat during this observation. With a magnitude of $J=18.12$, GU Psc b is $\sim1~$mag fainter than the other targets in our survey and we do not detect significant variability in its $\sim3.5~$hr lightcurve. 
 The sensitivity plot of GU Psc b presented in Appendix 3 (available online) shows that this observation is insensitive to variability with amplitudes $<10\%$, and thus we cannot say whether the lightcurve has evolved from the variable epoch detected by \citet{Naud2017}. While this remains a prime target for variability monitoring, long observations and high photometric precision are needed to confirm and characterise its variability.

  \textit{2MASS J16154255+4953211} --- 2M1615+49 is identified as a young object by \citet{Cruz2007, Kirkpatrick2008, Allers2013}, although it has not been identified as a member of a young moving group. \citet{Kirkpatrick2008} tentatively assign this object an age estimate of $\sim100~$Myr based on its optical spectrum. \citet{Metchev2015a} obtained $21~$hr of \textit{Spitzer} $3.6~\mu$m and $4.5~\mu$m variability monitoring, finding significant variability in the $14~$hr $3.6~\mu$m sequence but not in the $7~$hr $4.5~\mu$m sequence. The authors estimate a period of $\sim24~$hr for 2M1615+49. We do not observe any significant variability in our UKIRT $J$-band observation of 2M1615+49 taken on Jul 10 2016. The sensitivity plot shown in Appendix 3 (available online) indicates that we are not sensitive to periods longer than $\sim10~$hr, so it is unsurprising that we did not detect variability in this long-period variable.

 \textit{HN Peg B} --- Discovered by \citet{Luhman2007}, HN Peg B is a T2.5 dwarf companion to the $300~$Myr old star HN Peg. \citet{Metchev2015a} report significant variability in both the \textit{Spitzer} $3.6~\mu$m and $4.5~\mu$m bands and estimate a period of $\sim18~$hr. \added{More recently, \citet{Zhou2018} obtained \textit{HST} WFC3 near-IR monitoring of HN Peg B, observing significant variability at all wavelengths in the range $1.1-1.7~\mu$m. They estimate a period of $15.4\pm0.5~$hr and measure a $J$-band amplitude of $1.28\pm0.3\%$.}
 We observed HN Peg B four times in total, twice with both the NTT and UKIRT telescopes, taking care to keep the primary HN Peg A off-frame. Although HN Peg A was kept off-frame for these observations, diffraction spikes still affected the quality of all of our observations. For our NTT data taken on October 8 2014 and August 7 2015, contaminated frames had to be removed from the lightcurve where the diffraction spikes coincided with the position of HN Peg B on the detector. For our UKIRT observations taken on July 11 2016 and July 13 2016 photometry from one nod position had to be removed from the data. Although the NTT lightcurves have lower $\sigma_{\mathrm{pt}}$, their short duration ($<2.5~$hr) means that they are insensitive to trends on timescales $>5~$hr. During two $\sim5~$hr observations with UKIRT, we do not detect any significant variability. Longer duration observations will be needed to characterise the variability of this young companion.
 
\begin{landscape}
\begin{table*}
\centering
\caption{Signatures of Youth in Sample}
\label{tab:sampleprops}
{\renewcommand{\arraystretch}{1.0}
\hspace*{-3cm}
\begin{tabular}{llllllllllll}
\hline \hline
Name                     & SpT (Opt)  & Ref  & SpT (IR)    & Ref         & K05 Class       & K05 Ref & AL13 Ind & AL13 Class &AL13 Ref & Signs of Youth & Ref  \\
\hline
2MASS J00011217+1535355   	&...	&... & L4   & G15a   & $\beta$          & G15a       & 1211          & {\sc int-g}               & G15a  & OTR            & G15a  \\
2MASS J00452143+1634446  	&L2	&C09 & L2   &AL13   & $\beta$          & G14a      & 1221          &{\sc vl-g}               & AL13 & OITRH          & G14a  \\
2MASS J01033203+1935361  	&L6 &F12 & L6   &M03   & $\beta$          & G14a      & 1n11          & {\sc int-g}              & AL13 & OITR           & G14a  \\
GU Psc b                 			&...	&... & T3.5  &N14  & ...              	& N14       &      ...         & ...                    &...      &RM                & N14     \\
2MASS J01174748-3403258  	&L2	&C03 & L1 &AL13  & $\gamma$         & G14a      & 1121          & {\sc int-g}               & AL13 & TRM            & G14a  \\
2MASS J02340093-6442068  	&L0	&K10 & L0     &F16  & $\gamma$         & G14a      &  2211             &  {\sc vl-g}                   & F16      & OR             & G14a  \\
2MASS J03032042-7312300  	&L2	&K10 &...     &...  & $\gamma$         & G14a      & ...              & ...                    &...      & OR             & G14a  \\
2MASS J03101401-2756452  	&L5	&C07 & ...      &... & $\gamma$         & G14a      &    ...           &      ...               & ...     & RHL           & G14a  \\
2MASS J03231002-4631237  	&L0	&C09 & L0     &F16  & $\gamma$         & G14a      &   2222            &  {\sc vl-g}                   &F16      & ORL            & G14a  \\
2MASS J03264225-2102057  	&L5	&G15a & L5    &F16  & $\beta / \gamma$ & G15a &    0n01          &         {\sc fld-g}           & F16     & RL        & G14a  \\
2MASS J03421621-6817321  	&L4	&G15a & ...     & ... & $\gamma$         &  G15a &   ...            &    ...                 &  ...    & R              & G15a  \\
PSO J057.2893+15.2433    	&...	& ...& L7     &B15  &   ...               	& B15       &...               &  ...                   &...      & R              & B15  \\
2MASS J03552337+1133437  	&L5	&C09 & L3      &AL13 & $\gamma$         & G14a      & 2122          & {\sc vl-g}                & G15a  & OITRL          & G14a  \\
2MASS J03572695-4417305  	&M9+L1.5	&M06 & ... & ... & $\beta$          & G14a      &      ...         &  ...                   &      & OR             & G14a  \\
2MASS J04185879-4507413  	&...	& ... & L3      & G15a & $\gamma$         & G15a       & 2211          & {\sc vl-g}                & G15a  & OR             & G15a  \\
2MASS J04210718-6306022  	&L5	&C09 & L5      &F16 & $\gamma$         & G14a      &    0n11          &{\sc int-g}                     &F16      & OIRL           & G14a  \\
PSO J071.8769-12.2713   	&...	 & ... & T2      & B15	&  ...                	& B15       &       ...        &          ...           &    ...  &       ...         &...      \\
2MASS J05012406-0010452  	&L4	&C09 & L3      &AL13 & $\gamma$         & G14a      & 2112          & {\sc vl-g}                & AL13 & OTRL           & G14a  \\
2MASS J05120636-2949540   	&L5	&K08 & L5   &G15a  & $\beta$          	& G15a       & 1n01          & {\sc int-g}              & G15a  & R              & G15a  \\
2MASS J05184616-2756457  	&L1	&C07 & L1    &AL13   & $\gamma$         & G15a       & 2222          & {\sc vl-g}               & AL13 & OITRU          & G14a  \\
2MASS J05361998-1920396  	&L2	&C07 & L2     &AL13  & $\gamma$         & G14a      & 2212          & {\sc vl-g}                & AL13 & OTR            & G14a  \\
SDSS J111010.01+011613.1 	&...	& & T5.5  &G15c   &                  	& G15c      &  ...             &  ...                   &  ...    & MR             & G15c \\
2MASS J12074836-3900043  	&L0	&G14b & L1      &G14b & $\delta$         & G15a       & 2222          & {\sc vl-g}                & G15a  & OITR           & G15a  \\
2MASS J12563961-2718455  	&...	&... & L4      &AL13 & $\beta$          & G15a       & 2021          & {\sc vl-g}                & G15a  & TR             & G15a  \\
2MASS J14252798-3650229   	&L3	&R08 & L4      &G15a & $\gamma$         & G15a       & 11?1          & {\sc int-g}               & G15a  & TR             & G15a  \\
2MASS J16154255+4953211  	&L4	&C07 & L3      &AL13 & $\gamma$         & G15a       & 2022          & {\sc vl-g}                & AL13 & OITRL          & G14a  \\
WISE J174102.78-464225.5 	&...	& ...& L7   &S14 & $\gamma$         & S14       &       ...        &      ...               &    ...  & ITRM           & S14  \\
PSO J272.4689-04.8036    	&...	&... & T1     &B15  &         ...         	& B15       &   ...            &      ...               &   ...   &    ...            &   ...   \\
2MASS J20025073-0521524  	&L6	&C07 & L5-7  &G15a   & $\gamma$         & G15a       & --             & --                   & F16  &    ...            &...      \\
2MASS J20113196-5048112   	&...	&... & L3      &G15a & $\gamma$         & G15a       & 2222          & {\sc vl-g}                & G15a  & OT             & G15a  \\
PSO J318.5338-22.8603   	&...	& ... & L7      &L13 & $\gamma$         & G15a       & XXX2, 2X21    & {\sc vl-g}                & L13  & ITRM           & G14a  \\
2MASS J21324036+1029494   	&...	& ...& L4   &  & $\beta$          	& G15a       & 00?1          & {\sc fld-g  }             & G15a  & TR             & G15a  \\
HN Peg B                 			&...	&... & T2.5 & L07   &            ...      	& L07       &   ...            &  ...                   &   ...   &   I             & L07      \\
SIMP J215434.5-105530.8  	&L4	&G14c& L5      &F16 & $\beta$          	& G14c     & 0n11          & {\sc int-g}               & G15a  & ITR            & G15a  \\
2MASS J2244316+204343    	&L6.5	&K08 & L6   &AL13 & $\gamma$         & G15a       & 2n21          &{\sc vl-g}                & AL13 & ITRLM          & G14a  \\
2MASS J23225299-6151275  	&L2	&C09 & L3     &F16 & $\gamma$         & G14a      & 1221          & {\sc vl-g}                & G15a  & OR             & G14a \\\hline 

\end{tabular}

}
\begin{flushleft}
 \textbf{References:}
AL13: \citet{ Allers2013}, 
B15 \citet{Best2015}, 
C03; \citet{Cruz2003},
C07; \citet{Cruz2007},
C09; \citet{Cruz2009},
F12: \citet{Faherty2012},
F16: \citet{Faherty2016}, 
G14a: \citet{Gagne2014a}, 
G14b: \citet{Gagne2014b},
G14c: \citet{Gagne2014c},
G15b \citet{Gagne2015b}, 
G15c: \citet{Gagne2015c}, 
K05: \citet{Kirkpatrick2005}
K08: \citet{Kirkpatrick2008},
K10: \citet{Kirkpatrick2010},
L13: \citet{Liu2013}, 
M03: \citet{Mclean2003},
M06: \citet{Martin2006},
N14: \citet{Naud2014a}, 
L07: \citet{Luhman2007}, 
R08: \citet{Reid2008},
S14 \citet{Schneider2014}\\
$^\mathrm{a}$ A capital letter means the object displays the associated sign of youth. 
O: lower-than normal equivalent width of atomic species in the optical spectrum, 
I: same but in the NIR spectrum, 
T: a triangular-shaped H-band continuum, 
R: redder-than-normal colors for given spectral type, 
U: over luminous, 
H: H$\alpha$ emission, 
L: Li absorption, 
M: signs of low gravity from atmospheric models fitting. 
\end{flushleft}
\end{table*}
\end{landscape}




\section{Assessing Evidence of Youth in the Sample}\label{sec:assess_sample}
\subsection{Analysing Sample Spectra}\label{sec:spectra}

Our targets have been identified as potentially young in the literature, through indications of low-gravity in their spectra and/or identification as probable members of young moving groups \citep[e.g. ][]{Cruz2009, Allers2013, Gagne2015c, Best2015}. In this section, we consider gravity-sensitive features in the spectra of our targets.
\citet{Kirkpatrick2005} present a spectral classification scheme for L0-L5 brown dwarfs that includes three gravity classes based on gravity sensitive features in their optical spectra. The three gravity subtypes $\alpha$, $\beta$ and $\gamma$, denote objects of normal gravity, intermediate gravity and very low gravity respectively. The $\delta$ suffix is used to designate objects with an even younger age (typically less than a few Myr) and lower surface gravity than those associated with the $\gamma$ suffix \citep{Kirkpatrick2006}.
 \citet{Gagne2015c} use optically anchored IR spectral average templates for classifying the gravity subtype for L0-L9 dwarfs. This method assigns $\alpha$, $\beta$ and $\gamma$ subtypes for each object. 
 \citet{Allers2013} present an index-based infrared gravity classification method that is based on FeH, VO, K {\sc i}, Na {\sc i} and $H$-band continuum shape in the IR. A score of 0 indicates that the feature is consistent with field gravity objects, 1 indicates intermediate gravity and 2 indicates very low gravity. A score of ``n'' is assigned if either the spectrum does not cover the wavelength range of the index or the feature is not gravity-sensitive at the object's spectral type. A score of ``?'' indicates that an index hints at low gravity, but the uncertainty in the calculated index is too large. The final gravity classification ({\sc fld-g, int-g, vl-g}) is assigned based on the median of the individual gravity scores, ignoring ``n'' or ``?'' scores. 

We present the spectral types, gravity subtypes and the specific signatures of low-gravity exhibited by each object in the sample in Table \ref{tab:sampleprops}.

\begin{table*}
\caption{Kinematic Information of Variability Sample}
\label{tab:kinematicdata}
 \begin{tabular}{llllllll}
 \hline \hline
Name                     	& $\mu_{\alpha}\cos\delta$  	& $\mu_{\delta}$ 	& Ref 		& RV    		 & Ref 		& $\pi$ &   Ref  \\
\hline
2M0001+15  		& $135.2  \pm 10.7$      		& $-169.6 \pm 13.7$       &F16         	&   ...          		&    ...     		&   ...          &  ...   \\
2M0045+16  		& $355 \pm 10$        			& $-40    \pm  10 $        & F16     	& $3.16 \pm 0.83$  & F16     	& $65.9\pm 1.3$    & L16 \\
2M0103+19  		& $293.0 \pm 4.6$      		& $ 27.7 \pm  4.7$        & F12     	&    ...        &   ...      			& $46.9 \pm 7.6$    & F12 \\
GU Psc b                 	& $90\pm 6$         			& $-102 \pm  6 $         & N14     	& $-1.6 \pm  0.4$   & N14     	&     ...       &  ...   \\
2M0117--34  		& $84 \pm 15$        			& $-45 \pm  8    $     & F16     		&($ 3.96 \pm  2.09 $) & F16     	& $26.1\pm1.9$      &L16     \\
2M0234--64  		& $88 \pm 12$        			& $-15\pm  12   $   & F16     		& $11.762 \pm 0.721$ &F16         	& ($21 \pm 5) $     & F16 \\
2M0303--73  		& $43\pm 12 $       			& $3  \pm  12   $	& F16    		&  ...            & ...        			&    ...           &...     \\
2M0310--27  		& $-119 \pm 18$       			&$ -47\pm  16$         & C08     		&   ...            &   ...      			&      ...         &..     \\
2M0323--46 		& $66 \pm 8$         			&$ 1 \pm  16 $        & F16     		& $13.001 \pm  0.045$ & F16    & ($17 \pm 3$)      & F16 \\
2M0326--21  		& $108 \pm14 $       			&$ -146 \pm  15$         & F16     	& ($22.91  \pm  2.07 $) & F16     & ($41\pm 1$)      & F16 \\
2M0342--68 		& $65.3 \pm 2.8 $      		& $18.5  \pm  9.1$        & F16     	& ($13.87  \pm  2.62$)  & F16     & ($21  \pm9 $)     & F16 \\
PSO 057.2+15  	&$ 68 \pm 11$        			& $-127 \pm  12$         & B15     	&     ...     & ...       				&...            &...     \\
2M0355+11		& $225  \pm13.2$      		&$ -630\pm  15  $       & F16     		& $11.92  \pm  0.22 $ & F16     & $109.5 \pm 1.4$    & F16 \\
2M0357--44 		&$ 64 \pm 13$        			& $-20 \pm  19$         & F16     		& $10.73 \pm  4.6$   & F16     	&   ...          &...     \\
2M0418--45		& $53.3\pm 8.4$       			&$ -8.2\pm  12.6 $      & F16     		&         ...   &   ...      			&  ...           & ...    \\
2M0421--63 		& $146 \pm 8$         			&$ 191\pm  18$         & F16     		& $14.7   \pm  0.33$ & F16     	&  ...             &   ...  \\
PSO 071.8--12    	& $20 \pm 19  $      			&$ -89 \pm  19$         & B15     		&     ...          &       ...  		&...               &    ... \\
2M0501--00  		& $190.3\pm 9.5$       		&$ -142.8 \pm 12.5  $     & F16     	& $21.77  \pm  0.66$  & F16     & $48.4  \pm 1.4$    & F16 \\
2M0512--29   		& $-10 \pm13$       			& $80 \pm  15 $        & F16     		&            ...  &    ...     		&     ...         &  ...   \\
2M0518--27  		& $28.6\pm4.2$       			& $-16\pm  4 $         & F16     		& $24.35  \pm  0.19$  & F16     & $18.4\pm1.1 $   & F16 \\
2M0536--19  		& $24.6 \pm5.3$       			&$ -30.6 \pm 5$          & F16    		& $22.065 \pm  0.695$ & F16    & $21.1  \pm 1.6$    & F16 \\
SDSS 1110+0		& $-217.1 \pm 0.7$       		&$-280.9 \pm  0.6$        & G15a   	& $7.5   \pm  3.8$   & G15a    	& $52.1  \pm 1.2$    & D12 \\
2M1207--39		& $-57.2 \pm7.9 $      		& $-24.8 \pm  10.5 $      & F16     	& ($9.48 \pm  1.91$)  & F16     	& ($15   \pm 3 $)     & F16 \\
2M1256--27 		& $-67.4 \pm 10.2$      		& $-56.5\pm  12.7  $     & F16     	&...             & ...        			&     ...         &  ...   \\
2M1425--36  		& $-284.89 \pm1.4$       		& $-463.08 \pm  1 $         & F16     	& $5.37 \pm 0.25$  & F16     	& $86.45 \pm0.83$  & F16 \\
2M1615+49  		& $-80 \pm 12 $       			& $-18  \pm  12  $       & F16     		& $-25.59 \pm  3.18$  & F16     & $32    \pm 1$      & L16 \\
WISE 1741--46		& $-20.4 \pm 9.2 $      		& $-343 \pm  13.7$       & F16     	& $-5.7   \pm  5.1$   & F16     	&   ...           &...     \\
PSO 272.4--04    	& $-46 \pm 4 $        			& $-400 \pm  13$         & B15     	& ...              & ...        			& ...              &...     \\
2M2002--05  		& $-98  \pm 5  $      			 & $-110\pm  8   $       & F16     		&  ...             & ...        			&   ...            & ...    \\
2M2011--50		& $21.3 \pm 8.1  $    			 & $-71.3\pm  14.5$       & F16     	&    ...           &    ...     			&   ...            & ...    \\
PSO 318.5-22  		& $137.3 \pm 1.3 $     		 & $-138.7\pm  1.4  $      & F16     	& $-6.0  \pm  0.95 $ & A16     	& $45.1  \pm 1.7$    & L16 \\
2M2132+10   		& $107.8  \pm16.4 $     		& $29.7   \pm  18.1$       & F16     	&...               &  ...       			&...              & ...    \\
HN Peg B                 	&        ...           				 &   ...                 			&  ...      	 &  ...            	&   ...      		&     ...          & ...    \\
SIMP 2154--10  		& $175     \pm12$        		&$ 9     \pm  12   $      & F16     		&   ...            &      ...  		 	& $32.6\pm1.0$    & L16 \\
2M2244+20   		& $252   \pm14    $    		& $-214 \pm  11 $        & F16     	& $-16.0  \pm  0.85$  & V17     & $58.7  \pm 1.0   $ & L16 \\
2M2322--61  		&$ 62\pm10    $    		& $85 \pm  9    $      & F16     		& $6.747\pm0.75$  & F16     & 	($22\pm1$)			&F16 \\ 
\hline
\end{tabular}
 \begin{flushleft} \textbf{References: }
 C08: \citet{Casewell2008}
 D12: \citet{Dupuy2012}
 G15a: \citet{Gagne2015b}
 F12: \citet{Faherty2012}
 F16: \citet{Faherty2016}
 N14: \citet{Naud2014a}
 V17: \citet{Vos2017} 
 \end{flushleft}
 \end{table*}

\subsection{Assessing Group Membership}\label{sec:groupmem}
Following a similar method to \citet{Faherty2016}, we investigate the likelihood that each object in the survey is a member of a young moving group using four methods of assessing group membership using kinematic data; the convergent point analysis of \citet{Rodriguez2013}, the BANYAN I tool of \citet{Malo2013}, the BANYAN $\Sigma$ method in \citet{Gagne2018b} and the LACEwING analysis of \citet{Riedel2017}.

Convergent point analysis estimates the probability of membership using the perpendicular motion of the candidate member and the convergent point location of a given moving group, but does not take into account radial velocity or parallax. This method considers six potential moving groups: TWA, THA, $\beta$ Pic, AB Dor, CarN and Col.
BANYAN I uses a Bayesian statistical analysis to identify members of kinematic groups. BANYAN I minimally requires the position, proper motion, magnitude and colour of a star but radial velocity and distance measurements can be added. In addition to the groups considered by the convergent point analysis of \citet{Rodriguez2013}, BANYAN I investigates membership in the Argus association.
BANYAN $\Sigma$ is a new Bayesian algorithm for identifying members of young moving groups that includes 27 young associations. This algorithm improves upon BANYAN I and II \citep{Malo2013, Gagne2014a} by using analytical solutions when marginalising over radial velocity and distance, using multivariate Gaussian models for the young moving groups and removing several approximations in the calculation of Bayesian likelihood. BANYAN $\Sigma$ does not include the Argus association in its analysis, as it is likely that this association suffers from a high level of contamination \citep{Bell2015}. Proper motions, radial velocities and parallaxes used in this analysis are shown in Table \ref{tab:kinematicdata} 

\begin{table*}
\centering
\caption{Moving Group Membership Probabilities}

\label{tab:groupmem}
{\footnotesize
 \begin{tabular}{lllllllllll}
\hline \hline
Name                    & Convergence$^{\mathrm{a}}$& P$(\%)$ & BANYAN I & P$(\%)$ & BANYAN $\Sigma$ & P$(\%)$ & LACEwING    & P$(\%)$ & Mem$^{\mathrm{b}}$  & Decision$^{\mathrm{d}}$  \\
\hline
2M0001+15   & AB Dor           & 43.2        & AB Dor   & 99.04       & AB Dor        & 76.0        & AB Dor      & 46          & AM            	& Young \\
2M0045+16  & CarN             & 74.2       	 & Arg      & 99.97       & CarN          & 89.0        & Argus       	& 99          & AM               & Young\\
2M0103+19  & CarN     	& 15.7        	& Old      & 96          & Field         & 76.1        & $\beta$Pic   	 & 44          & AM              	& Young \\
Gu Psc b     & ABDor            &99.5             & ABDor         & 99.88            &ABDor               &89    &ABDor     &82             & BM          	& Young  \\
2M0117-34  & THA              & 99.6        	& THA      & 92.4      & THA           & 79.3        	& THA     		& 51          & AM              &Young\\    			
2M0234-64  & THA              & 79.5        	& THA      & 99.99       & THA           & 96.3          & THA     		& 86          & HLM              &Young  \\
2M0303-73  & THA              & 97.6       	 & Old      & 92.5        & FLD           & 99.9        & None      		&0             & NM               	&Young\\ 
2M0310-27  & CarN             & 99.9       	 & Old      & 100         & FLD           & 99.9        & None      		&0             & AM               	&Young  \\
2M0323-46  & THA              & 92.1        	& THA      & 99.97       & THA           & 51.2        & THA     		& 41          &AM              &Young \\  
2M0326-21  & ABDor            & 65.8        	& ABDor    & 99.37       & ABDor         & 91.9        & AB Dor      & 55        & AM            	&Young \\
2M0342-68  & THA              & 98.3        	& THA      & 98.4       & THA           &67.7        & None      	&0             &AM            & Young\\
PSO 057.2+15    & ABDor    & 82.4     	& $\beta$Pic     & 65.19       & ABDor   & 67.2    & None      	&0             & AM               &Uncertain  \\
2M0355+11  & ABDor            & 17.5        & ABDor    & 99.99       & ABDor         & 99.9        & AB Dor     	 & 100         & BM             &Young\\
2M0357-44  & THA         	& 62.1        & THA      & 53.07       & Field, THA    & 66.6, 17.5  & THA     		& 39          & AM            	&Young \\
2M0418-45   & ABDor            & 91.5        & ABDor    & 64          & ABDor         & 41.9        & AB Dor      	& 24          & AM               &Young \\
2M0421-63  & $\beta$Pic, CarN & 95.8, 95.9  & $\beta$Pic     & 92.8        & CarN          & 93.8   & CarN 	& 38          & AM              &Young \\
PSO 071.8-12   & ABDor            & 77.8        & ABDor    & 62.7        & ABDor         & 56.5        & Col     	& 32          & AM              &Uncertain \\
2M0501-00 & THA              & 98.7        	& Old      & 99.86       & Field         & 99.9        & AB Dor      	& 70          & AM              &Young \\
2M0512-29   & CarN             & 5.5         	& Old      & 89.49       & Field         & 99.9        & None      		& 0            & NM               &Young\\
2M0518-27  & CarN             & 22.1        & Col      & 86.5        & $\beta$Pic          & 62.7        & Col     		& 77          & AM              & Young\\
2M0536-19  & $\beta$Pic    & 77.5        & $\beta$Pic     & 57.28       & $\beta$Pic  & 81.1  & Col     		& 52          & AM               & Young\\
SDSS1110+01 & ABDor            & 17.2        & ABDor    & 99.91       & ABDor         & 99.3        & AB Dor      & 46          & BM               &Young\\
2M1207-39  & $\beta$Pic, TWA  & 100, 91.6   & TWA      & 99.14       & TWA           & 91.6        & TWA      & 98          & HLM             & Young\\
2M1256-27  & ABDor            & 75.8       	 & TWA      & 99.14       & Field         & 99.8        & None      	& 0            & AM              & Young\\
2M1425-36   & ABDor            & 39.3        	& ABDor    & 99.98       & ABDor         & 99.7        & AB Dor      & 100         & BM               &Young\\
2M1615+49  & ABDor            & 59.3        	& ABDor    & 95.12       & ABDor         & 84.0        & AB Dor      & 68          & AM               &  Young\\
WISE 1741-46 & $\beta$Pic, ABDor   & 94.1, 91.8  & $\beta$Pic     & 99.88       & ABDor, $\beta$Pic & 52.7, 45.9  & AB Dor & 71& AM & Young    \\
PSO 272.4-04    & TWA     & 61.8        & ABDor    & 86.45       & ABDor         & 89.0        & AB Dor      	& 29          & AM               &Uncertain \\
2M2002-05  & None             & 0           & Old      & 100         & Field 		    & 99.9        & None      		&0             & NM             &Young \\ 
2M2011-50 2 & Col              & 96.5        & THA      & 66.53       & Field         & 72.2        & None      		&0             & AM               &Young \\
PSO 318.5-22    & $\beta$Pic & 98.9      & $\beta$Pic     & 99.99       & $\beta$Pic & 99.6 & $\beta$ Pic    	& 69          & BM               & Young\\
2M2132+10   & CarN             & 92.8        & Arg      & 53.44       & Field         & 99.9        & None     		 &0             & AM               & Young\\
HN Peg B                 &   ...       &  ...           &...          &   ...          & ...              & ...            &  ...           &...             	& ...            		& Young \\
SIMP 2154-10  & CarN             & 28.8        & Old      & 89.97       & CarN          & 71.4        & None     	 &             & AM               &Young \\
2M2244+20 & ABDor            & 68.9        & AB Dor   & 99.99       & ABDor         & 99.8        & AB Dor      	& 100         & BM            & Young\\
2M2322-61  & THA              & 34.1        & THA      & 99.78      & THA           & 79.7        & THA     		& 97          & AM             &Young\\
\hline
\end{tabular}
}
\begin{flushleft}
 \textbf{Notes: } \\
$^\mathrm{a}$ Moving groups: AB Dor: AB Doradus, Arg: Argus, $\beta$ Pic: $\beta$ Pictoris, CarN: CarN, Col: Col, THA: Tucana-Horologium, TWA: TW Hydrae.\\
$^\mathrm{b}$ BM: bona fide member, HLM: high-likelihood member, AM: ambiguous member, NM: non-member.\\
$^\mathrm{c}$ Final decision based on spectral signatures of youth, membership probabilities and colour-magnitude diagrams. Young: Object is a high likelihood member of a young-moving group and/or has clear indications of youth in its IR and/or optical spectrum. Uncertain: This object does not have sufficient evidence of youth and is thus excluded from the final statistical sample. \\
\end{flushleft}
\end{table*}


The results of each membership tool should be evaluated differently. \citet{Malo2013} and \citet{Gagne2018b} use a threshold of $90\%$ to confirm membership for the  BANYAN I and $\Sigma$ tools respectively. We use this threshold probability of $90\%$ for the Convergent Point tool \citep{Rodriguez2013}. \citet{Riedel2017} find that a membership probability $>66\%$ indicates a high membership likelihood using LACEwING.

We present the results of each method in Table \ref{tab:groupmem}. To assess the membership probability of each object based on the results of the kinematic analysis, we use the categories outlined in \citet{Faherty2016}:
\begin{enumerate}
  \item \textit{Non-member} (NM): an object that is rejected from nearby associations due to its kinematics.
  \item \textit{Ambiguous member} (AM): an object requiring higher precision kinematics because it is classified as a candidate to more than one group or cannot be differentiated from field objects.
  \item \textit{High-likelihood member} (HLM): an object that does not have full kinematic information (proper motion, radial velocity and parallax) but is regarded as high confidence ($>90\%$ for BANYAN I, BANYAN $\Sigma$ and Convergent Point analysis, $>66\%$ in LACEwING) in at least three out of four algorithms.
  \item \textit{Bona fide member} (BM): an object regarded as a high-likelihood member with full kinematic information.
\end{enumerate}

\citet{Faherty2016} carried out this analysis on a larger sample of potential young objects, using Convergent Point analysis, BANYAN I, BANYAN II and LACEwING.
Our results, which substitutes BANYAN $\Sigma$ for BANYAN II, are mostly consistent with those found in \citet{Faherty2016} with a few exceptions.
2M0303--73 drops from an ambiguous member of Tucana-Horologium in \citet{Faherty2016} to a non-member in our analysis. 2M0045+16 had been previously identified as a bona fide member of Argus \citep{Faherty2016}, however given the uncertainty in the Argus group, \citet{Gagne2018b} excluded this group from the analysis. The convergent point method and LACEwING assign it to the Argus association while BANYAN I and $\Sigma$ assign it to the Carina-Near association. 
The object 2M0117--34 drops from a high-likelihood member to an ambiguous member. Compared to analysis in \citet{Faherty2016}, we include a parallax measurement from \citet{Liu2016} for this object. All moving group tools favour the Tucana-Horologium association, however BANYAN $\Sigma$ and LACEwING probabilities are below $90\%$ and $66\%$ respectively. 
2M0323--46, 2M0342--68 and 2M2322--61 all drop from a high-likelihood member to an ambiguous member due to lower membership probabilities calculated in BANYAN $\Sigma$ compared to BANYAN II.
2M0326--21 is classified as an ambiguous member of AB Doradus because the Convergent Point tool and LACEwING predict membership probabilities of $66\%$ and $55\%$ respectively.
As can be seen in Table \ref{tab:groupmem}, our sample is composed of six bona fide members, two high-likelihood members, twenty-four ambiguous candidates and three non-members.

\begin{figure}
  \hspace*{-.75cm}
 \subfloat{
\includegraphics[width=0.52\textwidth]{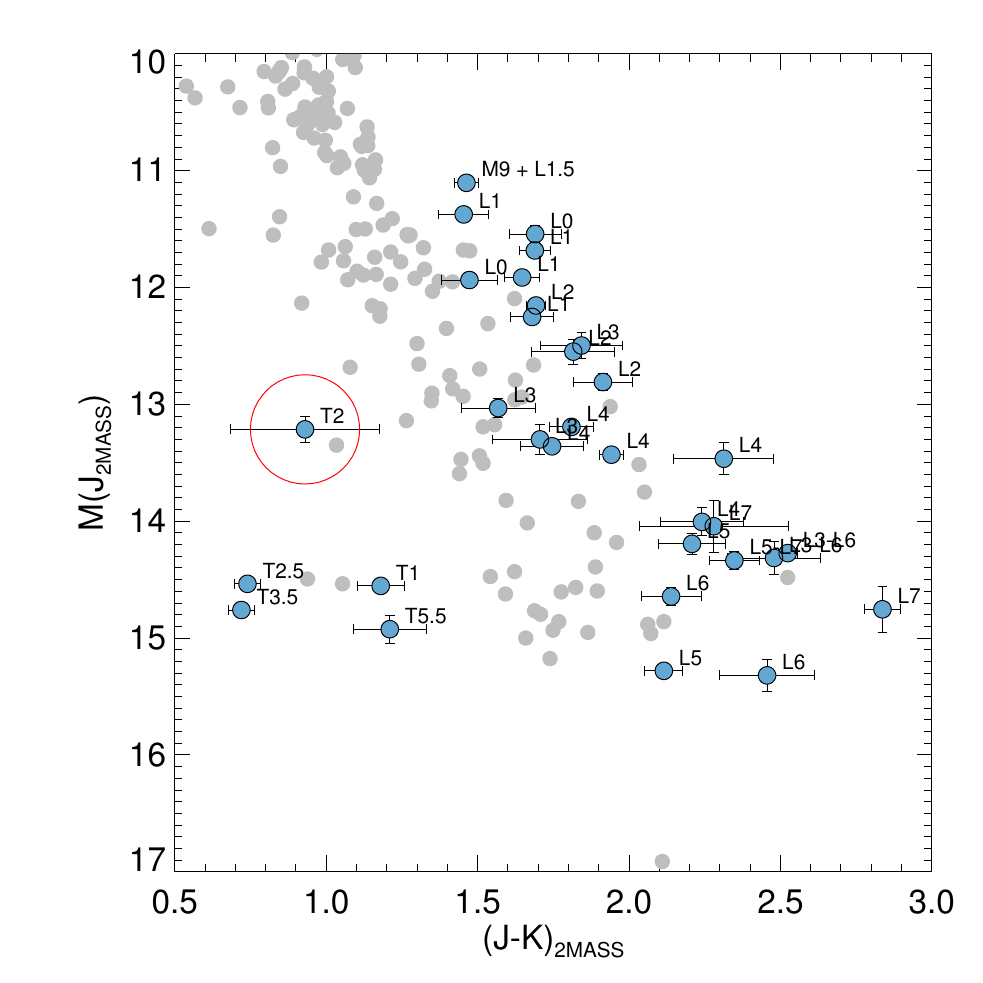}} \\
  \hspace*{-.75cm}
 \subfloat{
\includegraphics[width=0.52\textwidth]{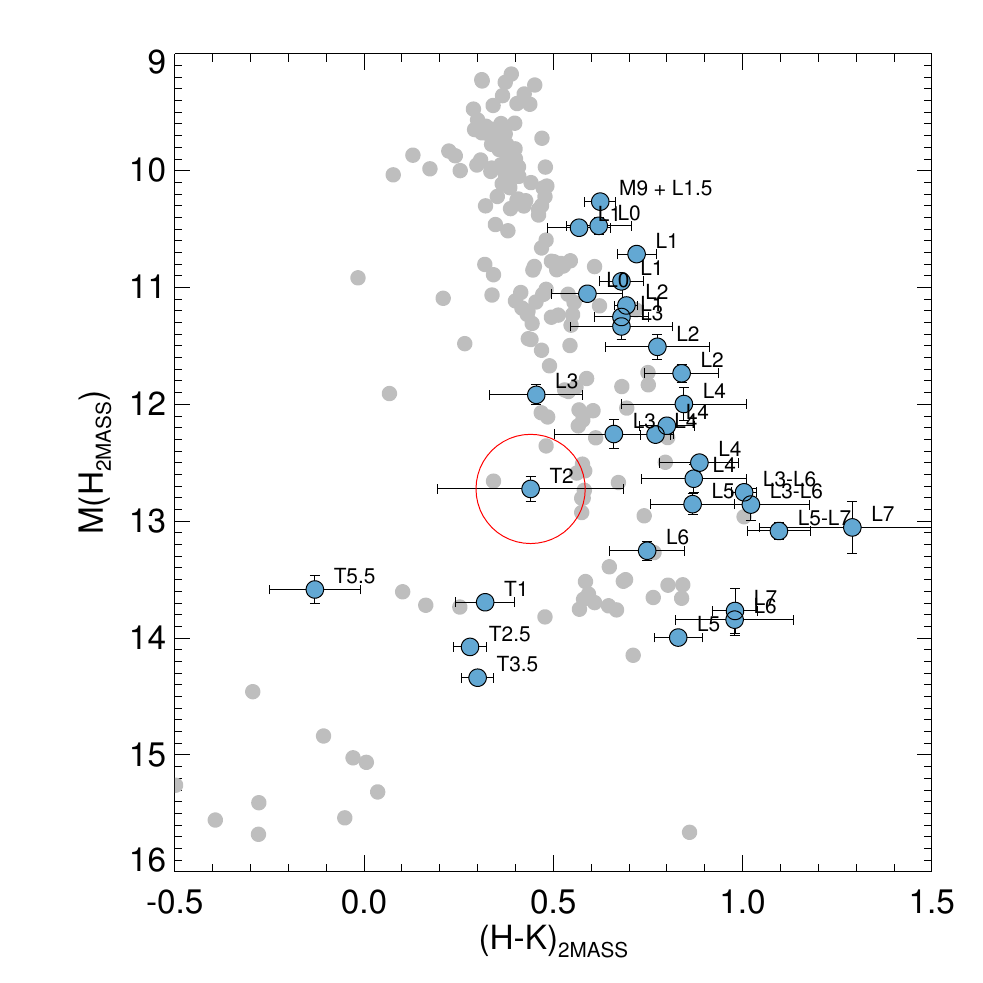}}\\
\caption{Colour-magnitude diagrams showing the field brown dwarf population (grey points) and our full sample of objects showing signs of low-gravity. Absolute magnitudes were calculated either using measured parallaxes or kinematic distances. From this analysis it is clear that the T2 object PSO 071.8-12 (shown in red circle) is an outlier in the sample. Assuming AB Doradus membership for PSO 071.8-12 results in magnitudes that are $\sim1~$mag brighter than other T dwarfs in the field and young populations.}
\label{fig:colmags}
 \end{figure}

We additionally look at our sample on a colour-magnitude diagram to check that they follow the general trends seen in intermediate and low-gravity objects to date \citep{Liu2016, Faherty2016}. Many objects in the sample have measured parallaxes from \citet{Dupuy2012, Faherty2012, Faherty2016, Liu2016}. 
When parallaxes were not available we used their estimated distance from kinematic group membership. We plot absolute magnitude against colour in Figure \ref{fig:colmags}. Overall, the survey objects appear redder and more luminous than the field brown dwarf population, as seen in a larger sample of young objects by \citet{Liu2016}. However the object PSO 071.8--12, shown by red circle in Figure \ref{fig:colmags}, appears to be an outlier in this sequence. PSO 071.8--12 was discovered by \citet{Best2015}, who find that it is a high probability candidate of $\beta$ Pictoris using BANYAN II. However, our group membership assigns PSO 071.8--12 a moderate probability candidacy of the AB Doradus moving group, with very low probability candidacy to $\beta$ Pictoris. The estimated kinematic distance of $45\pm7~$pc assuming AB Doradus membership results in an absolute magnitude that is $\sim1~$mag brighter than both T-type field brown dwarfs and T-type low-gravity objects. This over-luminosity could be explained is PSO 071.8--12 is a binary, however \citet{Best2015} do not identify PSO 071.8--12 as a possible binary based on its spectrum. If PSO 071.8--12 is not a binary then we estimate that PSO 071.8--12 must lie at a distance of $\sim20-30~$pc, and thus is not a member of AB Doradus. Intriguingly, $\beta$ Pic membership would imply a distance of $19\pm4~$pc according to BANYAN $\Sigma$, which would result in magnitudes consistent with other T dwarfs. More kinematic data is needed to robustly assess the binarity and youth of PSO 071.8--12.

\begin{figure}
\centering
\includegraphics[width=0.5\textwidth]{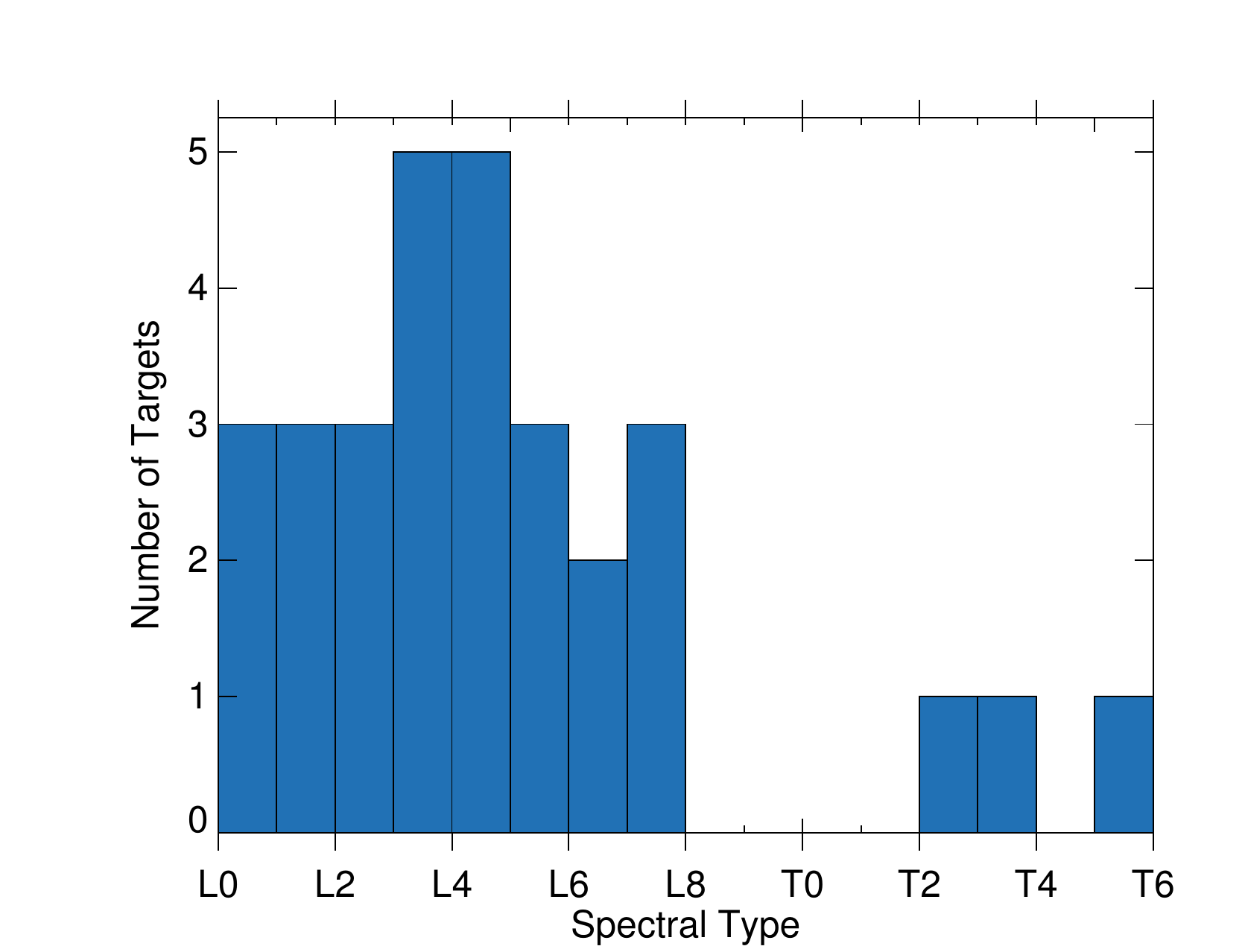}\\
\caption[Histogram showing the distribution of spectral types in our final sample of 30 low-gravity brown dwarfs.]{Histogram showing the distribution of spectral types in our final sample of 30 low-gravity brown dwarfs. Few young, low-gravity T dwarfs bright enough for variability observations are known, so our sample is predominantly composed of objects with spectral types $<\mathrm{L}8.5$}
\label{fig:samplehist}
 \end{figure}

Combining the available kinematic information and spectral information for each target in the survey, we make a final call on whether the objects presented in Tables \ref{tab:sampleprops} and \ref{tab:groupmem} are likely low-gravity. We exclude 3 objects from the original survey on the basis that there is insufficient evidence of youth. These objects are classified as `Uncertain' in Table \ref{tab:groupmem}. We discuss the excluded objects below.
The object PSO 057.2+15 is excluded from the survey. This L7 object appears redder than the field population and exhibits a triangular shaped $H$-band, however a gravity class could not be determined due to its low S/N spectrum \citep{Best2015}.
The moving group tools suggest possible membership in AB Doradus or $\beta$ Pictoris but are not consistent with each other. Updated kinematics and/or spectral analysis are needed to confirm the possible youth of this object. We thus exclude PSO 057.2+15 from the `Young' sample.
PSO 272.4--04 is a low-probability AB Doradus member using 3/4 membership tools. Gravity sensitive indices only apply to objects with spectral types $\geq$L7, since low-gravity spectral signatures are not very well established for late-L and T-type objects. Thus, signatures of youth for the T1 objects PSO 272.4--04 could not be analysed \citep{Best2015}.
 We thus exclude it from the `Young' sample.
Gravity indices could not be analysed for the T2 object PSO 071.8--22 for the same reason.
PSO 071.8--12  also has uncertain membership status, as discussed above, and we thus exclude it from the final statistical analysis of the survey.  
We additionally exclude the binary 2M0357--44 from the final survey. Although 2M0357--44 is likely young \citep{Cruz2009, Gagne2015a}, we have a reduced likelihood of detecting variability in either component of the binary, since the non-variable component would effectively dilute the variability signal.
2M2322--61 and 2M0412--63 are also left out of the survey because they were observed during poor weather conditions which prohibited us from determining meaningful constraints on their variability properties. Their lightcurves are shown in Appendix 2 (available online).
In total we exclude 6 objects that we observed in our survey from the final sample of 30 young, low-gravity objects used in our analysis. We show the distribution of spectral types in our 30 object sample in Figure \ref{fig:samplehist}.

\section{Variability Statistics}

\begin{figure}
\includegraphics[width=\columnwidth]{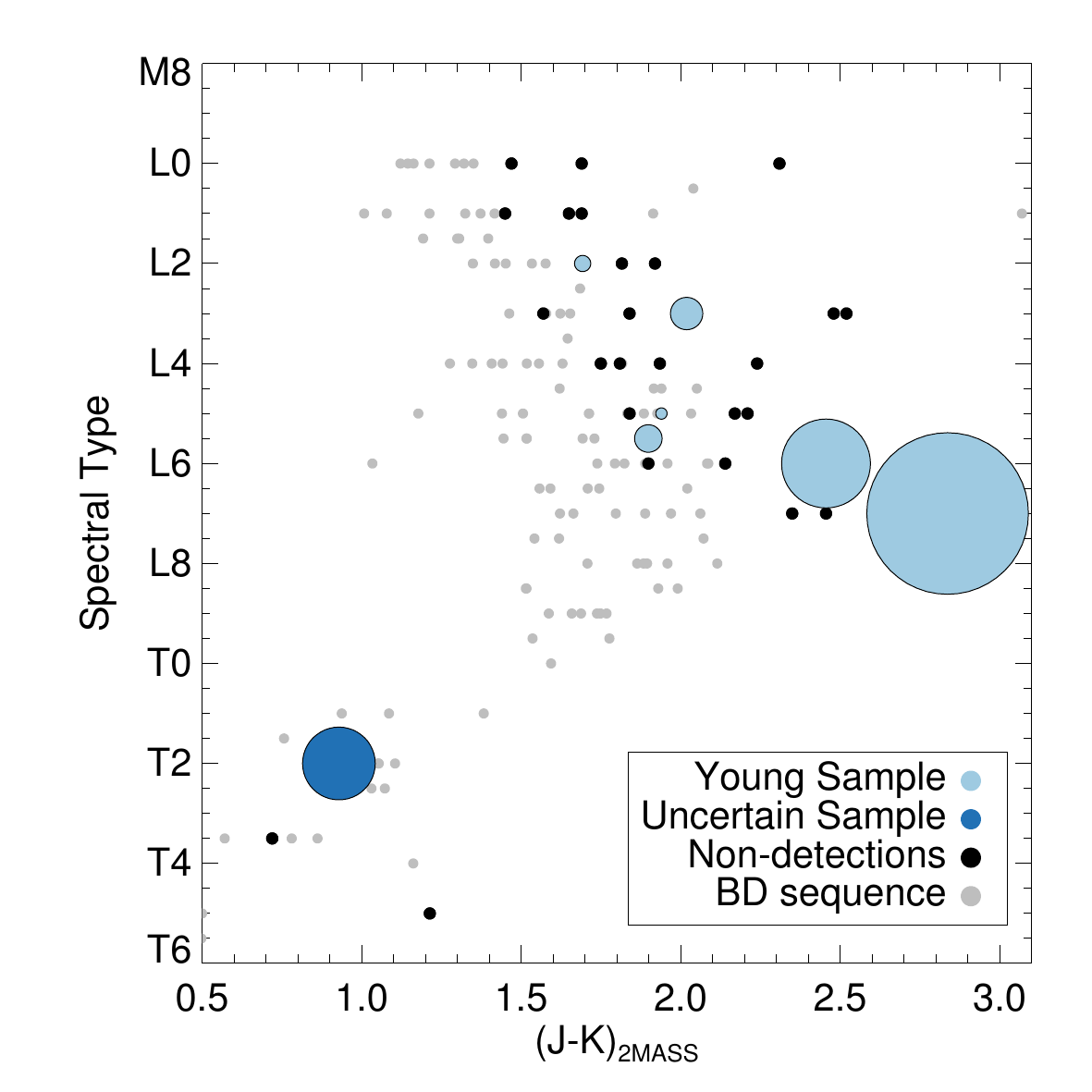}
\caption{Spectral type of variable objects plotted against $(J-K)_{\mathrm{2MASS}}$ colour. Blue symbols represent young objects displaying significant photometric variability, where the radius is proportional to the variability amplitude. Dark blue symbols denote the objects that are highly likely to be young while light blue circles denote objects whose youth is more uncertain.}
\label{fig:col_spt}
\end{figure}

Figure \ref{fig:col_spt} shows the spectral type of our sample plotted against $(J-K)_{2MASS}$ colour. Blue symbols correspond to variability detections, where the symbol size is proportional to the variability amplitude. Although many of our measured variability amplitudes are only a lower limit estimate, we see evidence for increasing $J$-band amplitude along the L sequence, something that is noted in \citet{Metchev2015a} for mid-IR variability.

To explore the effect of low surface gravity on variability properties,
we compare our results to the results of variability surveys of the field brown dwarf population reported by \citet{Radigan2014,Radigan2014a}. Our methods for determining variability significance are very similar to the methods presented by \citet{Radigan2014}. In both papers the significance of a variability detection is based on the Lomb-Scargle periodogram power. In \citet{Radigan2014}, the $1\%$ FAP periodogram power is empirically increased in the same way as our analysis (Section \ref{sec:idvar}). Furthermore, the same method is used to create the sensitivity plots in both surveys -- by injecting simulated sinusoidal signals into reference star lightcurves and measuring the recovery rates. Thus, the two surveys can be robustly compared.

We find that 6/30 ($20\%$) of objects in the full statistical sample exhibit significant variability, similar to the 9/57 ($16\%$) reported by \citet{Radigan2014} for a similar high-gravity sample of field dwarfs.
However, with only three young objects with spectral types $>$L9 included in our sample, we are lacking in L/T transition and T spectral type objects compared to the  \citet{Radigan2014} and \citet{Radigan2014a} samples and thus we cannot obtain a robust comparison between the low-gravity and high-gravity populations as a function of spectral type. We thus consider objects with spectral types of  L0-L8.5 in both samples. We find that 6/27 of L0-L8.5 low-gravity objects appear variable while \citet{Radigan2014a} report 2/34 variables in the field brown dwarf sample of L0-L8.5 objects

\subsection{Statistical Formalism}
Our formalism for the statistical analysis of this survey is based on the method described in \citet{Lafreniere2007} and \citet{Vigan2012}. We consider the observation of $N$ targets enumerated by $j=1...N$. We note $f$, the fraction of objects that exhibit variability with amplitude and rotational period in the interval $[a_{\mathrm{min}},a_{\mathrm{max}}]\cap[r_{\mathrm{min}},r_{\mathrm{max}}]$, and $p_j$, the probability that such variability would be detected from our observations. With this notation, the probability of detecting variability in target $j$ is $(fp_j)$ and the probability of not detecting variability is $(1-fp_j)$. Denoting ${d_j}$ the detections made by the observations, such that $d_j=1$ for a variability detection in target $j$ and 0 otherwise, the likelihood of the data given $f$ is
\begin{equation}\label{eq:likelihood}
L \big({d_j}|f \big) = \prod\limits_{j=1}^N(1-fp_j)^{1-d_j}(fp_j)^{d_j}
\end{equation} 
According to Baye's theorem, from the a priori probability density $p(f)$, or prior distribution, and the likelihood function $L$, we can calculate the posterior distribution $p(f|{d_j})$, the probability density updated in light of the data:
\begin{equation}\label{eq:posterior}
p\big( f|{d_j} \big) = \frac{L \big({d_j}|f \big) p(f)}{\int_{0}^{1}L \big({d_j}|f \big) p(f)\mathrm{d}f}
\end{equation}
This is the frequency of variable objects, or variability occurrence rate of objects in the survey.

\subsection{Estimating the Frequency of Variable Objects}
\begin{figure}
  \includegraphics[width=\columnwidth]{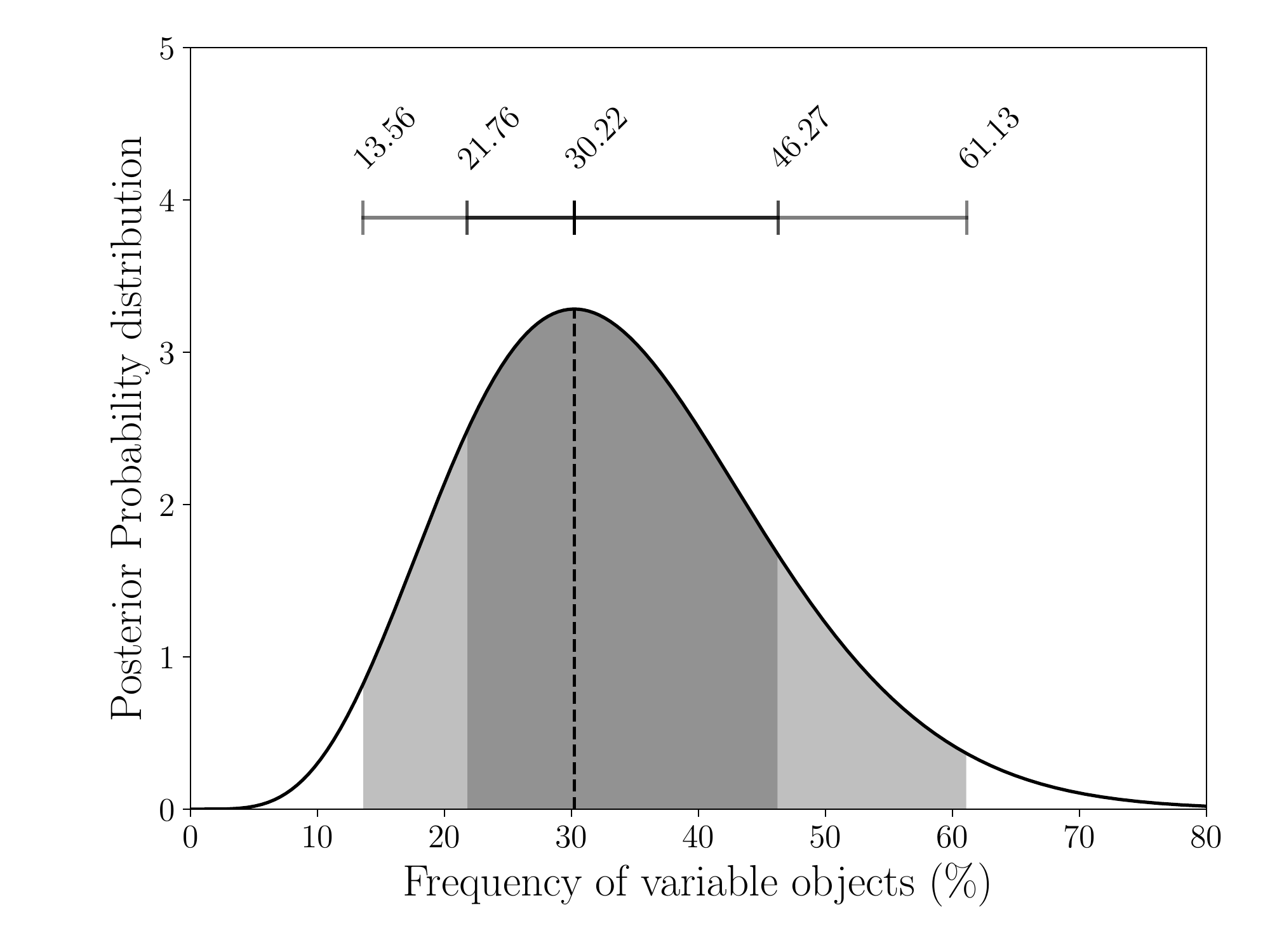}
\caption{Probability distribution of the frequency of variable objects in the 'Young' sample. The dark grey area shows the $1\sigma$ regions while the light grey area shows the $3\sigma$ region.}
\label{fig:PDF_lg}
\end{figure}

\begin{figure}
  \includegraphics[width=\columnwidth]{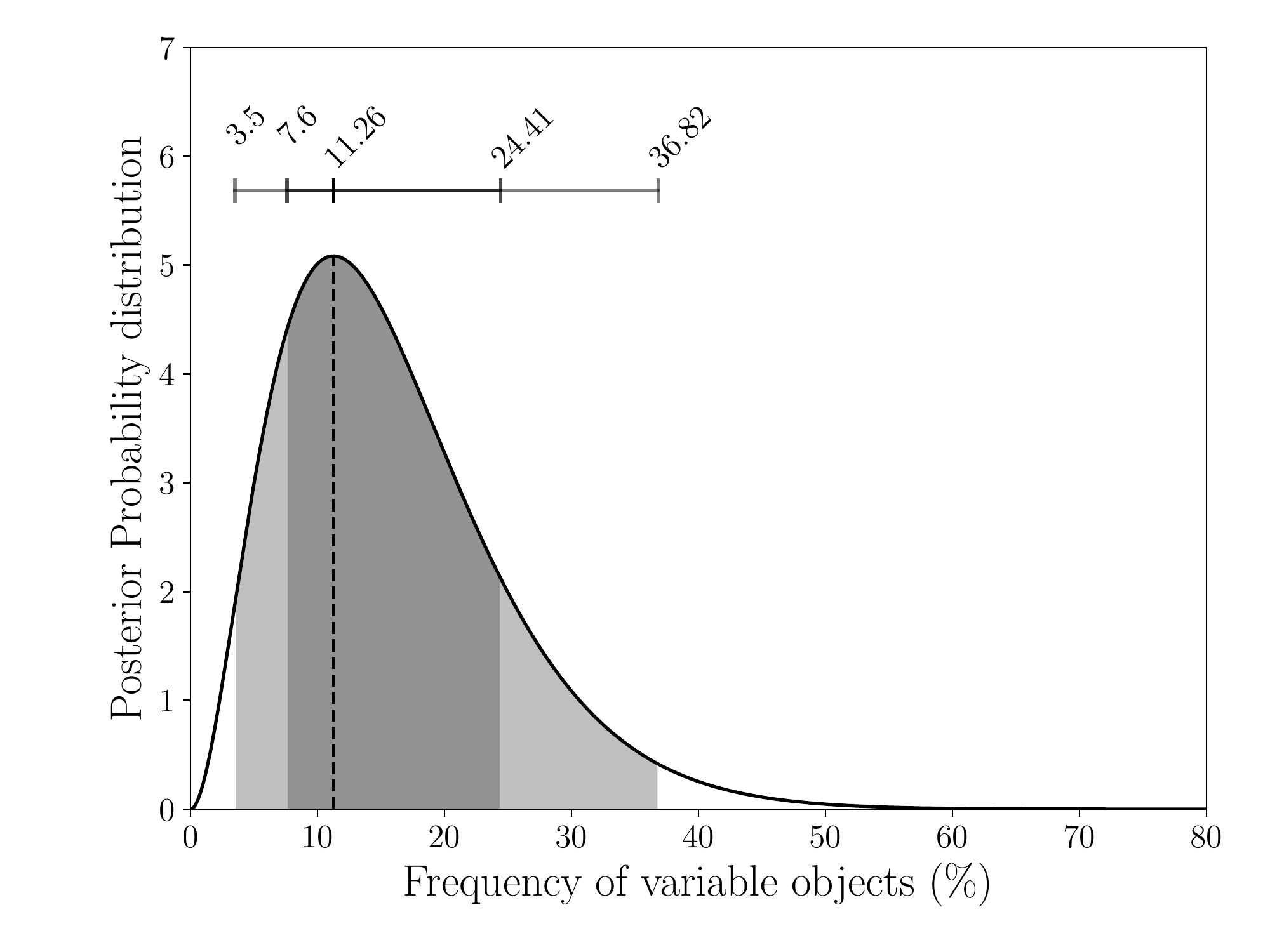}
\caption{Probability distribution of the frequency of variable objects in the field brown dwarf sample \citep{Radigan2014a}. The dark grey area shows the $1\sigma$ regions while the light grey area shows the $3\sigma$ region.}
\label{fig:PDF_field}
\end{figure}

We use a modified version of the Quick Multi-purpose Exoplanet Simulation System  \citep[QMESS;][]{Bonavita2013, Bonavita2016} to calculate the posterior distribution of the frequency of variable objects in they survey.
QMESS is a grid-based, non-Monte Carlo simulation code that uses direct-imaging sensitivity plots to estimate the frequency of giant planets. We use QMESS to estimate the fraction of objects that display variability using the statistical framework discussed above. We use the sensitivity plots obtained for each observation (described in Section \ref{sec:sensplot}) as $p_j$ in Equation \ref{eq:likelihood} to calculate the likelihood, $L$ for values of $f$ between 0 and 1. For the \citet{Radigan2014a} sample, we use the average sensitivity plot from the \citet{Radigan2014} survey. This is reasonable since the reported photometric precision and observation lengths are comparable for both surveys \citep{Radigan2014a}.
Since we have no prior knowledge of the variability frequency of low-gravity brown dwarfs, we use a uniform prior distribution of $p(f) =1$ in Equation \ref{eq:posterior} to calculate the posterior probability density, $p( f|{d_j})$. This is the probability density function (PDF) of the frequency of variable objects.
The PDFs of the frequency of variable objects for the `Young' sample and field brown dwarf sample \citep{Radigan2014a} are plotted in Figures \ref{fig:PDF_lg} and \ref{fig:PDF_field}. For the low-gravity sample analysed in this work, we find the frequency of variables objects is $30^{+16}_{-8\%}$, which is higher than the rate of $11^{+13}_{-4}\%$ that we find for the \citet{Radigan2014a} survey.


\deleted{We use the Kolmogorov-Smirnov (K-S) test to quantitatively compare the PDFs of the young and field brown dwarf samples. The K-S test estimates the probability that two distributions are drawn from the same distribution function. We find a probability $<0.5\%$ that the PDFs are drawn from the same underlying distribution. Thus, we have found the first quantitative indication that the L-type low-gravity objects are more likely to be variable than the higher mass field dwarf counterparts.}

\added{We additionally employ a second method to analyse how statistically significant the correlation between low-gravity and frequent variability is. To do this we use a Bayesian framework to analyse the $2\times2$ contingency table shown in Table \ref{tab:contingency}, following the method described by \citet{Biller2011} to determine the probability that the samples are drawn from different distributions. We denote $y_1$ as the number of young objects with detected variability and $y_2$ as the number of field objects with detected variability. We model the number of variable objects as a binomial function: }

\begin{align}
y_1 \sim \mathrm{Binom}(n_1,\theta_1) \\
y_2 \sim \mathrm{Binom}(n_2,\theta_2) 
\end{align}

\begin{figure*}
\subfloat{ 
\includegraphics[width=0.5\textwidth]{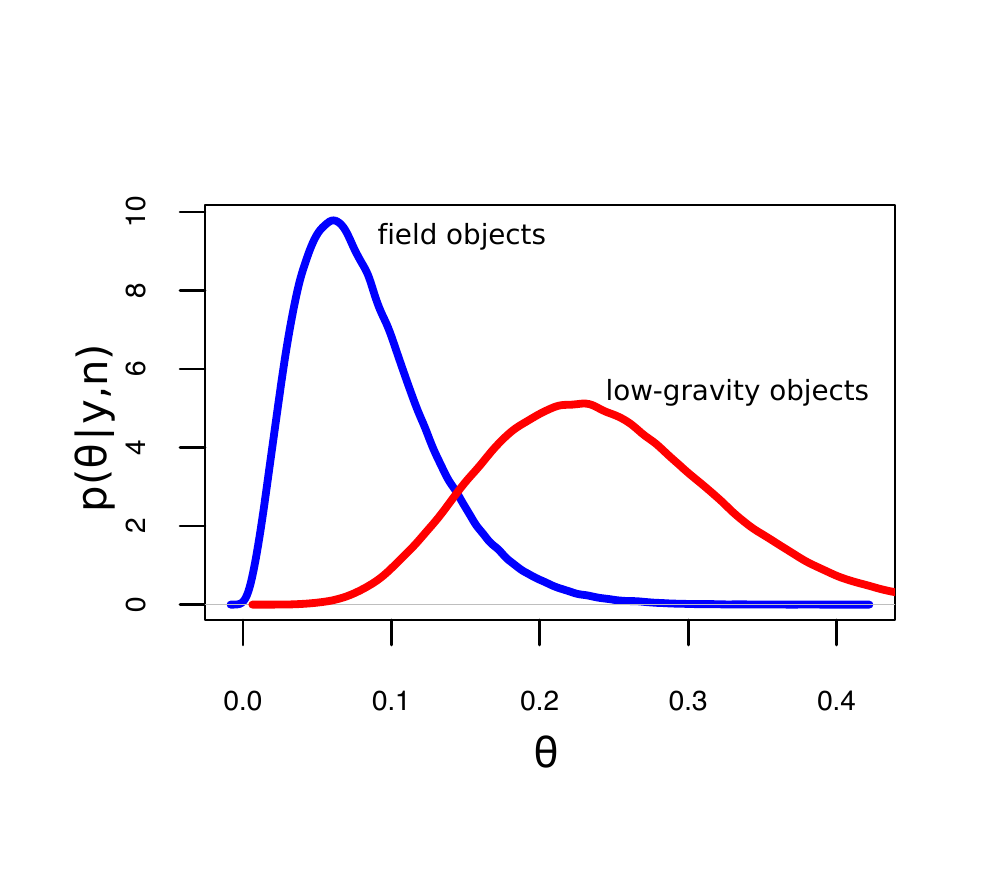}}
\subfloat{  \hspace*{-0.65cm}
\includegraphics[width=0.5\textwidth]{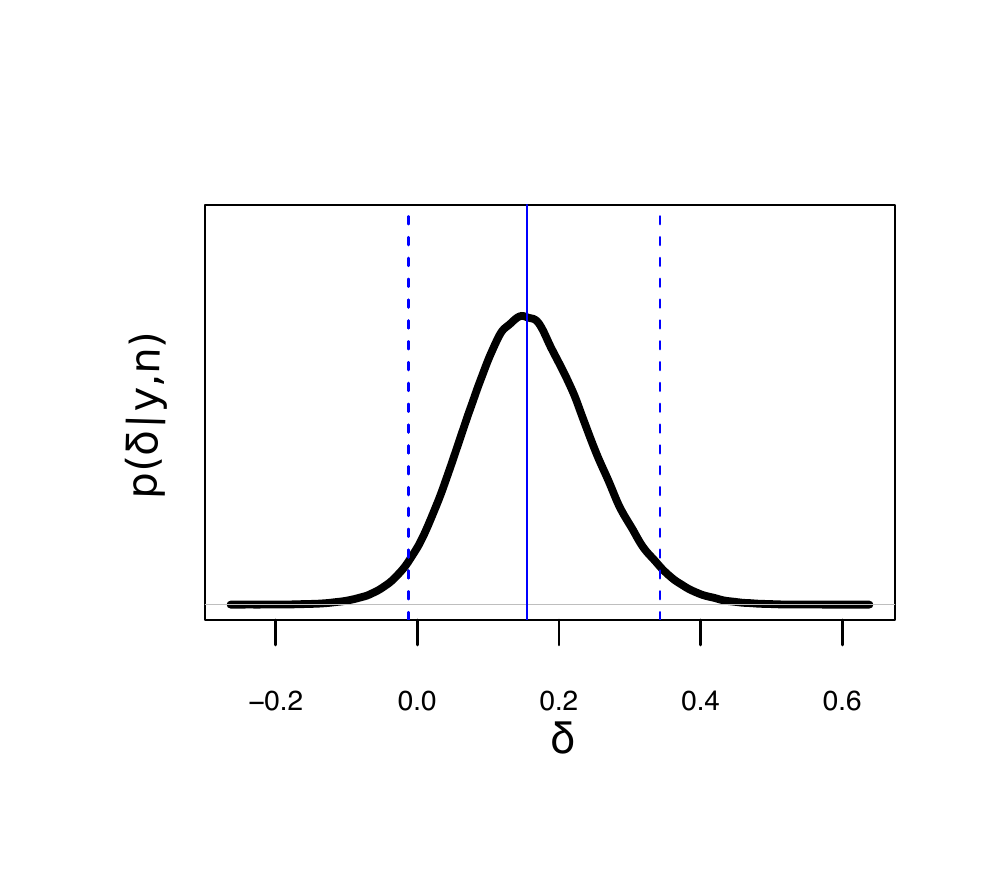}}\\
\caption{Left panel shows probability distributions for the variability occurrence rate of the field brown dwarf sample (blue) from \citet{Radigan2014} and the low-gravity sample (red) from this survey, assuming binomial statistics and a uniform prior. The right panel shows the difference between these distributions. We find a $98\%$ probability that the planetary-mass sample has a higher variability occureence rate than the field brown dwarf sample.}
\label{fig:bayes}
\end{figure*}

\added{where $n_1, n_2$ are the total sample sizes and $\theta_1, \theta_2$ are the variability occurrence rates of the low-gravity and field objects respectively. We use uniform priors on the fraction of variable for each population:}
\begin{align}
\theta_1 \sim \mathrm{Unif}(0,1) = \mathrm{Beta}(1,1)\\
\theta_2 \sim \mathrm{Unif}(0,1) = \mathrm{Beta}(1,1)
\end{align}
\added{Since the Beta distribution is a conjugate prior to the binomial distribution we can analytically compute the posteriors of the variability occurrence rate for each sample:}
\begin{align}
p(\theta_1| y_1, n_1) =  \int^{+\infty}_{-\infty} \mathrm{Beta}(\theta_1| y_1 + 1, n_1 - y_1 +1)\\
p(\theta_2| y_2, n_2) =  \int^{+\infty}_{-\infty} \mathrm{Beta}(\theta_2| y_2 + 1, n_2 - y_2 +1)
\end{align}
\added{We plot the probability distributions of the variability occurrence rates in the left panel of Figure \ref{fig:bayes}. We define the difference between the variability occurrence rates as $\delta = \theta_1 - \theta_2$. We then draw 10000 simulations from the joint posterior $p(\theta_1, \theta_2 | y_1, n_1, y_2, n_2)$ and estimate the probability that $\delta>0$ by the fraction of samples, $m$, where $\theta_1^{m}>\theta_2^{m}$. We plot the distribution $\delta$ in the right panel of Figure \ref{fig:bayes}. We find a $98\%$ probability that the variability occurrence rates of the field brown dwarf and low-gravity populations are drawn from different distributions. Thus, our survey strongly suggests that the low-gravity L-type objects appear more variable than their higher mass counterparts. }

\begin{table}
\centering
\caption{Contingency table showing the number of variability detections and non-detections in the \citet{Radigan2014a} survey  (field objects) and this survey (low-gravity).}
\label{tab:contingency}
\begin{tabular}{lll}
\hline \hline
              & Variable & Non-Variable \\
              \hline
Field objects & 2        & 32           \\
Low-gravity   & 6        & 21          \\
\hline
\end{tabular}
\end{table}

Due to a low number of young T dwarfs in our sample, our survey cannot place strong constraints on the variability properties of low-gravity T-type objects. However there are a number of high-amplitude variability detections in young T dwarfs that suggest that this trend between low-gravity and high amplitude variability may extend into the T dwarfs.
\citet{Metchev2015a} report \textit{Spitzer} $3.6~\mu$m and $4.5~\mu$m in the intermediate-gravity T2.5 companion HN Peg B, the only low-gravity T dwarf in the survey.
The T2.5 known variable object SIMP 0136 was recently found to be a likely planetary-mass member of the Carina-Near moving group \citep{Gagne2017}. With its $1-6\%$ $J$-band variability, SIMP 0136 exhibits one of the highest variability amplitudes of the known variable T dwarfs. 
\citet{Gagne2018a} recently reported that the T2 object 2M1324+63 is a planetary-mass member of the AB Doradus moving group.
This object is known to exhibit high-amplitude variability in the optical and the mid-IR \citep{Heinze2015, Apai2017} and also provides a good spectrophotometric match to the directly-imaged planet HR8799b \citep{Bonnefoy2016}.
Finally, we report high-amplitude variability in PSO 071.8--22 in this survey. Although it does not have sufficient evidence of youth to be classed as `Young' in our sample, additional kinematic information may confirm PSO 071.8--12 as a young object. As we identify more low-gravity T dwarfs it will become clearer whether the link between low-gravity and enhanced variability holds for cooler T-type objects.

\section{Follow-up Observations of Variable Objects} \label{sec:followup}
When possible, we obtained follow-up observations of objects found to be variable in their first epoch. These observations were carried out so that we could confirm variability and also look for evidence of lightcurve evolution for our variable objects \citep{Apai2017, Vos2018}. We obtained follow-up observations of 2M0045+16, PSO 071.8--12, 2M0501--00, 2M1425--36 and PSO 318.5--22. We discuss each object below.

\textit{2M0045+16 ---} We observed 2M0045+16 on November 11 2014 and August 17 2015 with the NTT and November 13 2016 with UKIRT. 2M0045+16 was found to be variable in two out of three epochs - the NTT November 11 2014 and UKIRT November 13 2016 observation. As can be seen from the sensitivity plots in Figure \ref{fig:3panels}, both lightcurves exhibit a similar shape, with periodograms indicating a period of $\sim3-6~$hr. We measure amplitudes of $1.0\pm0.1\%$ and $0.9\pm0.1\%$ for the 2014 and 2016 lightcurves respectively, thus we do not see any indication of lightcurve evolution in this case. The NTT August 17 2015 lightcurve shows a similar trend, however the periodogram peak power does not fall above our significance threshold. According to the sensitivity plot of the August 17 2015 observation (shown in Appendix 3 which is available online), we can place an upper limit on the variability amplitude of this epoch of $\sim2\%$ for a rotational period of $\sim3-6~$hr. Thus, we did not reach the photometric precision necessary to robustly detect a $1\%$ modulation in the lightcurve in this observation.

\textit{PSO 071.8--12 ---} We reobserved the variable object PSO 071.8--12 with UKIRT on December 8 2017. During this $4~$hr observation we do not detect significant variability. The sensitivity plot shown in Appendix 3 (available online) rules out significant variability $>5\%$ for short periods, however we believe that PSO 071.8--12 has a somewhat longer period. For a rotational period of $5-8~$hr we place an upper limit of $6-8\%$ on the variability amplitude of PSO 071.8--12 in this epoch. 

\textit{2M0501--00 ---} We observed 2M0501--00 a total of four times with the NTT. We detect significant variability on November 11 2014 and October 19 2016 and do not detect variability on August 16 2015 and March 12  2017. In the two variable epochs (Figures \ref{fig:variables1} and \ref{fig:3panels}), which are separated by almost two years we observe a similar lightcurve shape - a slowly decreasing relative flux over the entire observation. Both periodograms favour a period $>5~$hr and the Levenberg-Marquardt least-squares fits give amplitudes of $1-2\%$, although both the rotational period and variability amplitude are very uncertain since we did not observe a maximum or minimum in either lightcurve.  We do not detect variability during two $\sim2~$hr observations on August 16 2015 and March 12 2017 (shown in Appendix 3; available online). As these observations are shorter than the variable epochs, they are less sensitive to long period variability. The sensitivity plot from August 16 2015 shows that this observation is sensitive to amplitudes $>6\%$ for a rotational period of $5~$hr. The March 12 2017 lightcurve is noisier than the other epochs due to poor weather conditions, and the sensitivity plot indicates that the observation is not sensitive to periods of $\geq5~$hr in this epoch. Thus, our observations do not show evidence for an evolving lightcurve in this case.

\textit{2M1425--36 ---} We obtained two epochs of variability monitoring of 2M1425--36 with NTT/SofI. The initial observation on August 17 2015 shows a low-amplitude ($\sim0.7\%$) trend with a period $>2.5~$hr. Our second epoch observation, obtained using the NTT on March 14 2017 (shown in Appendix 4; available online) suffered from poor weather conditions, with the seeing ranging from $0.9-1.7\arcsec$. While the sensitivity plot suggests sensitivity to very low variability amplitudes, the periodograms of the reference stars display significant trends due to changing weather conditions.

\textit{PSO 318.5--22 ---} As part of the initial survey observations, we obtained three epochs of NTT variability monitoring for the L7 object PSO 318.5--22. On October 9 2014 we observed significant $J_S$ variability, with an amplitude of $10\pm1.3\%$, and a period $>5~$hr (Figure \ref{fig:3panels}). The $J_S$ lightcurve obtained on November 9 2014 again shows similar variability, this time varying with an amplitude of $4.8\pm0.7\%$ over a $\sim3~$hr observation. Finally, we observed PSO 318.5--22 in the $K_S$ band on November 10 2014. Since PSO 318.5--22 is much brighter in $K_S$, we attain higher photometric precision in this band. The lightcurve shows a smooth upward trend with an amplitude of $2.2\pm0.6\%$  (Figure \ref{fig:3panels}). All three lightcurves were originally presented in \citet{Biller2015}.

\begin{figure*}
\includegraphics[width=0.78\textwidth]{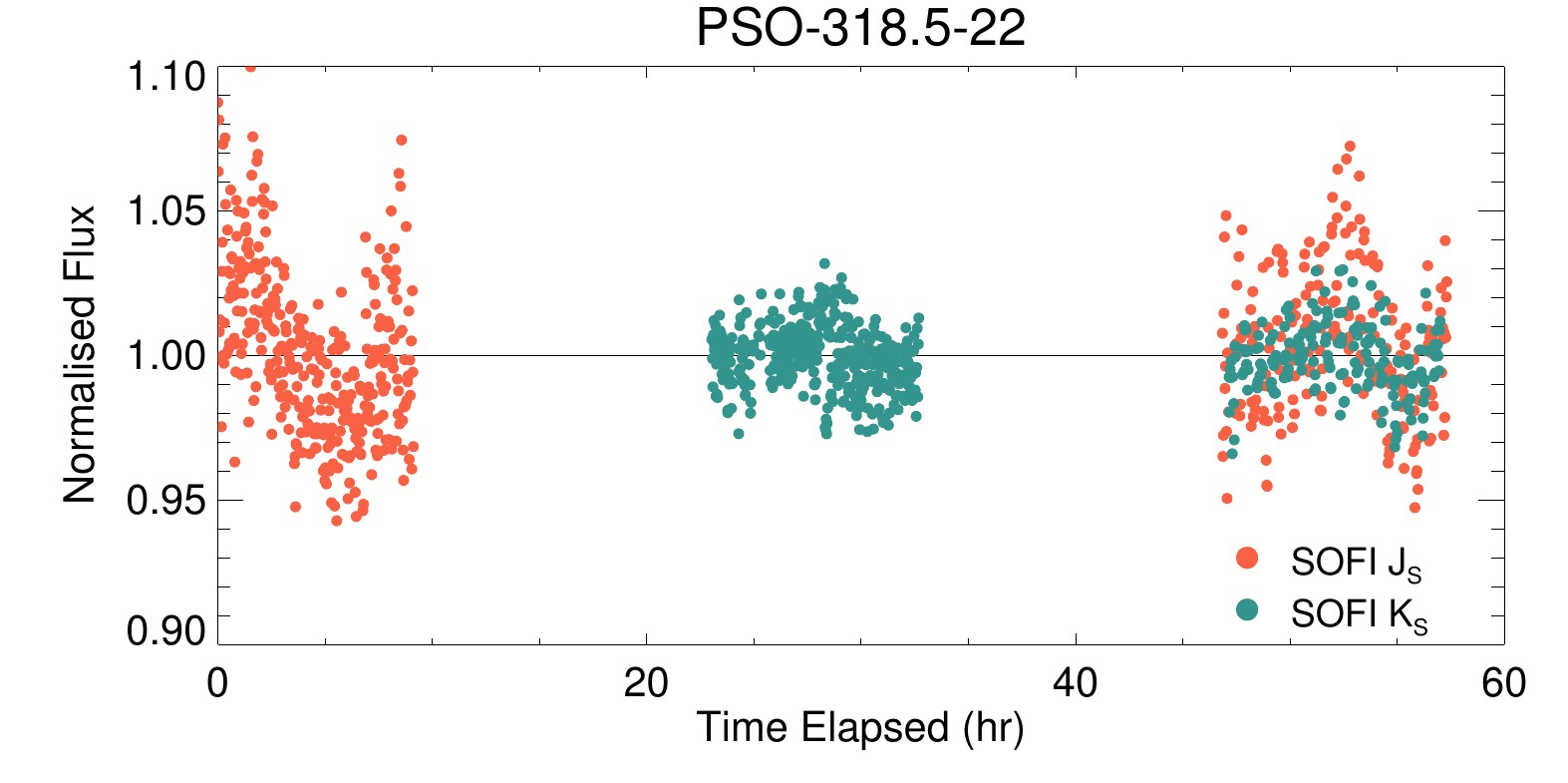}
\caption{Follow-up observations of PSO 318.5--22 taken over three consecutive nights. Orange points show observations taken with the $J_S$ filter and the teal points show the $K_S$ filter observation. These three nights allow us to constrain the rotational period of PSO 318.5--22 to $\sim8.5~$hr, in agreement with the $8.6~$hr period reported by \citet{Biller2018}.}
\label{fig:PSO-3night}
\end{figure*}

\section{Follow-up Observations of PSO 318.5--22}\label{sec:followup_PSO}

Additional observations of PSO 318.5--22 were taken in August 2016, to more accurately constrain the rotational period and to investigate the wavelength dependence of the variability. On August 9 2016 we observed using the $J_S$ filter, followed by the $K_S$ filter on August 10 2016. Finally, on August 11 2016, we observed with both filters, swapping over every 20 minutes. The data were reduced and analysed as described in Sections \ref{reduc1}, \ref{sec:lcanalysis} and \ref{sec:idvar}. The corrected lightcurves for all three nights are shown in Figure \ref{fig:PSO-3night}. We detect significant variability in both bands in all three epochs.

The $J_S$ lightcurve obtained on August 9 2016 shows significant variability with an amplitude of $2.4\pm0.2\%$. This is the highest amplitude detected over the course of the three nights but is much lower than the initial variability detection on October 9 2014. This suggests that we are observing a quiescent phase of the variability of PSO 318.5--22, as has been observed previously in a number of variable brown dwarfs \citep{Artigau2009,Radigan2012,Apai2017}. The $K_S$ lightcurve obtained on August 10 2016 also shows significant variability. We fit a sinusoid to the lightcurve to obtain a variability amplitude of $0.48\pm0.08\%$.

The simultaneous $J_S$ and $K_S$ monitoring obtained on August 11 2016 allows us to directly compare the variability in both bands during a single rotational phase. We observe significant variability in both filters and we measure a $A_K/A_J$ ratio of $0.36\pm0.25$. This is similar to the ratios previously observed for the population of field brown dwarfs \citep{Artigau2009, Radigan2012}, suggesting that the variability mechanism for the exoplanet analogues is similar to that of the field brown dwarfs. 
\citet{Biller2018} obtained simultaneous \textit{HST} WFC3 and \textit{Spitzer} IRAC variability monitoring of PSO 318.5--22, detecting variability amplitudes of $\sim3\%$ in the \textit{Spitzer} $3.6~\mu$m band and$\sim4-6\%$ in the near-IR bands ($1.07-1.67~\mu$m). The variability amplitude was found to decrease with increasing wavelength and we observe this same trend with the high $J_S$ amplitude and lower $K_S$ amplitude.

Additionally, we find that the $J_S$ and $K_S$ variability is in phase.  \citet{Biller2018} report large phase shifts between the near-IR and mid-IR lightcurves of PSO 318.5--22, and tentative phase shifts between the near-IR spectral bands. Phase changes have been attributed to different wavelengths probing different heights in the atmosphere \citet{Buenzli2012, Biller2013}, so this suggests that the $J_S$ and $K_S$ bands probe similar heights in the photosphere, while the mid-IR band is sensitive to surface homogeneities that are located at higher atmospheric levels.

 \begin{figure*}
\includegraphics[width=0.9\textwidth]{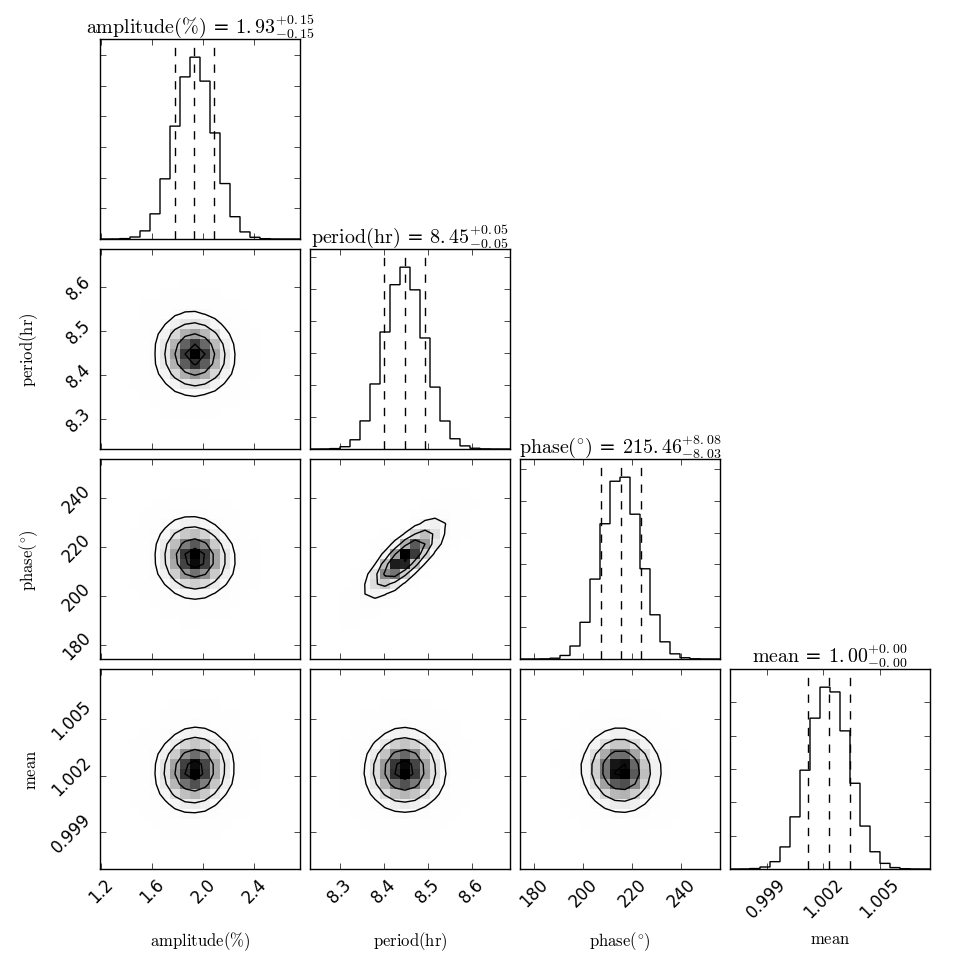}
\caption{PSO 318.5-22 posterior distributions for amplitude, period, phase and the mean obtained from MCMC analysis.}
\label{fig:triangle}
\end{figure*}

The long baseline of this observation allows us to constrain the rotational period of PSO J318.5-22 using Monte Carlo analysis. 
We use the \textsc{emcee} package \citep{fm2013} to obtain the full posterior probability distributions for each parameter of the sinusoidal model. We use 500 walkers with 
$20000$ steps and discard an initial burn-in sample of $1000$ steps to explore the four-dimensional parameter space to model the light curve. Figure \ref{fig:triangle} shows the posterior distributions of the amplitude, period, phase and constant parameters of the fit. Each parameter is well-constrained, and the MCMC method gives a rotational period of $8.45\pm0.05~$hr for PSO J318.5--22. The derived error of $0.05~$hr on the period is very small, which shows the advantage of having long temporal coverage. However it is important to keep in mind that the above fit assumes a sinusoidal model that did not change between nights, and in reality the full light curve may resemble a more complex or evolving function.
In any case, our estimated rotational period is consistent with the $8.6\pm0.1~$hr period reported by \citet{Biller2018} using $16~$hr of \textit{Spitzer} monitoring. This further supports our choice of a sinusoidal model and the derived posterior parameters.



\section{Conclusions}
We report the first large survey for photometric variability  in young low-gravity brown dwarfs with NTT/SofI and UKIRT/WFCAM. We monitored a total of 36 objects continuously for $\sim 2 - 6~$hr, detecting significant ($p>99\%$) variability in seven objects. We assess the spectral indicators of youth and moving group membership of each object in the sample, finding that three objects have rather uncertain ages and are thus left out of the survey analysis. We also leave one unresolved binary out of the survey and lose two objects due to poor weather conditions. We detect variability in six objects that are likely to be young, four of which are new detections of variability.

In the `Young' sample, we detect variability in 6/30 ($20\%$) objects, which is consistent with the $16\%$ variability fraction reported by \citet{Radigan2014} for the higher mass, field dwarfs. However, since we are lacking in objects with spectral types $>$L9 compared to earlier surveys of field L and T dwarf population, we focus our analysis on the L0-L8.5 objects in our sample.
We find that the frequency of variable L0-L8.5 objects in this survey is $30^{+16}_{-8}\%$, which is higher than the frequency of variable objects of $11^{+13}_{-4}\%$ that we find for the field brown dwarf population \citep{Radigan2014a}. We find that the PDFs of the variability occurrence rates of our two samples are drawn from different underlying distributions with a probability of \added{$98\%$.} Thus, we have found the first quantitative indication that the L-type low-gravity objects are more likely to be variable than the higher mass field dwarf counterparts.

We additionally present 3 consecutive nights of photometric monitoring of the highly-variable L7 spectral type object PSO 318.5--22 with NTT/SofI. We find no evidence of phase shifts between the $J_S$ and $K_S$ bands and find a $A_K/A_J$ ratio of $0.36\pm0.25$, consistent with previous amplitude ratios of field brown dwarfs \citep{Artigau2009, Radigan2012}. This suggests that the underlying variability mechanism is the same for both populations.
We perform MCMC analysis on the $J_S$ light curves to measure a rotational period of $8.45\pm0.05~$h for PSO 318.5--22.

\section*{Acknowledgements}
Based on observations collected at the European Organisation for Astronomical Research in the Southern Hemisphere under ESO programmes 194.C-0827, 095.C-0590  097.C-0693 and 098.C-0546 and observations collected at the United Kingdom Infrared Telescope.
JV acknowledges the support of the University of Edinburgh via the Principal's Career Development Scholarship. 
BB gratefully acknowledges support from STFC grant ST/M001229/1.
KA acknowledges support from the Isaac J. Tressler Fund for Astronomy at Bucknell University. 
UKIRT is owned by the University of Hawaii (UH) and operated by the UH Institute for Astronomy; operations are enabled through the cooperation of the East Asian Observatory. When the data reported here were acquired, UKIRT was supported by NASA and operated under an agreement among the University of Hawaii, the University of Arizona, and Lockheed Martin Advanced Technology Center.
The authors wish to recognise and acknowledge the very significant cultural role and reverence that the summit of Mauna Kea has always had within the indigenous Hawaiian community. We are most fortunate to have the opportunity to conduct observations from this mountain.




\bibliographystyle{mnras}
\bibliography{Full}

\begin{thebibliography}{}
\makeatletter
\relax
\def\mn@urlcharsother{\let\do\@makeother \do\$\do\&\do\#\do\^\do\_\do\%\do\~}
\def\mn@doi{\begingroup\mn@urlcharsother \@ifnextchar [ {\mn@doi@}
  {\mn@doi@[]}}
\def\mn@doi@[#1]#2{\def\@tempa{#1}\ifx\@tempa\@empty \href
  {http://dx.doi.org/#2} {doi:#2}\else \href {http://dx.doi.org/#2} {#1}\fi
  \endgroup}
\def\mn@eprint#1#2{\mn@eprint@#1:#2::\@nil}
\def\mn@eprint@arXiv#1{\href {http://arxiv.org/abs/#1} {{\tt arXiv:#1}}}
\def\mn@eprint@dblp#1{\href {http://dblp.uni-trier.de/rec/bibtex/#1.xml}
  {dblp:#1}}


\bibitem[\protect\citeauthoryear{Allers \& Liu}{Allers \&
  Liu}{2013}]{Allers2013}
Allers K.~N.,  Liu M.~C.,  2013, \mn@doi [The Astrophysical Journal]
  {10.1088/0004-637X/772/2/79}, 772, 79

\bibitem[\protect\citeauthoryear{Allers, Gallimore, Liu  \& Dupuy}{Allers
  et~al.}{2016}]{Allers2016}
Allers K.~N.,  Gallimore J.~F.,  Liu M.~C.,   Dupuy T.~J.,  2016, \mn@doi [The
  Astrophysical Journal] {10.3847/0004-637X/819/2/133}, 819, 133

\bibitem[\protect\citeauthoryear{Apai, Radigan, Buenzli, Burrows, Reid  \&
  Jayawardhana}{Apai et~al.}{2013}]{Apai2013}
Apai D.,  Radigan J.,  Buenzli E.,  Burrows A.,  Reid I.~N.,   Jayawardhana R.,
   2013, \mn@doi [The Astrophysical Journal] {10.1088/0004-637X/768/2/121},
  768, 121

\bibitem[\protect\citeauthoryear{Apai et~al.,}{Apai et~al.}{2016}]{Apai2016}
Apai D.,  et~al., 2016, \mn@doi [The Astrophysical Journal]
  {10.3847/0004-637X/820/1/40}, 820, 40

\bibitem[\protect\citeauthoryear{Apai et~al.,}{Apai et~al.}{2017}]{Apai2017}
Apai D.,  et~al., 2017, \mn@doi [Science] {10.1126/science.aam9848}, 357, 683

\bibitem[\protect\citeauthoryear{Artigau, Bouchard, Doyon  \&
  Lafreni{\`{e}}re}{Artigau et~al.}{2009}]{Artigau2009}
Artigau {\'{E}}.,  Bouchard S.,  Doyon R.,   Lafreni{\`{e}}re D.,  2009,
  \mn@doi [The Astrophysical Journal] {10.1088/0004-637X/701/2/1534}, 701, 1534

\bibitem[\protect\citeauthoryear{Barenfeld, Bubar, Mamajek  \& Young}{Barenfeld
  et~al.}{2013}]{Barenfeld2013}
Barenfeld S.~A.,  Bubar E.~J.,  Mamajek E.~E.,   Young P.~A.,  2013, \mn@doi
  [The Astrophysical Journal] {10.1088/0004-637X/766/1/6}, 766, 6

\bibitem[\protect\citeauthoryear{Bell, Mamajek  \& Naylor}{Bell
  et~al.}{2015}]{Bell2015}
Bell C. P.~M.,  Mamajek E.~E.,   Naylor T.,  2015, \mn@doi [Monthly Notices of
  the Royal Astronomical Society] {10.1093/mnras/stv1981}, 454, 593

\bibitem[\protect\citeauthoryear{Best et~al.,}{Best et~al.}{2015}]{Best2015}
Best W. M.~J.,  et~al., 2015, \mn@doi [The Astrophysical Journal]
  {10.1088/0004-637X/814/2/118}, 814, 118

\bibitem[\protect\citeauthoryear{Biller, Allers, Liu, Close  \& Dupuy}{Biller
  et~al.}{2011}]{Biller2011}
Biller B.~A.,  Allers K.,  Liu M.~C.,  Close L.~M.,   Dupuy T.~J.,  2011,
  \mn@doi [The Astrophysical Journal] {10.1088/0004-637X/730/1/39}, 730, 39

\bibitem[\protect\citeauthoryear{Biller et~al.,}{Biller
  et~al.}{2013}]{Biller2013}
Biller B.~A.,  et~al., 2013, \mn@doi [The Astrophysical Journal]
  {10.1088/2041-8205/778/1/L10}, 778, L10

\bibitem[\protect\citeauthoryear{Biller et~al.,}{Biller
  et~al.}{2015}]{Biller2015}
Biller B.~A.,  et~al., 2015, \mn@doi [Astrophysical Journal Letters]
  {10.1088/2041-8205/813/2/L23}, 813, 1

\bibitem[\protect\citeauthoryear{Biller et~al.,}{Biller
  et~al.}{2018}]{Biller2018}
Biller B.~A.,  et~al., 2018, \mn@doi [The Astronomical Journal]
  {10.3847/1538-3881/aaa5a6}, 155, 95

\bibitem[\protect\citeauthoryear{Bonavita, de Mooij  \& Jayawardhana}{Bonavita
  et~al.}{2013}]{Bonavita2013}
Bonavita M.,  de Mooij E. J.~W.,   Jayawardhana R.,  2013, \mn@doi
  [Publications of the Astronomical Society of the Pacific] {10.1086/671758},
  125, 849

\bibitem[\protect\citeauthoryear{Bonavita, Desidera, Thalmann, Janson, Vigan,
  Chauvin  \& Lannier}{Bonavita et~al.}{2016}]{Bonavita2016}
Bonavita M.,  Desidera S.,  Thalmann C.,  Janson M.,  Vigan A.,  Chauvin G.,
  Lannier J.,  2016, \mn@doi [Astronomy {\&} Astrophysics]
  {10.1051/0004-6361/201628231}, 593, A38

\bibitem[\protect\citeauthoryear{Bonnefoy et~al.,}{Bonnefoy
  et~al.}{2016}]{Bonnefoy2016}
Bonnefoy M.,  et~al., 2016, \mn@doi [Astronomy {\&} Astrophysics]
  {10.1051/0004-6361/201526906}, 587, A58

\bibitem[\protect\citeauthoryear{Bouy, Brandner, Martı, Delfosse  \&
  Allard}{Bouy et~al.}{2003}]{Bouy2003}
Bouy H.,  Brandner W.,  Martı E.~L.,  Delfosse X.,   Allard F.,  2003, \mn@doi
  [The Astronomical Journal] {10.1086/377343}, pp 1526--1554

\bibitem[\protect\citeauthoryear{Buenzli et~al.,}{Buenzli
  et~al.}{2012}]{Buenzli2012}
Buenzli E.,  et~al., 2012, \mn@doi [The Astrophysical Journal]
  {10.1088/2041-8205/760/2/L31}, 760, L31

\bibitem[\protect\citeauthoryear{Buenzli, Apai, Radigan, Reid  \&
  Flateau}{Buenzli et~al.}{2014}]{Buenzli2014}
Buenzli E.,  Apai D.,  Radigan J.,  Reid I.~N.,   Flateau D.,  2014, \mn@doi
  [The Astrophysical Journal] {10.1088/0004-637X/782/2/77}, 782, 77

\bibitem[\protect\citeauthoryear{Buenzli, Saumon, Marley, Apai, Radigan, Bedin,
  Reid  \& Morley}{Buenzli et~al.}{2015}]{Buenzli2015a}
Buenzli E.,  Saumon D.,  Marley M.~S.,  Apai D.,  Radigan J.,  Bedin L.~R.,
  Reid I.~N.,   Morley C.~V.,  2015, \mn@doi [The Astrophysical Journal]
  {10.1088/0004-637X/798/2/127}, 798, 127

\bibitem[\protect\citeauthoryear{Casali et~al.,}{Casali
  et~al.}{2007}]{Casali2007}
Casali M.,  et~al., 2007, \mn@doi [Astronomy {\&} Astrophysics]
  {10.1051/0004-6361:20066514}, 467, 777

\bibitem[\protect\citeauthoryear{Casewell, Jameson  \& Burleigh}{Casewell
  et~al.}{2008}]{Casewell2008}
Casewell S.~L.,  Jameson R.~F.,   Burleigh M.~R.,  2008, \mn@doi [Monthly
  Notices of the Royal Astronomical Society]
  {10.1111/j.1365-2966.2008.13855.x}, 390, 1517

\bibitem[\protect\citeauthoryear{Croll, Muirhead, Lichtman, Han, Dalba  \&
  Radigan}{Croll et~al.}{2016}]{Croll2016a}
Croll B.,  Muirhead P.~S.,  Lichtman J.,  Han E.,  Dalba P.~A.,   Radigan J.,
  2016

\bibitem[\protect\citeauthoryear{Cruz, Reid, Liebert, Kirkpatrick  \&
  Lowrance}{Cruz et~al.}{2003}]{Cruz2003}
Cruz K.~L.,  Reid I.~N.,  Liebert J.,  Kirkpatrick J.~D.,   Lowrance P.~J.,
  2003, \mn@doi [The Astronomical Journal] {10.1086/378607}, 126, 2421

\bibitem[\protect\citeauthoryear{Cruz et~al.,}{Cruz et~al.}{2007}]{Cruz2007}
Cruz K.~L.,  et~al., 2007, \mn@doi [The Astronomical Journal] {10.1086/510132},
  133, 439

\bibitem[\protect\citeauthoryear{Cruz, Kirkpatrick  \& Burgasser}{Cruz
  et~al.}{2009}]{Cruz2009}
Cruz K.~L.,  Kirkpatrick J.~D.,   Burgasser A.~J.,  2009, \mn@doi [The
  Astronomical Journal] {10.1088/0004-6256/137/2/3345}, 137, 3345

\bibitem[\protect\citeauthoryear{Dupuy \& Liu}{Dupuy \& Liu}{2012}]{Dupuy2012}
Dupuy T.~J.,  Liu M.~C.,  2012, \mn@doi [The Astrophysical Journal Supplement
  Series] {10.1088/0067-0049/201/2/19}, 201, 19

\bibitem[\protect\citeauthoryear{Eriksson}{Eriksson}{2016}]{Eriksson2016}
Eriksson S.,  2016, MSc thesis, Stockholm University,  \url
  {http://su.diva-portal.org/smash/record.jsf?pid=diva2{\%}3A1178019{\&}dswid=2943}

\bibitem[\protect\citeauthoryear{Faherty et~al.,}{Faherty
  et~al.}{2012}]{Faherty2012}
Faherty J.~K.,  et~al., 2012, \mn@doi [The Astrophysical Journal]
  {10.1088/0004-637X/752/1/56}, 752, 56

\bibitem[\protect\citeauthoryear{Faherty et~al.,}{Faherty
  et~al.}{2016}]{Faherty2016}
Faherty J.~K.,  et~al., 2016, \mn@doi [The Astrophysical Journal Supplement
  Series] {10.3847/0067-0049/225/1/10}, 225, 1

\bibitem[\protect\citeauthoryear{Foreman-Mackey, Hogg, Lang  \&
  Goodman}{Foreman-Mackey et~al.}{2013}]{fm2013}
Foreman-Mackey D.,  Hogg D.~W.,  Lang D.,   Goodman J.,  2013, \mn@doi
  [Publications of the Astronomical Society of the Pacific] {10.1086/670067},
  125, 306

\bibitem[\protect\citeauthoryear{Gagn{\'{e}}, Lafreni{\`{e}}re, Doyon, Malo  \&
  Artigau}{Gagn{\'{e}} et~al.}{2014a}]{Gagne2014a}
Gagn{\'{e}} J.,  Lafreni{\`{e}}re D.,  Doyon R.,  Malo L.,   Artigau {\'{E}}.,
  2014a, \mn@doi [The Astrophysical Journal] {10.1088/0004-637X/783/2/121},
  783, 121

\bibitem[\protect\citeauthoryear{Gagn{\'{e}}, Faherty, Cruz, Lafreni{\`{e}}re,
  Doyon, Malo  \& Artigau}{Gagn{\'{e}} et~al.}{2014b}]{Gagne2014b}
Gagn{\'{e}} J.,  Faherty J.~K.,  Cruz K.,  Lafreni{\`{e}}re D.,  Doyon R.,
  Malo L.,   Artigau {\'{E}}.,  2014b, \mn@doi [The Astrophysical Journal]
  {10.1088/2041-8205/785/1/L14}, 785, L14

\bibitem[\protect\citeauthoryear{Gagn{\'{e}}, Lafreni{\`{e}}re, Doyon, Artigau,
  Malo, Robert  \& Nadeau}{Gagn{\'{e}} et~al.}{2014c}]{Gagne2014c}
Gagn{\'{e}} J.,  Lafreni{\`{e}}re D.,  Doyon R.,  Artigau E.,  Malo L.,  Robert
  J.,   Nadeau D.,  2014c, \mn@doi [Astrophysical Journal Letters]
  {10.1088/2041-8205/792/1/L17}, 792

\bibitem[\protect\citeauthoryear{Gagn{\'{e}} et~al.,}{Gagn{\'{e}}
  et~al.}{2015a}]{Gagne2015c}
Gagn{\'{e}} J.,  et~al., 2015a, \mn@doi [The Astrophysical Journal Supplement
  Series] {10.1088/0067-0049/219/2/33}, 219, 33

\bibitem[\protect\citeauthoryear{Gagn{\'{e}}, Lafreni{\`{e}}re, Doyon, Malo  \&
  Artigau}{Gagn{\'{e}} et~al.}{2015b}]{Gagne2015a}
Gagn{\'{e}} J.,  Lafreni{\`{e}}re D.,  Doyon R.,  Malo L.,   Artigau {\'{E}}.,
  2015b, \mn@doi [The Astrophysical Journal] {10.1088/0004-637X/798/2/73}, 798,
  73

\bibitem[\protect\citeauthoryear{Gagn{\'{e}}, Burgasser, Faherty,
  Lafreni{\'{e}}re, Doyon, Filippazzo, Bowsher  \& Nicholls}{Gagn{\'{e}}
  et~al.}{2015c}]{Gagne2015b}
Gagn{\'{e}} J.,  Burgasser A.~J.,  Faherty J.~K.,  Lafreni{\'{e}}re D.,  Doyon
  R.,  Filippazzo J.~C.,  Bowsher E.,   Nicholls C.~P.,  2015c, \mn@doi [The
  Astrophysical Journal] {10.1088/2041-8205/808/1/L20}, 808, L20

\bibitem[\protect\citeauthoryear{Gagn{\'{e}} et~al.,}{Gagn{\'{e}}
  et~al.}{2017}]{Gagne2017}
Gagn{\'{e}} J.,  et~al., 2017, \mn@doi [The Astrophysical Journal]
  {10.3847/2041-8213/aa70e2}, 841, L1

\bibitem[\protect\citeauthoryear{Gagn{\'{e}}, Allers, Theissen, Faherty,
  Gagliuffi  \& Artigau}{Gagn{\'{e}} et~al.}{2018a}]{Gagne2018a}
Gagn{\'{e}} J.,  Allers K.~N.,  Theissen C.~A.,  Faherty J.~K.,  Gagliuffi
  D.~B.,   Artigau {\'{E}}.,  2018a, \mn@doi [The Astrophysical Journal]
  {10.3847/2041-8213/aaacfd}, 854, L27

\bibitem[\protect\citeauthoryear{Gagn{\'{e}} et~al.,}{Gagn{\'{e}}
  et~al.}{2018b}]{Gagne2018b}
Gagn{\'{e}} J.,  et~al., 2018b, \mn@doi [The Astrophysical Journal]
  {10.3847/1538-4357/aaae09}, 856, 23

\bibitem[\protect\citeauthoryear{Gelino, Marley, Holtzman, Ackerman  \&
  Lodders}{Gelino et~al.}{2002}]{Gelino2002}
Gelino C.~R.,  Marley M.~S.,  Holtzman J.~A.,  Ackerman A.~S.,   Lodders K.,
  2002, \mn@doi [The Astrophysical Journal] {10.1086/342150}, 577, 433

\bibitem[\protect\citeauthoryear{Heinze, Metchev  \& Kellogg}{Heinze
  et~al.}{2015}]{Heinze2015}
Heinze A.~N.,  Metchev S.,   Kellogg K.,  2015, \mn@doi [Astrophysical Journal]
  {10.1088/0004-637X/801/2/104}, 801

\bibitem[\protect\citeauthoryear{Hodgkin, Irwin, Hewett  \& Warren}{Hodgkin
  et~al.}{2009}]{Hodgkin2009}
Hodgkin S.~T.,  Irwin M.~J.,  Hewett P.~C.,   Warren S.~J.,  2009, \mn@doi
  [Monthly Notices of the Royal Astronomical Society]
  {10.1111/j.1365-2966.2008.14387.x}, 394, 675

\bibitem[\protect\citeauthoryear{Irwin et~al.,}{Irwin et~al.}{2004}]{Irwin2004}
Irwin M.~J.,  et~al., 2004, \mn@doi [Optimizing Scientific Return for Astronomy
  through Information Technologies. Edited by Quinn] {10.1117/12.551449}, 5493,
  411

\bibitem[\protect\citeauthoryear{Irwin, Hodgkin, Aigrain, Bouvier, Hebb  \&
  Moraux}{Irwin et~al.}{2008}]{Irwin2008}
Irwin J.,  Hodgkin S.,  Aigrain S.,  Bouvier J.,  Hebb L.,   Moraux E.,  2008,
  \mn@doi [Monthly Notices of the Royal Astronomical Society]
  {10.1111/j.1365-2966.2007.12669.x}, 383, 1588

\bibitem[\protect\citeauthoryear{Kirkpatrick}{Kirkpatrick}{2005}]{Kirkpatrick2005}
Kirkpatrick J.~D.,  2005, \mn@doi [Annual Review of Astronomy and Astrophysics]
  {10.1146/annurev.astro.42.053102.134017}, 43, 195

\bibitem[\protect\citeauthoryear{Kirkpatrick et~al.,}{Kirkpatrick
  et~al.}{2000}]{Kirkpatrick2000}
Kirkpatrick J.~D.,  et~al., 2000, \mn@doi [The Astronomical Journal]
  {10.1086/301427}, 120, 447

\bibitem[\protect\citeauthoryear{Kirkpatrick, Barman, Burgasser, McGovern,
  McLean, Tinney  \& Lowrance}{Kirkpatrick et~al.}{2006}]{Kirkpatrick2006}
Kirkpatrick J.~D.,  Barman T.~S.,  Burgasser A.~J.,  McGovern M.~R.,  McLean
  I.~S.,  Tinney C.~G.,   Lowrance P.~J.,  2006, \mn@doi [Astrophysical
  Journal] {10.1086/499622}, 639, 1120

\bibitem[\protect\citeauthoryear{Kirkpatrick et~al.,}{Kirkpatrick
  et~al.}{2008}]{Kirkpatrick2008}
Kirkpatrick J.~D.,  et~al., 2008, \mn@doi [The Astrophysical Journal]
  {10.1086/592768}, 689, 1295

\bibitem[\protect\citeauthoryear{Kirkpatrick et~al.,}{Kirkpatrick
  et~al.}{2010}]{Kirkpatrick2010}
Kirkpatrick J.~D.,  et~al., 2010, \mn@doi [The Astrophysical Journal Supplement
  Series] {10.1088/0067-0049/190/1/100}, 190, 100

\bibitem[\protect\citeauthoryear{Lafreniere et~al.,}{Lafreniere
  et~al.}{2007}]{Lafreniere2007}
Lafreniere D.,  et~al., 2007, \mn@doi [The Astrophysical Journal]
  {10.1086/522826}, 670, 1367

\bibitem[\protect\citeauthoryear{Lew et~al.,}{Lew et~al.}{2016}]{Lew2016}
Lew B. W.~P.,  et~al., 2016, \mn@doi [The Astrophysical Journal]
  {10.3847/2041-8205/829/2/L32}, 829, L32

\bibitem[\protect\citeauthoryear{Liu et~al.,}{Liu et~al.}{2013}]{Liu2013}
Liu M.~C.,  et~al., 2013, \mn@doi [The Astrophysical Journal]
  {10.1088/2041-8205/777/2/L20}, 777, L20

\bibitem[\protect\citeauthoryear{Liu, Dupuy  \& Allers}{Liu
  et~al.}{2016}]{Liu2016}
Liu M.~C.,  Dupuy T.~J.,   Allers K.~N.,  2016, \mn@doi [The Astrophysical
  Journal] {10.3847/1538-4357/833/1/96}, 833, 96

\bibitem[\protect\citeauthoryear{Luhman et~al.,}{Luhman
  et~al.}{2007}]{Luhman2007}
Luhman K.~L.,  et~al., 2007, \mn@doi [The Astrophysical Journal]
  {10.1086/509073}, 654, 570

\bibitem[\protect\citeauthoryear{Malo, Doyon, Lafreni{\`{e}}re, Artigau,
  Gagn{\'{e}}, Baron  \& Riedel}{Malo et~al.}{2013}]{Malo2013}
Malo L.,  Doyon R.,  Lafreni{\`{e}}re D.,  Artigau {\'{E}}.,  Gagn{\'{e}} J.,
  Baron F.,   Riedel A.,  2013, \mn@doi [The Astrophysical Journal]
  {10.1088/0004-637X/762/2/88}, 762, 88

\bibitem[\protect\citeauthoryear{Mamajek \& Bell}{Mamajek \&
  Bell}{2014}]{Mamajek2014}
Mamajek E.~E.,  Bell C. P.~M.,  2014, \mn@doi [Monthly Notices of the Royal
  Astronomical Society] {10.1093/mnras/stu1894}, 445, 2169

\bibitem[\protect\citeauthoryear{Manjavacas et~al.,}{Manjavacas
  et~al.}{2018}]{Manjavacas2018}
Manjavacas E.,  et~al., 2018, \mn@doi [The Astronomical Journal]
  {10.3847/1538-3881/aa984f}, 155, 11

\bibitem[\protect\citeauthoryear{Mart{\'{i}}n, Brandner, Bouy, Basri, Davis,
  Deshpande  \& Montgomery}{Mart{\'{i}}n et~al.}{2006}]{Martin2006}
Mart{\'{i}}n E.~L.,  Brandner W.,  Bouy H.,  Basri G.,  Davis J.,  Deshpande
  R.,   Montgomery M.~M.,  2006, \mn@doi [Astronomy {\&} Astrophysics]
  {10.1051/0004-6361:20054186}, 456, 253

\bibitem[\protect\citeauthoryear{McLean, McGovern, Burgasser, Kirkpatrick,
  Prato  \& Kim}{McLean et~al.}{2003}]{Mclean2003}
McLean I.~S.,  McGovern M.~R.,  Burgasser A.~J.,  Kirkpatrick J.~D.,  Prato L.,
    Kim S.~S.,  2003, \mn@doi [The Astrophysical Journal] {10.1086/377636},
  596, 561

\bibitem[\protect\citeauthoryear{Metchev et~al.,}{Metchev
  et~al.}{2015}]{Metchev2015a}
Metchev S.~A.,  et~al., 2015, \mn@doi [The Astrophysical Journal]
  {10.1088/0004-637X/799/2/154}, 799, 154

\bibitem[\protect\citeauthoryear{Miles-P{\'{a}}ez, Metchev, Heinze  \&
  Apai}{Miles-P{\'{a}}ez et~al.}{2017}]{Miles-Paez2017}
Miles-P{\'{a}}ez P.~A.,  Metchev S.~A.,  Heinze A.,   Apai D.,  2017, \mn@doi
  [The Astrophysical Journal] {10.3847/1538-4357/aa6f11}, 840, 83

\bibitem[\protect\citeauthoryear{Morales-Calderon et~al.,}{Morales-Calderon
  et~al.}{2006}]{Morales-Calderon2006}
Morales-Calderon M.,  et~al., 2006, \mn@doi [The Astrophysicsl Journal]
  {10.1086/507866}, 43, 544

\bibitem[\protect\citeauthoryear{Mortier, Faria, Correia, Santerne  \&
  Santos}{Mortier et~al.}{2015}]{Mortier2015}
Mortier A.,  Faria J.~P.,  Correia C.~M.,  Santerne A.,   Santos N.~C.,  2015,
  101, 1

\bibitem[\protect\citeauthoryear{Naud et~al.,}{Naud et~al.}{2014}]{Naud2014a}
Naud M.-E.,  et~al., 2014, \mn@doi [The Astrophysical Journal]
  {10.1088/0004-637X/787/1/5}, 787, 5

\bibitem[\protect\citeauthoryear{Naud, Artigau, Doyon, Malo, Gagn{\'{e}},
  Lafreni{\`{e}}re, Wolf  \& Magnier}{Naud et~al.}{2017}]{Naud2017}
Naud M.-e.,  Artigau {\'{E}}.,  Doyon R.,  Malo L.,  Gagn{\'{e}} J.,
  Lafreni{\`{e}}re D.,  Wolf C.,   Magnier E.~A.,  2017, \mn@doi [The
  Astronomical Journal] {10.3847/1538-3881/aa826b}, 154, 129

\bibitem[\protect\citeauthoryear{Radigan}{Radigan}{2014}]{Radigan2014a}
Radigan J.,  2014, \mn@doi [The Astrophysical Journal]
  {10.1088/0004-637X/797/2/120}, 797, 120

\bibitem[\protect\citeauthoryear{Radigan, Jayawardhana, Lafreni{\`{e}}re,
  Artigau, Marley  \& Saumon}{Radigan et~al.}{2012}]{Radigan2012}
Radigan J.,  Jayawardhana R.,  Lafreni{\`{e}}re D.,  Artigau {\'{E}}.,  Marley
  M.,   Saumon D.,  2012, \mn@doi [The Astrophysical Journal]
  {10.1088/0004-637X/750/2/105}, 750, 105

\bibitem[\protect\citeauthoryear{Radigan, Lafreni{\`{e}}re, Jayawardhana  \&
  Artigau}{Radigan et~al.}{2014}]{Radigan2014}
Radigan J.,  Lafreni{\`{e}}re D.,  Jayawardhana R.,   Artigau E.,  2014,
  \mn@doi [The Astrophysical Journal] {10.1088/0004-637X/793/2/75}, 793, 75

\bibitem[\protect\citeauthoryear{Reid, Cruz, Kirkpatrick, Allen, Mungall,
  Liebert, Lowrance  \& Sweet}{Reid et~al.}{2008}]{Reid2008}
Reid I.~N.,  Cruz K.~L.,  Kirkpatrick J.~D.,  Allen P.~R.,  Mungall F.,
  Liebert J.,  Lowrance P.,   Sweet A.,  2008, \mn@doi [The Astronomical
  Journal] {10.1086/517914}, 136, 1290

\bibitem[\protect\citeauthoryear{Riedel, Blunt, Lambrides, Rice, Cruz  \&
  Faherty}{Riedel et~al.}{2017}]{Riedel2017}
Riedel A.~R.,  Blunt S.~C.,  Lambrides E.~L.,  Rice E.~L.,  Cruz K.~L.,
  Faherty J.~K.,  2017, \mn@doi [The Astronomical Journal]
  {10.3847/1538-3881/153/3/95}, 153, 1

\bibitem[\protect\citeauthoryear{Rodriguez, Zuckerman, Kastner, Bessell,
  Faherty  \& Murphy}{Rodriguez et~al.}{2013}]{Rodriguez2013}
Rodriguez D.~R.,  Zuckerman B.,  Kastner J.~H.,  Bessell M.~S.,  Faherty J.~K.,
    Murphy S.~J.,  2013, \mn@doi [The Astrophysical Journal]
  {10.1088/0004-637X/774/2/101}, 774, 101

\bibitem[\protect\citeauthoryear{Schneider, Cushing, Kirkpatrick, Mace, Gelino,
  Faherty, Fajardo-Acosta  \& Sheppard}{Schneider et~al.}{2014}]{Schneider2014}
Schneider A.~C.,  Cushing M.~C.,  Kirkpatrick J.~D.,  Mace G.~N.,  Gelino
  C.~R.,  Faherty J.~K.,  Fajardo-Acosta S.,   Sheppard S.~S.,  2014, \mn@doi
  [The Astronomical Journal] {10.1088/0004-6256/147/2/34}, 147, 34

\bibitem[\protect\citeauthoryear{Shkolnik, Allers, Kraus, Liu  \&
  Flagg}{Shkolnik et~al.}{2017}]{Shkolnik2017}
Shkolnik E.~L.,  Allers K.~N.,  Kraus A.~L.,  Liu M.~C.,   Flagg L.,  2017,
  \mn@doi [The Astronomical Journal] {10.3847/1538-3881/aa77fa}, 154, 69

\bibitem[\protect\citeauthoryear{Snellen, Brandl, de Kok, Brogi, Birkby  \&
  Schwarz}{Snellen et~al.}{2014}]{Snellen2014}
Snellen I.,  Brandl B.,  de Kok R.,  Brogi M.,  Birkby J.,   Schwarz H.,  2014,
  \mn@doi [Nature] {10.1038/nature13253}, 509, 63

\bibitem[\protect\citeauthoryear{Torres, Quast, Melo  \& Sterzik}{Torres
  et~al.}{2008}]{Torres2008}
Torres C. a.~O.,  Quast G.~R.,  Melo C. H.~F.,   Sterzik M.~F.,  2008, Handbook
  of Star Forming Regions, II, 757

\bibitem[\protect\citeauthoryear{Vigan et~al.,}{Vigan et~al.}{2012}]{Vigan2012}
Vigan A.,  et~al., 2012, \mn@doi [Astronomy {\&} Astrophysics]
  {10.1051/0004-6361/201218991}, 544, A9

\bibitem[\protect\citeauthoryear{Vos, Allers  \& Biller}{Vos
  et~al.}{2017}]{Vos2017}
Vos J.~M.,  Allers K.~N.,   Biller B.~A.,  2017, \mn@doi [The Astrophysical
  Journal] {10.3847/1538-4357/aa73cf}, 842, 78

\bibitem[\protect\citeauthoryear{Vos, Allers, Biller, Liu, Dupuy, Gallimore,
  Adenuga  \& Best}{Vos et~al.}{2018}]{Vos2018}
Vos J.~M.,  Allers K.~N.,  Biller B.~A.,  Liu M.~C.,  Dupuy T.~J.,  Gallimore
  J.~F.,  Adenuga I.~J.,   Best W. M.~J.,  2018, \mn@doi [Monthly Notices of
  the Royal Astronomical Society] {10.1093/mnras/stx2752}, 474, 1041

\bibitem[\protect\citeauthoryear{Wilson, Rajan  \& Patience}{Wilson
  et~al.}{2014}]{Wilson2014}
Wilson P.~A.,  Rajan A.,   Patience J.,  2014, \mn@doi [Astronomy {\&}
  Astrophysics] {10.1051/0004-6361/201322995}, 566, A111

\bibitem[\protect\citeauthoryear{{Zapatero Osorio}, Martin, Bouy, Tata,
  Deshpande  \& Wainscoat}{{Zapatero Osorio} et~al.}{2006}]{ZapateroOsorio2006}
{Zapatero Osorio} M.~R.,  Martin E.~L.,  Bouy H.,  Tata R.,  Deshpande R.,
  Wainscoat R.~J.,  2006, \mn@doi [The Astrophysical Journal] {10.1086/505484},
  647, 1405

\bibitem[\protect\citeauthoryear{Zhou, Apai, Schneider, Marley  \&
  Showman}{Zhou et~al.}{2016}]{Zhou2016}
Zhou Y.,  Apai D.,  Schneider G.~H.,  Marley M.~S.,   Showman A.~P.,  2016,
  \mn@doi [The Astrophysical Journal] {10.3847/0004-637X/818/2/176}, 818, 176

\bibitem[\protect\citeauthoryear{Zhou et~al.,}{Zhou et~al.}{2018}]{Zhou2018}
Zhou Y.,  et~al., 2018

\bibitem[\protect\citeauthoryear{Zuckerman, Bessell, Song  \& Kim}{Zuckerman
  et~al.}{2006}]{Zuckerman2006}
Zuckerman B.,  Bessell M.~S.,  Song I.,   Kim S.,  2006, \mn@doi [The
  Astrophysical Journal] {10.1086/508060}, 649, L115

\makeatother
\end{thebibliography}







\bsp	
\label{lastpage}
\end{document}